\documentclass[sigconf,nonacm,fleqn]{acmart}
\usepackage{fancyhdr}
\usepackage{xcolor}
\usepackage{colortbl}
\usepackage{amsmath}
\usepackage{enumitem}
\usepackage{subfigure}
\usepackage{diagbox}
\usepackage[ruled,vlined,linesnumbered,noend]{algorithm2e}
\usepackage{tabu}
\usepackage{pifont}
\usepackage{array}
\usepackage{multirow}

\definecolor{Gray}{gray}{0.90}
\definecolor{Yellow}{rgb}{1,1,0.5}
\definecolor{Cyan}{rgb}{0.9,1,1}
\definecolor{Green}{rgb}{0.93,1,0.93}

\newcommand{\cg}{\cellcolor{Gray}}

\newcommand\vldbpagestyle{plain} 

\begin{document}

\newcommand\blfootnote[1]{%
    \begingroup
    \renewcommand\thefootnote{}
\footnote{#1}%
    \addtocounter{footnote}{-1}%
    \endgroup
}

\newcommand{\proofsketch}{{\sc Proof: }}
\newcommand{\myproof}{\noindent{\bf Proof: }}

\newcommand{\nthesection}{\arabic{section}}
\newcommand{\eot}{\hspace*{\fill}\mbox{$\Box$}}
\newcommand{\eop}{\hspace*{\fill}\mbox{$\Box$}}

\newcommand{\stitle}[1]{\vspace{0.2ex} \noindent{{\bf #1}}}

\newcommand{\itemsection}[1]{\vspace{0ex} \noindent{\textbf{$\bullet$ ~ #1}}}
\newcommand{\sstitle}[1]{\vspace{0.2ex} \noindent{{\textit{ #1}}}}
\newcommand{\ssstitle}[1]{\vspace{0.2ex} \noindent{\textit{ #1}}}

\newcommand{\red}[1]{\color{red}{#1}}

\newcommand{\kw}[1]{{\ensuremath {\mathsf{#1}}}\xspace}
\newcommand{\bkw}[1]{{\ensuremath {\mathsf{\textbf{#1}}}}\xspace}

\newcommand{\kwnospace}[1]{{\ensuremath {\mathsf{#1}}}}
\newcommand{\ltt}{\kw{LTT}}
\newcommand{\arr}{\kw{arrive}}
\newcommand{\vp}{\kw{p}}

\newcommand{\bellf}{{\sc Bellman-Ford}\xspace}
\newcommand{\bfalgo}{{\sc OR}\xspace}

\newcommand{\aalgo}{{\sc KDXZ}\xspace}

\newcommand{\dalgo}{{\sc Two-Step-LTT}\xspace}

\newcommand{\dtalgo}{{\sc DOT}\xspace}

\newcommand{\genfunc}{{\sl timeRefinement}\xspace}
\newcommand{\pathc}{{\sl pathSelection}\xspace}
\newcommand{\fifo}{{\sl FIFO}\xspace}

\newcommand{\sti}{starting-time interval\xspace}
\newcommand{\stsi}{starting-time subinterval\xspace}
\newcommand{\stsis}{starting-time subintervals\xspace}
\newcommand{\stis}{starting-time intervals\xspace}
\newcommand{\ti}{time interval\xspace}
\newcommand{\tis}{time intervals\xspace}
\newcommand{\ttime}{travel time\xspace}
\newcommand{\ttimea}{travel time\xspace}
\newcommand{\at}{arrival time\xspace}
\newcommand{\ata}{arrival-time\xspace}
\newcommand{\atf}{arrival-time function\xspace}
\newcommand{\ati}{arrival-time interval\xspace}
\newcommand{\ats}{arrival times\xspace}
\newcommand{\ed}{edge delay\xspace}
\newcommand{\eda}{edge-delay\xspace}
\newcommand{\edf}{edge-delay function\xspace}
\newcommand{\eds}{edge delays\xspace}
\newcommand{\wt}{waiting time\xspace}

\newcommand{\g}{\overline{g}}
\newcommand{\iend}{\tau}

\newcommand{\argmin}{\operatornamewithlimits{argmin}}

%%%%%%%%%%%%%%%%%%%%%%%%%%%%%%%%%%%%%%%%%% Defined in XML Paper %%%%%%%%%%%%%%%%%%%%%%%%%%%%%%%%%%%%%%%%%%%%%%
\newcommand{\myhead}[1]{\vspace{.05in} \noindent {\bf #1.}~~}
\newcommand{\cond}[1]{(\emph{#1})~}
\newcommand{\op}[1]{(\emph{#1})~}

\newcommand{\qc}{\ensuremath{Q^c}}

\newcommand{\rewrite}{\kw{XPathToReg}}

\newcommand{\upparen}[1]{\ensuremath{\mathrm{(}}{#1}\ensuremath{\mathrm{)}}}
\newcommand{\func}[2]{\funcname{#1}\upparen{\ensuremath{#2}}}
\newcommand{\funcname}[1]{\ensuremath{\mathit{#1}}}

\newcommand\AS{\textbf{as}\ }
\newcommand{\xsltsize}{\small}

\newcommand{\X}{{\cal X}}
\newcommand{\sem}[1]{[\![#1]\!]}
\newcommand{\NN}[2]{#1\sem{#2}}
\newcommand{\pcdata}{{\tt str}\xspace}

\newcommand{\exa}[2]{{\tt\begin{tabbing}\hspace{#1}\=\hspace{#1}\=\+\kill #2\end{tabbing}}}
\newcommand{\ra}{\rightarrow}
\newcommand{\la}{\leftarrow}
\newcommand{\rsa}{\_} % {\rightsquigarrow}
\newcommand{\Ed}[2]{E_{{\scriptsize \mbox{#1} \rsa \mbox{#2}}}}
\newenvironment{bi}{\begin{itemize}
        \setlength{\topsep}{0.5ex}\setlength{\itemsep}{0ex}\vspace{-0.6ex}}
        {\end{itemize}\vspace{-1ex}}
\newenvironment{be}{\begin{enumerate}
        \setlength{\topsep}{0.5ex}\setlength{\itemsep}{0ex}\vspace{-0.6ex}}
        {\end{itemize}\vspace{-1ex}}
\newcommand{\ei}{\end{itemize}}
\newcommand{\ee}{\end{enumerate}}

\newcommand{\mat}[2]{{\begin{tabbing}\hspace{#1}\=\+\kill #2\end{tabbing}}}
\newcommand{\m}{\hspace{0.05in}}
\newcommand{\ls}{\hspace{0.1in}}
\newcommand{\beqn}{\begin{eqnarray*}}
\newcommand{\eeqn}{\end{eqnarray*}}

\newcounter{ccc}
\newcommand{\bcc}{\setcounter{ccc}{1}\theccc.}
\newcommand{\icc}{\addtocounter{ccc}{1}\theccc.}

\newcommand{\oneurl}[1]{\texttt{#1}}
\newcommand{\tabstrut}{\rule{0pt}{4pt}\vspace{-0.1in}}
\newcommand{\tabstruct}{\rule{0pt}{8pt}\\[-2ex]}
\newcommand{\stab}{\rule{0pt}{8pt}\\[-2.2ex]}
\newcommand{\sstab}{\rule{0pt}{8pt}\\[-2.2ex]}

\newcommand{\eat}[1]{}

\newfloat{tcm}{thp}{loa}
\floatname{tcm}{Recursive \sql}

%\def\subfigcapskip{2pt}
% \def\subfigtopskip{0pt}
% \def\subfigbottomskip{0pt}

%%%%%%%%%%%%%%%%%%%%%%%%%%%%%%%%%%%%%%%%%% Defined in XML Paper (END) %%%%%%%%%%%%%%%%%%%%%%%%%%%%%%%%%%%%%%%%%%%%%%

\newcommand{\rdms}{{\sc rdbms}\xspace}
\newcommand{\sql}{{\sc sql}\xspace}
\newcommand{\dbms}{{\sc dbms}\xspace}

\newcommand{\cfig}{Figure~}
\newcommand{\ctab}{Table~}
\newcommand{\csec}{Section~}
\newcommand{\cdef}{Definition~}
\newcommand{\cthm}{Theorem~}
\newcommand{\clem}{Lemma~}
\newcommand{\cequ}[1]{Equation~(#1)}
\newcommand{\SG}{\mathbf{SG}}
\newcommand{\SA}{\mathbf{SA}}
\renewcommand{\AA}{\mathbf{AA}}

\newcommand{\xml}{{\sl XML}\xspace}
\newcommand{\xlink}{{\sl XLink}\xspace}
\newcommand{\xpath}{{\sl XPath}\xspace}
\newcommand{\xpointer}{{\sl XPointer}\xspace}
\newcommand{\rdf}{{\sl RDF}\xspace}
\newcommand{\tc}{{\sl TC}\xspace}
\newcommand{\dfs}{{\sl DFS}\xspace}
\newcommand{\DAG}{{\sl DAG}\xspace}
\newcommand{\DAGs}{{\sl DAG}s\xspace}
\newcommand{\grail}{{\sl GRAIL}\xspace}
\newcommand{\yesgrail}{{\sl Yes-GRAIL}\xspace}
\newcommand{\code}{\kw{code}}
\newcommand{\sit}{\kw{sit}}
\newcommand{\psit}{{\cal P}_{sit}}
\newcommand{\yescode}{{\sl Yes-Label}\xspace}
\newcommand{\nocode}{{\sl No-Label}\xspace}
\newcommand{\entry}{\kw{entry}\xspace}
\newcommand{\yngindex}{{\sl YNG-Index}\xspace}
\newcommand{\rqrun}{{\sl RQ-Run}\xspace}
\newcommand{\citeseerx}{{\sl citeseerx}\xspace}
\newcommand{\gouniprot}{{\sl go-uniprot}\xspace}
\newcommand{\uniprot}{{\sl uniprot150}\xspace}

\long\def\comment#1{}

\newcommand{\scc}{strongly connected component\xspace}
\newcommand{\sccs}{strongly connected components\xspace}
\newcommand{\sscc}{\kw{SCC}}
\newcommand{\ssccs}{\kwnospace{SCC}s\xspace}
\newcommand{\sccg}{\kwnospace{SCC}\textrm{-}\kw{Graph}}
\newcommand{\strongc}{\leftrightarrow}
\newcommand{\nstrongc}{\nleftrightarrow}
\newcommand{\emscc}{\kwnospace{EM}\textrm{-}\kw{SCC}}
\newcommand{\dfsscc}{\kwnospace{DFS}\textrm{-}\kw{SCC}}
\newcommand{\dfstree}{\kwnospace{DFS}\textrm{-}\kw{Tree}}
\newcommand{\len}{\kw{len}}
\newcommand{\dep}{\kw{depth}}
\newcommand{\tdep}{\kw{drank}}
\newcommand{\tlink}{\kw{dlink}}
\newcommand{\vedges}{up-edges\xspace}
\newcommand{\vedge}{up-edge\xspace}
\newcommand{\cvedge}{Up-Edge\xspace}

\newcommand{\drsscc}{\kwnospace{1P}\textrm{-}\kw{SCC}}
\newcommand{\drssccb}{\kwnospace{1PB}\textrm{-}\kw{SCC}}

\newcommand{\Bdrsscc}{\kwnospace{B}\textrm{-}\kwnospace{BR'}\textrm{-}\kw{SCC}}

\newcommand{\deprtree}{depth-ranked tree\xspace}
\newcommand{\cdeprtree}{Depth-Ranked Tree\xspace}
\newcommand{\drtree}{\kwnospace{BR}\textrm{-}\kw{Tree}}
\newcommand{\drplustree}{\kwnospace{BR}$^+$\textrm{-}\kw{Tree}}
\newcommand{\drscc}{\kwnospace{2P}\textrm{-}\kw{SCC}}
\newcommand{\updatedrank}{\kwnospace{update}\textrm{-}\kw{drank}}

\newcommand{\drtreeconstruct}{\kwnospace{Tree}\textrm{-}\kw{Construction}}
\newcommand{\drtreesearch}{\kwnospace{Tree}\textrm{-}\kw{Search}}
\newcommand{\depthrerank}{\kw{pushdown}}
\newcommand{\itrerank}{\kwnospace{iterative}\textrm{-}\kw{rerank}}
\newcommand{\drr}{\Downarrow}
\newcommand{\reach}{\kw{Rset}}
\newcommand{\earlyrejection}{\kwnospace{early}\textrm{-}\kw{rejection}}
\newcommand{\earlyacceptance}{\kwnospace{early}\textrm{-}\kw{acceptance}}
\newcommand{\greduce}{\earlyacceptance}
\newcommand{\drea}{\kwnospace{1P}\textrm{/}\kw{ER}}
\newcommand{\myinf}{\kw{INF}}

%==========defined in GCloud ========
\newcommand{\gcloud}{\kw{GCloud}\xspace}
\newcommand{\degree}{\kw{Degree}\xspace}
\newcommand{\subgraph}{\kw{Subgraph}\xspace}
\newcommand{\pagerank}{\kw{PageRank}\xspace}
\newcommand{\bfs}{\kw{BFS}\xspace}
\newcommand{\keysearch}{\kw{KWS}\xspace}
\newcommand{\cc}{\kw{CC}\xspace}
\newcommand{\msf}{\kw{MSF}\xspace}
\newcommand{\dmax}{\kw{rmax}}
\newcommand{\mystar}{\kw{star}\xspace}
\newcommand{\twitter}{{\sl{Twitter-2010}}\xspace}
\newcommand{\friendster}{{\sl{Friendster}}\xspace}
\definecolor{lgray}{gray}{0.85}
\definecolor{llgray}{gray}{0.9}
\newcommand{\mycc}{CC\xspace}
\newcommand{\myccs}{CCs\xspace}
\newcommand{\mymsf}{MSF\xspace}
\newcommand{\oneroundmsf}{\kw{OneRoundMSF}}
\newcommand{\multiroundmsf}{\kw{MultiRoundMSF}}

\newcommand{\hashtomin}{\kw{HashToMin}\xspace}
\newcommand{\hashgreatertomin}{\kw{HashGToMin}\xspace}
\newcommand{\pramsimulation}{\kwnospace{PRAM}\textrm{-}\kw{Simulation}\xspace}

\newcommand{\pagerankpig}{\kwnospace{PageRank}\textrm{-}\kw{Pig}\xspace}
\newcommand{\bfspig}{\kwnospace{BFS}\textrm{-}\kw{Pig}\xspace}
\newcommand{\keysearchpig}{\kwnospace{KWS}\textrm{-}\kw{Pig}\xspace}

%=============defined in SubgEnum======
\newcommand{\ttwig}{\kwnospace{TwinTwig}\xspace}
\newcommand{\ttwigs}{\kwnospace{TwinTwig}s\xspace}
\newcommand{\ttjoin}{\kwnospace{TwinTwig}\kw{Join}}
\newcommand{\sdec}{\kwnospace{SDEC}\xspace}
\newcommand{\subgenum}{\kwnospace{SubgraphEnum}\xspace}
\newcommand{\mymap}{\kwnospace{map}\xspace}
\newcommand{\myreduce}{\kwnospace{reduce}\xspace}
\newcommand{\cascadejoin}{\kwnospace{Edge}\kw{Join}}
\newcommand{\starjoin}{\kwnospace{Star}\kw{Join}}
\newcommand{\multiwayjoin}{\kwnospace{Multiway}\kw{Join}}
\newcommand{\cost}{\kwnospace{cost}\xspace}
\newcommand{\mysize}{\kwnospace{card}\xspace}
\newcommand{\er}{\kwnospace{ER}\xspace}
\newcommand{\optdec}{\kwnospace{Optimal}\textrm{-}\kwnospace{Decomp}\xspace}

\newcommand{\ttone}{\kwnospace{TT1}\xspace}
\newcommand{\tttwo}{\kwnospace{TT2}\xspace}
\newcommand{\ttthree}{\kwnospace{TT3}\xspace}

\newcommand{\alEdge}{\kwnospace{Edge}\xspace}
\newcommand{\alMul}{\kwnospace{Mul}\xspace}
\newcommand{\alStar}{\kwnospace{Star}\xspace}

\newcommand{\alTTBO}{\kwnospace{TTBS}\xspace}
\newcommand{\alTTNLB}{\kwnospace{TTOA}\xspace}
\newcommand{\alTTLB}{\kwnospace{TTLB}\xspace}
\newcommand{\alTTFil}{\kwnospace{TT}\xspace}

%=============defined for reference======
\newcommand{\reffig}[1]{Figure~\ref{fig:#1}}
\newcommand{\refsec}[1]{Section~\ref{sec:#1}}
\newcommand{\reftable}[1]{Table~\ref{tab:#1}}
\newcommand{\refalg}[1]{Algorithm~\ref{alg:#1}}
\newcommand{\refeq}[1]{Equation~\ref{eq:#1}}
\newcommand{\refdef}[1]{Definition~\ref{def:#1}}
\newcommand{\refthm}[1]{Theorem~\ref{thm:#1}}
\newcommand{\reflem}[1]{Lemma~\ref{lem:#1}}
\newcommand{\refcor}[1]{Corollary~\ref{cor:#1}}
\newcommand{\refex}[1]{Example~\ref{ex:#1}}
\newcommand{\refassum}[1]{Assumption~\ref{assum:#1}}
\newcommand{\refpro}[1]{Property~\ref{pro:#1}}
\newcommand{\refobv}[1]{Observation~\ref{obv:#1}}
\newcommand{\refproof}[1]{Proof~\ref{proof:#1}}

\makeatletter
\newcommand{\rmnum}[1]{\romannumeral #1}
\newcommand{\Rmnum}[1]{\expandafter\@slowromancap\romannumeral #1@}
\makeatother

%===============defined for DivKClique======

%================defined for dynamic========
\newcommand{\rebuild}{\kw{Rebuild}}
\newcommand{\BatchUpdatingAlgorithm}{\text{Batch Updating Algorithm}}
\newcommand{\BatchUpdatingProcessing}{\text{Batch Updating Processing}}
\newcommand{\BU}{\text{BU}}
\newcommand{\CO}{\text{Contraction Operation}}
\newcommand{\dynch}{\kw{DynCH}}
\newcommand{\dynchvcs}{\kw{DCH_{vcs}}}
\newcommand{\dynchvcb}{\kw{DCH_{vcb}}}
\newcommand{\dynchscs}{\kw{DCH_{scs}}}
\newcommand{\dynchscswinc}{\kwnospace{DCH_{scs}}\textrm{-}\kw{WInc}}
\newcommand{\dynchscswincp}{\kwnospace{DCH_{scs}}\textrm{-}\kw{WInc}}
\newcommand{\dynchscswincpp}{\kwnospace{DCH_{scs}^+}\textrm{-}\kw{WInc}}
\newcommand{\dynchscswincn}{\kwnospace{DCH_{scs}}\textrm{-}\kw{WIncDirect}}
\newcommand{\dynchscswdec}{\kwnospace{DCH_{scs}}\textrm{-}\kw{WDec}}
\newcommand{\dynchscswdecpp}{\kwnospace{DCH_{scs}^+}\textrm{-}\kw{WDec}}
\newcommand{\dynchscbwdec}{\kwnospace{DCH_{scb}}\textrm{-}\kw{WDec}}
\newcommand{\dynchscbwdecp}{\kwnospace{DCH_{scb}^+}\textrm{-}\kw{WDec}}
\newcommand{\dynchscbwdecn}{\kwnospace{DCH_{scb}}\textrm{-}\kw{WDecDirect}}
\newcommand{\dynchscbwinc}{\kwnospace{DCH_{scb}}\textrm{-}\kw{WInc}}
\newcommand{\dynchscbwincp}{\kwnospace{DCH_{scb}^+}\textrm{-}\kw{WInc}}
\newcommand{\dynchscb}{\kw{DCH_{scb}}}
\newcommand{\dynchscbn}{\kwnospace{DCH_{scb}}\textrm{-}\kw{Naive}}
\newcommand{\dynchscbp}{\kw{DCH^+_{scb}}}
\newcommand{\CHI}{\text{CH Index}\xspace}
\newcommand{\scgraph}{\text{$G_{sc}$}\xspace}
\newcommand{\RedOrd}{\text{reduce order}\xspace}
\newcommand{\ch}{\kw{CH}}
\newcommand{\gp}{{G'_{\oplus(e, k)}}}
\newcommand{\minw}{\kw{minWeight}}
\newcommand{\minwp}{\kw{minWeight^+}}
\newcommand{\ContractionHierarchies}{\text{Contraction Hierarchies}}
\newcommand{\RedPos}{\text{Pos}\xspace}
\newcommand{\ContractionHierarchy}{\text{Contraction Hierarchy}}
\newcommand{\shortcutindex}{\emph{shortcut index}\xspace}
\newcommand{\ssgraph}{\text{SS}\textrm{-}\text{Graph}\xspace}

\newcommand{\PartRelationshipEdge}{\text{part relationship edge}\xspace}
\newcommand{\UpdatingRelationshipGraph}{\text{updating relationship Graph}\xspace}
\newcommand{\OnLineAlgorithm}{\text{on-line query algorithm}\xspace}
\newcommand{\OffLineAlgorithm}{\text{index algorithm}\xspace}
\newcommand{\RoadUpdatingProcessing}{\text{road updating processing}\xspace}
\newcommand{\ShortcutRelationshipGraph}{\text{shortcut relationship graph}\xspace}
\newcommand{\SRGraph}{\text{SR-Graph}\xspace}
\newcommand{\cun}{\kw{CUN}}
\newcommand{\nbr}{\kw{nbr}\xspace}
\newcommand{\nbrin}{\kw{nbr^-}}
\newcommand{\nbrout}{\kw{nbr^+}}
\newcommand{\mydeg}{\kw{deg}}
\newcommand{\myid}{\kw{ID}}
\newcommand{\CompactShortcutRelationGraph}{\text{compact shortcut relation graph}\xspace}
\newcommand{\Effect}{\text{effect}\xspace}
\newcommand{\EffectSet}{\text{effect set}\xspace}
\newcommand{\GenerationSet}{\text{generation set}\xspace}
\newcommand{\DirectedRelation}{\text{directed relation}\xspace}
\newcommand{\UndirectedRelation}{\text{undirected relation}\xspace}
\newcommand{\RoadNetwork}{\text{road network}\xspace}
\newcommand{\DependentRelationship}{\text{dependent relationship}\xspace}
\newcommand{\AffectRelationship}{\text{affect relationship}\xspace}
\newcommand{\AffectRelationshipGraph}{\text{affect relationship graph}\xspace}
\newcommand{\CandidateRelationshipEdge}{\text{candidate relation edge}\xspace}

\newcommand{\ARG}{\text{$ARG$}\xspace}
\newcommand{\CalculateAffectSC}{\text{casc}\xspace}
\newcommand{\CalculateDependentSC}{\text{$cdsc$}\xspace}
\newcommand{\Calculation}{\text{Calculation}\xspace}
\newcommand{\DistancePreservedGraph}{\text{distance-preserved graph}\xspace}
\newcommand{\AffectCalculationSet}{\text{affected calculation set}\xspace}
\newcommand{\UnaffectCalculationSet}{\text{unaffected calculation set}\xspace}
\newcommand{\ACS}{\text{$\mathcal{ACS}$}\xspace}
\newcommand{\UCS}{\text{$\mathcal{UCS}$}\xspace}
\newcommand{\DependentCalculationSet}{\text{dependent calculation set}\xspace}
\newcommand{\DCS}{\text{$\mathcal{DCS}$}\xspace}
\newcommand{\CHNeighbourSet}{\text{CH neightbour set}\xspace}
\newcommand{\CHNS}{\text{$\mathcal{CHNS}$}\xspace}
\newcommand{\DirectedAffectCalculationSet}{\text{Directed Affect set}\xspace}
\newcommand{\DirectedCandidateSet}{\text{directed candidate set}\xspace}
\newcommand{\DACS}{\text{$\mathcal{DAS}$}\xspace}
\newcommand{\ShortcutCombination}{\text{shortcut combination}\xspace}
\newcommand{\CandidateSet}{\text{candidate set}\xspace}
\newcommand{\CS}{\text{$\mathcal{CS}$}\xspace}
\newcommand{\AffectShortcut}{\text{affect shortcut}\xspace}

\newcommand{\collectw}{\kw{collectWeight}}
\newcommand{\updatew}{\kw{updateWeight}}
\newcommand{\notifyw}{\kw{notifyWeight}}

\newcommand{\IndexWeightRecomputing}{\text{Index Weight Recomputing}\xspace}
\newcommand{\IWR}{\text{IWR}\xspace}
\newcommand{\QCHIU}{\text{Quick \CHI  Updating}\xspace}
\newcommand{\QU}{\text{QU}\xspace}
\newcommand{\QUP}{\text{QU$^+$}\xspace}
\newcommand{\RA}{\text{RA}\xspace}

\newcommand{\dsny}{\textit{NY}\xspace}
\newcommand{\dsco}{\textit{COL}\xspace}
\newcommand{\dsfl}{\textit{FLA}\xspace}
\newcommand{\dsca}{\textit{CAL}\xspace}
\newcommand{\dseus}{\textit{E-US}\xspace}
\newcommand{\dswus}{\textit{W-US}\xspace}
\newcommand{\dscus}{\textit{C-US}\xspace}
\newcommand{\dsus}{\textit{US}\xspace}

%==============================defined for KeccDecom===================
\newcommand{\topcaption}{%
 \setlength{\abovecaptionskip}{4pt}%
 \setlength{\belowcaptionskip}{0pt}%
 \caption}

\newcommand{\decb}{\kwnospace{DecB}\textrm{-}\kw{LMSD}}
\newcommand{\kecc}{\kw{kECC}}
\newcommand{\kec}{\kw{kEC}}
\newcommand{\keccs}{\kwnospace{kECC}s\xspace}
\newcommand{\ecc}{\kw{ECC}}
\newcommand{\eccs}{\kwnospace{ECC}s\xspace}
\newcommand{\inputgraph}{\textit{Input Graph G}\xspace}
\newcommand{\sparsifiedgraph}{\textit{Sparsified Graph G'}\xspace}
\newcommand{\preservedgraph}{\textit{kECC Preserved Graph G'}\xspace}
\newcommand{\observone}{\textit{Community Structure}\xspace}
\newcommand{\observtwo}{\textit{Core-Periphery Structure}\xspace}
\newcommand{\spanforest}{\kw{DisjointForest}}
\newcommand{\contract}{\kwnospace{CE}\textrm{-}\kw{Disk}}
\newcommand{\recover}{\kw{Recover}}
\newcommand{\componentid}{\kwnospace{CID}}
\newcommand{\kcore}{\kw{k}\textrm{-}\kw(core)}
\newcommand{\rand}{\kwnospace{Random}\textrm{-}\kw{Decom}}
\newcommand{\exact}{\kwnospace{Exact}\textrm{-}\kw{Decom}}
\newcommand{\bu}{\kwnospace{BU}\textrm{-}\kw{Decom}}
\newcommand{\td}{\kwnospace{TD}\textrm{-}\kw{Decom}}
\newcommand{\bottomup}{\kwnospace{Bottom}\textrm{-}\kw{Up}}
\newcommand{\hybrid}{\kwnospace{Hybrid}\textrm{-}\kw{Decom}}
\newcommand{\mydegree}{\kwnospace{degree}}
\newcommand{\topdown}{\kwnospace{Top}\textrm{-}\kw{Down}}
\newcommand{\hybrids}{\kw{Hybrid}}
\newcommand{\mem}{\kwnospace{Mem}\textrm{-}\kw{Decom}}
\newcommand{\id}{\kwnospace{id}}
\newcommand{\dsdblp}{\textit{DBLP}\xspace}
\newcommand{\dslj}{\textit{LiveJournal}\xspace}
\newcommand{\dsorkut}{\textit{Orkut}\xspace}
\newcommand{\dshw}{\textit{Hollywood}\xspace}
\newcommand{\dsuk}{\textit{uk-2005}\xspace}
\newcommand{\dsit}{\textit{it-2004}\xspace}
\newcommand{\dstwitter}{\textit{twitter-2010}\xspace}
\newcommand{\dssk}{\textit{sk-2005}\xspace}
\newcommand{\re}{\kwnospace{RE}\xspace}
\newcommand{\ce}{\kwnospace{CE}\xspace}
\newcommand{\rn}{\kwnospace{RN}}
\newcommand{\cn}{\kwnospace{CN}}
\newcommand{\kpg}{$k$\textrm{-}\kw{PG}}
\newcommand{\kpgs}{$k$\textrm{-}\kwnospace{PG}s\xspace}
\newcommand{\pg}{\kw{PG}}
\newcommand{\pgn}{\kwnospace{PG}}
\newcommand{\kpgn}{$k$\textrm{-}\kwnospace{PG}}
\newcommand{\remove}{\kwnospace{RE}\textrm{-}\kw{Disk}}
\newcommand{\contractmem}{\kwnospace{CE}\textrm{-}\kw{Mem}}

\newcommand{\VCS}{\text{DCH$_{vcs}$}\xspace}
\newcommand{\WDEC}{\text{DCH$_{scs}$-WDec}\xspace}
\newcommand{\WDECP}{\text{DCH$^+_{scs}$-WDec}\xspace}
\newcommand{\WINC}{\text{DCH$_{scs}$-WInc}\xspace}
\newcommand{\WINCP}{\text{DCH$^+_{scs}$-WInc}\xspace}
\newcommand{\SCB}{\text{DCH$_{scb}$}\xspace}
\newcommand{\SCBP}{\text{DCH$^+_{scb}$}\xspace}

%============================== defined for HCSPEnum =============
\newcommand{\hcstp}{\kwnospace{HC}\textrm{-}\kwnospace{s}\textrm{-}\kw{t~path}}
\newcommand{\hcstps}{\kwnospace{HC}\textrm{-}\kwnospace{s}\textrm{-}\kw{t~paths}}
\newcommand{\join}{\kw{JOIN}}
\newcommand{\bcdfs}{\kwnospace{BC}\textrm{-}\kw{DFS}}
\newcommand{\disbfs}{\kw{DisBFSEnum}}
\newcommand{\bfsenum}{\kw{BFSEnum}}
\newcommand{\hpindex}{\kwnospace{HP}\textrm{-}\kw{Index}}
\newcommand{\dishpindex}{\kwnospace{DisHP}\textrm{-}\kw{Index}}
\newcommand{\hpi}{\kw{HPI}}
\newcommand{\dishpi}{\kw{DisHPI}}
\newcommand{\centralized}{\kw{DFSEnum}}
\newcommand{\hybridsearch}{\kw{HybridEnum}}
\newcommand{\hybridenum}{\kw{HybridEnum^+}}
\newcommand{\pextend}{\kw{PrunExte}}
\newcommand{\backprop}{\kw{BackProp}}
\newcommand{\jump}{\kw{Jump}}
\newcommand{\level}{\kw{level}}
\newcommand{\degout}{\kw{deg^+}}
\newcommand{\degin}{\kw{deg^-}}
\newcommand{\flowctrl}{\kw{FlowCtrl}}
\newcommand{\concat}{\kw{ConCat}}

%+++++++++++++++++++++++++++++++defined for LCSDist +++++++++++++++++++

%\DeclareMathOperator*{\argmax}{arg\,max}
%\DeclareMathOperator*{\argmin}{arg\,min}
%\DeclarePairedDelimiter\abs{\lvert}{\rvert}
%\newcommand{\dist}{\kw{dist}}
%\newcommand{\edp}{\kw{EDP}}
%\newcommand{\tdnaive}{\kw{NaiveQuery}}
%\newcommand{\lcdpindex}{\kwnospace{LSD}\textrm{-}\kw{Index}}
%\newcommand{\lcbpquery}{\kwnospace{LSD}\textrm{-}\kwnospace{Index}\textrm{-}\kw{Query}}
%\newcommand{\lcdconnect}{\kw{LSDSJoin}}
%\newcommand{\lcdprune}{\kw{LSDSPrune}}
%\newcommand{\lcbpcons}{\kwnospace{LSD}\textrm{-}\kwnospace{Index}\textrm{-}\kw{Cons}}
%\newcommand{\lcbppcons}{\kwnospace{LSD}\textrm{-}\kwnospace{Index}\textrm{-}\kw{ParCons}}
%\newcommand{\lcd}{\kw{\mathcal{S}_{\kw{LSD}}}}
%\newcommand{\lsd}{\kw{\mathcal{S}_{\kw{LSD}}}}
%\newcommand{\lsds}{\kw{LSDS}}
%\newcommand{\lsdss}{\kw{LSDSs}}

%+++++++++++++++++++++++++++++++defined for SPC +++++++++++++++++++

\newcommand{\hpspc}{\kwnospace{HP}\textrm{-}\kw{SPC}}
\newcommand{\ctspc}{\kwnospace{ST}\textrm{-}\kw{SPC}}
\newcommand{\spc}{\kw{spc}}
\newcommand{\sdgamma}{\kw{sd_{\scriptscriptstyle\mathcal{T}}}}
\newcommand{\rev}{\kw{rev}}
\newcommand{\ctdecom}{\kwnospace{ST}\textrm{-}\kw{Decom}}
\newcommand{\ctspccons}{\kwnospace{ST}\textrm{-}\kwnospace{SPC}\textrm{-}\kw{ConS}}
\newcommand{\ctspcconp}{\kwnospace{ST}\textrm{-}\kwnospace{SPC}\textrm{-}\kw{ConP}}
\newcommand{\ctspccon}{\kwnospace{ST}\textrm{-}\kwnospace{SPC}\textrm{-}\kw{Cons}}

%+++++++++++++++++++++++++++++++defined for KNN +++++++++++++++++++

\newtheorem{property}{Property}

\SetKwInput{KwInput}{Input}                
\SetKwInput{KwOutput}{Output}
\SetKwProg{myproc}{Procedure}{}{}

\newcommand{\tenindex}{$\kw{TEN}\textrm{-}\kw{Index\ }$}
\newcommand{\tenindexcnt}{\kwnospace{TEN}\textrm{-}\kw{Index}}
\newcommand{\tenindexno}{\kwnospace{TEN}\textrm{-}\kwnospace{Index}}
\newcommand{\tenknn}{\kwnospace{TEN}\textrm{-}\kwnospace{Index}\textrm{-}\kw{Cons}}

\newcommand{\glad}{\kwnospace{GLAD}}

\newcommand{\tenknncnt}{\kwnospace{TEN}\textrm{-}\kwnospace{Index}\textrm{-}\kwnospace{Cons}}

\newcommand{\dijk}{\kwnospace{Dijkstra}\textrm{-}\kw{Cons}}

\newcommand{\hhindex}{\kwnospace{H2H}\textrm{-}\kw{Index}}
\newcommand{\hhindexcnt}{\kwnospace{H2H}\textrm{-}\kw{Index}}
\newcommand{\knnindex}{\kwnospace{KNN}\textrm{-}\kw{Index}}
\newcommand{\pknnindex}{\kwnospace{PKNN}\textrm{-}\kw{Index}}
\newcommand{\pknnindexcnt}{\kwnospace{PKNN}\textrm{-}\kw{Index}}
\newcommand{\knnindexcnt}{\kwnospace{KNN}\textrm{-}\kw{Index}}

\newcommand{\knn}{\kw{KNN\ }}
\newcommand{\pknn}{\kw{PKNN}}

\newcommand{\naivecon}{$\kw{KNN}\textrm{-}\kw{Index}\textrm{-}\kw{Cons}$\ }
\newcommand{\optcon}{$\kw{KNN}\textrm{-}\kw{Index}\textrm{-}\kw{Cons}^+$\ }

\newcommand{\naiveconcnt}{$\kw{KNN}\textrm{-}\kw{Index}\textrm{-}\kw{Cons}$}
\newcommand{\optconcnt}{$\kw{KNN}\textrm{-}\kw{Index}\textrm{-}\kw{Cons}^+$}

\newcommand{\knnalg}{\kwnospace{V_k}}

\newcommand{\pknnalg}{\kwnospace{V_k^<}}

\newcommand{\knnalgex}{\kwnospace{V_5}}

\newcommand{\pknnalgex}{\kwnospace{V_5^<}}

\newcommand{\knnalgpro}{\kwnospace{V_k}}

\newcommand{\pknnalgpro}{\kwnospace{V_k^<}}

\newcommand{\subg}{\kwnospace{subG}}

\newcommand{\sd}{\kwnospace{dist}}
\newcommand{\vsd}{\kwnospace{vdist}}
\newcommand{\psd}{\kwnospace{dist_<}}

\newcommand{\spath}{\kwnospace{SPath}}
\newcommand{\vpath}{\kwnospace{VPath}}
\newcommand{\vspath}{SVPath}

\newcommand{\ppath}{\kwnospace{PPath}}
\newcommand{\pspath}{PSPath}
\newcommand{\sdgraph}{\kwnospace{BN}\textrm{-}\kw{Graph}}

%vertex cut
\newcommand{\vc}{\kwnospace{S}}

%essential vertex set
\newcommand{\ev}{\kwnospace{BNS}}

\newcommand{\sdg}{\kwnospace{BN}\textrm{-}\kw{Graph}}
\newcommand{\ldg}{LD-Graph\ }

\newcommand{\keyL}{\kwnospace{BNS^<}}
\newcommand{\keyU}{\kwnospace{BNS^>}}

\newcommand{\evl}{\kwnospace{V_k^<}}
\newcommand{\sdgl}{\kwnospace{G'^<}}
\newcommand{\sdgu}{\kwnospace{G'^>}}

\newcommand{\rank}{\kwnospace{Rank}}

\newtheorem{observation}{Observation}

\newcommand{\sdgraphgen}{\kwnospace{SD}\textrm{-}\kwnospace{Graph}\textrm{-}\kwnospace{Gen}}

\newcommand{\hl}[1]{{\color{blue}{#1}}}

\title{Simpler is More: Efficient Top-K Nearest Neighbors Search on Large Road Networks}

\author{
Yiqi Wang$^{1}$, Long Yuan$^{2}$,  Wenjie Zhang$^{1}$, Xuemin Lin$^{3}$, Zi Chen$^{1}$, Qing Liu$^{4}$}
\affiliation{$^1$ The University of New South Wales, $^2$  Nanjing University of Science and Technology}
\affiliation{$^3$  Shanghai Jiaotong University, $^4$  Data61, CSIRO, Australia}
\email{
yiqi.wang1@unsw.edu.au, longyuan@njust.edu.cn, zchen@sei.ecnu.edu.cn}
 \email{
 {zhangw,lxue}@cse.unsw.edu.au,q.liu@data61.csiro.au}
%\author{Yiqi Wang$^{1}$, Long Yuan$^{2}$,  Wenjie Zhang$^{1}$, Xuemin Lin$^{3}$, Zi Chen$^{4}$, Qing Liu$^{5}$}
%\affiliation{$^{1}$ The University of New South Wales, Sydney, Australia}
%\affiliation{$^{2}$ Nanjing University of Science and Technology, Nanjing, China}
%\affiliation{$^{3}$ East China Normal University, Shanghai, China,  $^{4}$Shanghai Jiaotong University, Shanghai, China}
%\affiliation{$^{5}$ Data61, CSIRO, Hobart, Australia}

%\email{yiqi.wang1@unsw.edu.au,longyuan@njust.edu.cn,wenjie.zhang@unsw.edu.au}
%\email{xuemin.lin@sjtu.edu.cn,zchen@sei.ecnu.edu.cn,q.liu@data61.csiro.au}

%%
%% The "author" command and its associated commands are used to define the authors and their affiliations.
\begin{abstract}
Top-\underline{$k$} \underline{N}earest \underline{N}eighbors ($k$NN) problem on road network has numerous applications on location-based services. As direct search using the \kw{Dijkstra}'s algorithm results in a large search space, a plethora of complex-index-based approaches have been proposed to speedup the query processing. However, even with the current state-of-the-art approach, long query processing delays persist, along with significant space overhead and prohibitively long indexing time. In this paper, we depart from the complex index designs prevalent in existing literature and propose a simple index named \knnindexcnt. With \knnindexcnt, we can answer a $k$NN query optimally and progressively with small and size-bounded index. To improve the index construction performance, we propose a bidirectional construction algorithm which can effectively share the common computation during the construction. Theoretical analysis and experimental results on real road networks demonstrate the superiority of \knnindex over the state-of-the-art approach in query processing performance, index size, and index construction efficiency.
\end{abstract}

\maketitle


\pagestyle{\vldbpagestyle}
\begingroup\small\noindent\raggedright\textbf{}\\
%{Yiqi Wang, Long Yuan,  Wenjie Zhang, Xuemin Lin, Zi Chen, Qing Liu}. \vldbtitle. PVLDB, \vldbvolume(\vldbissue): \vldbpages, \vldbyear.\\
%\href{https://doi.org/\vldbdoi}{doi:\vldbdoi}
\endgroup
%\begingroup
%\renewcommand\thefootnote{}\footnote{\noindent
%This work is licensed under the Creative Commons BY-NC-ND 4.0 International License. Visit \url{https://creativecommons.org/licenses/by-nc-nd/4.0/} to view a copy of this license. For any use beyond those covered by this license, obtain permission by emailing \href{mailto:info@vldb.org}{info@vldb.org}. Copyright is held by the owner/author(s). Publication rights licensed to the VLDB Endowment. \\
%%\raggedright Proceedings of the VLDB Endowment, Vol. \vldbvolume, No. \vldbissue\ %
%ISSN 2150-8097. \\
%%\href{https://doi.org/\vldbdoi}{doi:\vldbdoi} \\
%}\addtocounter{footnote}{-1}\endgroup

%%%% VLDB block end %%%
%
%%%% do not modify the following VLDB block %%
%%%% VLDB block start %%%

%\ifdefempty{\vldbavailabilityurl}{}{
%\vspace{.3cm}
%\begingroup\small\noindent\raggedright\textbf{PVLDB Artifact Availability:}\\
%The source code, data, and/or other artifacts have been made available at \url{https://github.com/xiaoqi-7/knn}.
%\endgroup

\section{Introduction}
\label{sec:intro}

Top $k$ nearest neighbors ($k$NN) search on road network is a fundamental operation in location-based services \cite{bhatia2010survey,abbasifard2014survey,nodarakis2017knn}. Formally,  given a road network $G(V, E)$, a set of candidate objects $\mathcal{M}$, and a query  vertex $u$, $k$NN search identifies $k$ objects in $\mathcal{M}$ with the shortest distance to $u$. $k$NN search finds many important real world applications. For example, in the accommodation booking platforms like \emph{Booking} \cite{booking}, \emph{Airbnb} \cite{airnb} and \emph{Trip} \cite{trip}, an important operation is to show several accommodations closest to the location provided by users. In  restaurant-review services, such as \emph{Yelp} \cite{yelp}, \emph{Dianping} \cite{dianping} and \emph{OpenRice} \cite{openrice},  platforms utilize $k$NN search to present several nearby restaurants to the user.  In ride-hailing services like \emph{Uber} \cite{uber} and \emph{Didi} \cite{didi}, several available vehicles near the pickup location are presented before users send the ride-hailing request.

\begin{figure}
    \centering
    \includegraphics[width=1\linewidth]{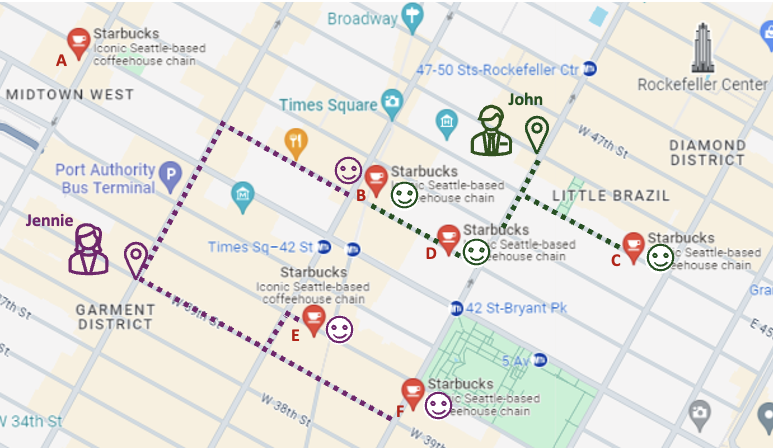}
    \vspace{-0.5cm}
    \caption{$k$NN Search in Location-based Service ($k = 3$)}
    \vspace{-0.5cm}
    \label{fig:realexam}
\end{figure}
 
\begin{example} 

\reffig{realexam} shows a $k$NN search example in location-based service. Assume that tourists in New York, such as "John" and "Jennie",  want to find Starbucks nearby to drink coffee, the location-based service providers like Google Map generally present several candidate stores based on the distance from  their locations, which can be modeled as  $k$NN search problem. In \reffig{realexam}, there are $6$ Starbucks stores (marked with $A, B, \cdots, F$), therefore, the candidate object set $\mathcal{M} = \{ A,B,C,D,E,F \}$. For  "Jennie", the  $3$NN search  returns $\{E, F, B\}$ while the $3$NN search for "John" returns $\{ C, D, B \}$.
\end{example}

 \stitle{Motivation.} Given a $k$NN query for vertex $u$, the query can be directly answered by exploring the vertices based on their distance to $u$ using \kw{Dijkstra}'s algorithm \cite{dijkstra2022note}. Nevertheless, this method is inefficient, especially when the road network is large and the candidate objects are far from $u$. Therefore, researchers resort to indexing-based solutions to accelerate query processing \cite{papadias2003query, demiryurek2009efficient, lee2010road, zhong2015g, shen2017v, luo2018toain, he2019efficient, ouyang2020progressive}.

Although existing index-based approaches have made strides in accelerating the query processing, they still suffer from the long query processing delay and their performance is far from optimal. Additionally, these solutions exhibit significant space overhead and prohibitively long indexing times, severely limiting their practical applicability. Take the state-of-the-art approach \tenindex  \cite{ouyang2020progressive} for $k$NN queries on road networks as an example. The size of \tenindex on the dataset \kw{USA} with only 23.95 million vertices and 58.33 million edges (nearly 442.5 MB if an edge is stored by two 4 byte integers) exceeds $160$ GB, and it takes more than $5.4$ hours to construct the corresponding index. Motivated by these, this paper aims to propose a new index-based solution for $k$NN query that can overcome the shortcomings of existing solutions in query processing performance, index size and index construction.

 \stitle{{A Minimalist $k$NN Index Design.}} Revisiting the existing solutions, they generally design a complex index to speedup the query processing. For example, \tenindex consists of three different parts. Incredibly, as an index for $k$NN query, one of the three parts is even a complete index structure for shortest distance query. The complex-index design leads to the drawbacks  of \tenindex as analyzed in \refsec{existing}. This drives us to ask: is this complex-index design thinking really suitable for $k$NN query?
 
In this paper, we adopt a completely opposite design approach. Going back to the essence of $k$NN search, it only needs to return the $k$ nearest neighbors for the query vertex. Moreover, the $k$ value of the $k$NN search used in real applications is typically not large as users often have limited attention spans and prefer to quickly obtain relevant information to reduce cognitive load and facilitate decision-making \cite{TooManyChoices,burChoice,sthapit2018more,chernev2015choice,schwartz2004paradox,luo2018toain}. For example, Yelp App \cite{yelp} provides customers with $20$ results every time when searching nearest specific place type, such as restaurant or  gas station. A similar strategy is also adopted in other Apps like OpenRice  \cite{openrice} and OpenTable \cite {opentable}. Therefore, our proposed new index named \knnindex only simply records the $k$ nearest neighbors of each vertex. The benefits of this minimalist $k$NN index design are twofold: regarding the query processing, the query can be answered \emph{progressively} \footnote{Progressive query processing outputs results gradually in a well-bounded delay, allowing users to obtain useful results even before query processing completes. It also provides a choice to early terminate the search when a user finds sufficient answers \cite{papadias2003optimal}.} in \emph{optimal} time. Regarding the space-consumption of the index, only the essential information directly to $k$NN query is stored in the index and the value of $k$ is small in practice, resulting in a \emph{well-bounded} index space. 

\stitle{{New Challenges.}}  \knnindex successfully addresses the issues of long query delays and oversized indexes by directly storing the $k$ nearest neighbors for each vertex. However, this strategy leaves the trouble to the index construction as the index structure intuitively implies that we have to explore all the query space before constructing it.  A straightforward approach is to compute the $k$ nearest neighbors for each vertex by \kw{Dijkstra}'s algorithm \cite{dijkstra2022note}. However, the time complexity of this approach is $O(n \cdot (m+ n \log n))$, where $n$ is the number of vertices and $m$ is the number of edges in the road network. Clearly,  this approach is impractical to handle large road networks. Another possible approach is to use the existing index like \tenindex to accelerate the computation of $k$ nearest neighbors for each vertex. Nevertheless, this approach unavoidably induces the drawbacks of existing approaches as discussed above. Overall, the efficiency of the  index construction algorithm determines the applicability of our index while it is challenging to design such an efficient index construction algorithm that could outperform existing solutions.

\stitle{{Our Idea.}} The above discussed approaches compute the $k$ nearest neighbors for each vertex independently, which miss the potential opportunities to re-use the intermediate results during the construction. Therefore, we adopt a computation sharing strategy to achieve the efficient index construction. To effectively share the computation, we introduce the concept of \emph{bridge neighbor set} for a vertex $v$ and reveal the hidden relationships between its bridge neighbor set and $k$ nearest neighbors. Following these findings, we design a bridge neighbor preserved graph  (\sdgraph) of the input road network with which the bridge neighbor set of a vertex can be easily obtained. Based on \sdgraph, we first propose a bottom-up index construction algorithm in which the intermediate results during the construction can be largely shared and further improve the performance by introducing a bidirectional construction algorithm. Additionally, the given candidate objects $\mathcal{M}$ may be updated in some cases \cite{ouyang2020progressive}, we also design efficient algorithm to incrementally maintain the index for these updates. 

\stitle{Contributions.} In this paper, we  make the following  contributions:

\sstitle{(1) A new attempt at an alternative $k$NN index design paradigm with a simple yet effective $k$NN index.} Recognizing the complex index in the existing solutions  leads to long query processing delay, oversized index and prohibitive indexing time, we embrace minimalism and design a simple $k$NN index that has a \emph{well-bounded} space and enables \emph{progressive} and \emph{optimal} query processing. To the best of our knowledge, this is the first work that systematically studies such simple yet effective index for $k$NN query.

\sstitle{(2) Efficient index construction and maintenance algorithms.} Following the designed index, we propose a novel  index construction algorithm with which the shortest distance computation regarding a vertex and its top $k$ nearest neighbors can be effectively shared.  We also propose index maintenance algorithms to handle  object insertion and deletion. We provide time complexity analysis for all proposed algorithms.  

\sstitle{(3) Extensive experiments on real-world road networks.} We extensively evaluate our proposed algorithms on real road networks. Compared with the state-of-the-art approach \tenindexcnt, experimental results demonstrate that our approach reduces the index space two order of magnitude, speeds up the query time up to two orders of magnitude, and achieves up to two orders of magnitude speedup in index construction. 

\stitle{Outline.} \refsec{pre} provides the problem definition. \refsec{existing} introduces the state-of-the-art algorithm. \refsec{indexquery}, \refsec{indexcon}, and \refsec{update} present the new indexing approach. \refsec{exp} evaluates our algorithms and \refsec{related} reviews the related work. \refsec{conclusion} concludes the paper. %The proofs of all the lemmas and theorems are omitted due to the limited space, and can be found in the technique report \cite{techniquereport}.

%\begin{table}[!tb]
%	{
%	\small{
%		\begin{center}
%		 %\setlength{\tabcolsep}{0.01em}
%		  \begin{tabular}{|l | c | c| c| }
%		  \hline % \diagbox[height=1.2\line]{}{}
%			\cg & \cg Query Time & \cg Index Size & \cg Index Construction \\
%		    \hline
%			\cg \tenindex &  $ O(h \cdot k)$ & $O(n \cdot \max\{h, k
%			\})$ & $O(n \cdot \log n + n \cdot h \cdot w)$ \\
%			 \hline
%			\cg ours & $ O(k)$ & $O(n \cdot k)$ & $O(n \cdot f^2 + n \cdot f \cdot k)$  \\
%			\hline
%		  \end{tabular}
%		\end{center}
%		%\vspace{-0.3cm}
%	  \caption{Comparison between \tenindex and ours}
%	  %\vspace{-0.1cm}
%	  \label{tab:compare}
%	  }
%	  }
%\end{table} 

\section{Preliminaries}
\label{sec:pre}

\begin{table}[htb]
    {
    \small{
    \setlength{\tabcolsep}{0.7em}
        \begin{center}
          \begin{tabular}{l l }
            \hline
             \multicolumn{1}{c}{ {Notations}} &  {Descriptions}  \\
           \hline
           {$\knnalg(u)$} & $k$ nearest neighbor set of $u$ \\
            %\hline
           {$\pknnalg(u)$} & decreasing rank partial $k$NN of $u$ \\
            %\hline
           {$G'$} & bridge neighbor preserved graph of $G$ \\
           %\hline
           {$\sdgl(u)$} & decreasing rank path subgraph of $u$ \\
            %\hline
           {$\sdgu(u)$} & increasing rank path subgraph of $u$ \\
            %\hline
           {$\ev(u)$} & neighbors of $u$ in $G'$ \\
            %\hline
           {$\keyL(u)$} & neighbors of $u$ in $G'$ with lower rank than $u$ \\
            %\hline
           {$\keyU(u)$} & neighbors of $u$ in $G'$ with higher rank than $u$ \\
            %\hline
           {$\psd(u,v)$} & length of decreasing rank shortest path between $u$ and $v$ \\
            \hline
          \end{tabular}
        \end{center}
        }
        
      }
      \vspace{0.12cm}
      \caption{Notations}
      \label{tab:notations}
      \vspace{-0.6cm}
\end{table}

%Let $G=(V,E)$ be a degree-bounded, connected and weighted graph to represent a real-world road network, where $V(G)$ and $E(G)$ refer to the set of vertices and edges in $G$, respectively. We use $n=|V(G)|$ (resp. $m=|E(G)|$) to denote the number of vertices (resp. edges) in $G$. For each vertex $v \in V(G)$, the neighbours of $v$, denoted by $\nbr(v,G)$, is defined as $\nbr(v,G)= \{ u|(u,v) \in E(G) \}$. The degree of a vertex $v$ is the number of neighbors of $v$, i.e., $\mydeg(v,G) = |\nbr(v,G)|$. The weight of an edge $e = (u,v)$ is denoted as $\phi(e, G)$ or $\phi((u,v), G)$. A path $p$ in $G$ is a sequence of vertices $p = (e_0,e_1,e_2,\dots, e_k)$, such that $e_i \in E(G)$ for each $0 \leq i \leq k$. The length of $p$, denoted by $\len(p)$, is the sum for the weight of edges in $p$, i.e., $\len(p) = \sum_{e \in p} \phi(e)$.  Given two vertices $u$ and $v$, the shortest path between $u$ and $v$ in $G$ is a path $p$ starting from $u$ and ending at $v$ such that $\len(p)$ is minimized, denoted as $SPath(u,v)$. The shortest distance between u and v in G, denoted as $dist(u,v)$, is the weight of the shortest path between them.     
Let $G=(V,E)$ be a connected and weighted graph to represent a real-world road network, where $V(G)$ and $E(G)$ is the set of vertices and edges in $G$, respectively. We use $n=|V(G)|$ (resp. $m=|E(G)|$) to denote the number of vertices (resp. edges) in $G$. For each vertex $v \in V(G)$, the neighbours of $v$, denoted by $\nbr(v,G)$, is defined as $\nbr(v,G)= \{ u|(u,v) \in E(G) \}$. The degree of a vertex $v$ is the number of neighbors of $v$, i.e., $\mydeg(v,G) = |\nbr(v,G)|$. The weight of an edge $(u,v)$ is denoted as $\phi((u,v), G)$. A path $p$ in $G$ is a sequence of vertices $p = (v_0,v_1,v_2,\dots, v_n)$, such that $v_i \in V(G)$ for each $0 \leq i \leq n$. The length of $p$, denoted by $\len(p)$, is the sum for the weight of edges in $p$, i.e., $\len(p) = \sum_{i=1}^n \phi(v_{i-1},v_{i})$.  Given two vertices $u$ and $v$, the shortest path between $u$ and $v$ in $G$ is a path $p$  from $u$ to  $v$ with smallest $\len(p)$. The distance between u and v in G, denoted as $\sd(u,v)$, is the weight of the shortest path between them.

Regarding the given set of candidate objects $\mathcal{M}$,  we assume all objects in $\mathcal{M}$ are on vertices following the previous works \cite{shen2017v, luo2018toain, he2019efficient,ouyang2020progressive}. In real-world road networks, each object $o \in \mathcal{M}$ may appear on any point of edges. For an object $o$ not on a vertex, we can see $o$ as a vertex object with an offset, and the distance between $o$ and a query vertex $v$ can be computed by mapping $o$ to an adjacent vertex with an offset following the previous works \cite{shen2017v, luo2018toain, he2019efficient,ouyang2020progressive} as well. Specifically, assume $o$ is on an edge $(u_o, u'_o)$ with a distance $\phi_o$ to $u'_o$, $q$ is on an edge $(u_q, u'_q)$ with a distance $\phi_{q}$. The distance between $q$ and $o$ is represented as $\sd(q, o) = \phi_q + \sd(u_q, u_o) + \phi_o$. We denote the $k$NN result of a vertex $u$ as $\knnalg(u)$ and define the problem of $k$NN search as follows.

\stitle{Problem Definition.}
Given a road network $G=(V,E)$, a query vertex $u$, an integer $k$, and a set of candidate objects $\mathcal{M}$ $(|\mathcal{M}|>k), \mathcal{M} \subseteq V(G)$), we aim to computes $k$ objects from $\mathcal{M}$, denoted by $\knnalg(u)$,  such that  $\forall v \in \knnalg(u),$ $ w \in \mathcal{M} \setminus \knnalg(u),$ $\sd(u,v) \leq \sd(u,w)$.

\begin{figure}
    \centering
    \vspace{0.2cm}
    \includegraphics[width=0.90\linewidth]{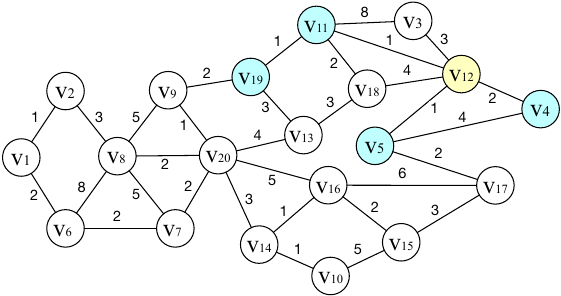}
    \vspace{-0.2cm}
    \caption{A Road Network}
    \vspace{-0.3cm}
    \label{fig:roadnetwork}
\end{figure}

\begin{example}
    Consider the graph $G$ in \reffig{roadnetwork} and assume all vertices are in the candidate object set. For a given query vertex $v_{12}$ and $k=5$, $V_5(v_{12}) = \{ v_{12}, v_5, v_{11}, v_4, v_{19} \}$. The corresponding distances between $v_{12}$ and vertex in $V_5(v_{12})$ are $0, 1, 1, 2$ and $2$ respectively.
\end{example}

%Note that $k$NN of each vertex may not be unique due to same distance values. In the discussion of this paper, we assume the distance of every pair of vertices is unique for ease of presentation. If there are many objects with the same distance to a query vertex $u$, our algorithm still returns the correct results. Due to space limitation, we omit our some proofs in this paper.

\section{The State-of-the-art Solution}
\label{sec:existing}

\tenindex \cite{ouyang2020progressive} is the state-of-the-art index-based approach for $k$NN queries on road networks. \tenindex designs an index based on tree decomposition \cite{robertson1984graph,xu2005tree} and \hhindex \cite{ouyang2018hierarchy}, which proves superiority over other existing approaches. Specifically,

\stitle{Index Structure.} \tenindex  decomposes the input road network into a tree-like structure by tree decomposition  \cite{xu2005tree}. Given the decomposed tree structure, each vertex $u$ has a child vertex set $\mathbb{T}(u)$ and an ancestor vertex set $\mathbb{A}(u)$. Apart from the decomposed tree structure, \tenindex contains the other two parts: $k$TNN for each vertex $u$ which stores the top $k$  nearest neighbors of $u$ in $\mathbb{T}(u)$ and \hhindex \cite{ouyang2018hierarchy} which is used to compute the shortest distance between $u$ and $v \in \mathbb{A}(u)$.

\stitle{Query Processing.} Given a query vertex $u$, for each vertex $v$ in the $k$NN of $u$, there exists a vertex $p$ such that $p \in \mathbb{A}(u) \cup \{u\}$ and $v$ in $k$TNN of $p$. Following this idea,  \tenindex answers the $k$NN query in $k$ rounds. In each $i$ round ($1 \leq i \leq k$), it outputs the top $i$-th result by iterating the vertices in $\mathbb{A}(u) \cup \{u\}$ and computing the corresponding shortest distance through \hhindex.  

\stitle{Index Construction.} To construct the index, \tenindex first decomposes the graph following \cite{xu2005tree}. With the decomposed tree, $\mathbb{A}(u)$ and $\mathbb{T}(u)$ can be obtained accordingly. After that, \tenindex builds the \hhindex based on \cite{ouyang2018hierarchy}. At last, the $k$TNN for each vertex is constructed by querying the shortest distance of corresponding vertex pairs through \hhindex.

\stitle{Drawbacks.} Although \tenindex accelerates the $k$NN query processing on road network, the following drawbacks limit its applicability in practice: 

\begin{itemize}[leftmargin=*]
\item {\textbf{Oversized Index.}} The size of \tenindex is generally huge in  practice. As verified in our experiments,  the size of \tenindex on \kw{USA} (only $23,947,347$ vertices and $58,333,344$ edges) exceeds $172.80$ GB, in which \hhindex takes $169.23$ GB space.  

\item {\textbf{Long Query Delay.}} To answer a $k$NN query regarding vertex $u$, \tenindex has to iterate the vertices in $\mathbb{A}(u)\cup\{u\}$ and compute the corresponding shortest distance in $k$ rounds. Moreover, the shortest distance computation is not free, and needs heavy exploration on the \hhindexcnt. These two factors lead to long query delay of \tenindexcnt. 
    
 \item {\textbf{Prohibitive Indexing Time.}} As shown in the above, to construct the index, \tenindex has to decompose the road network first, and then build the \hhindex and compute the $k$TNN accordingly. Obviously, the time cost of these procedures are expensive, especially the \hhindex construction. For the dataset USA, \tenindex takes $19666$s to construct the index, in which \hhindex consumes $19632$s.  
\end{itemize}

\section{Our Indexing Approach}
\label{sec:indexquery}

According to the above analysis,  although the use of \hhindex accelerates the query processing of \tenindexno,  heavily depending on the \hhindex directly leads to the drawbacks of \tenindexno. This raises  a natural question: why do we need an index for shortest distance such as \hhindex when addressing $k$NN problem?  Based on the logic of \tenindexno, partial $k$NN (namely $k$TNN) is maintained for each vertex and \hhindex is used to refine the partial $k$NN to obtain the final results when processing the query. This motivates us to further ask: Is it necessary to maintain the partial $k$NN? How about maintaining the $k$NN for each vertex directly as an index? In this way, the drawbacks regarding index size and query delay can be totally addressed. Following this idea, we propose the following index and query processing algorithm.
 
\subsection{Index Structure and Query Processing}
\label{sec:query}
Our index just simply records the $k$NN for each vertex in the graph,  which is formally defined as follows:

\begin{definition} 
    \textbf{(\knnindexcnt)} Given a graph $G$, an integer $k$ and a set of candidate objects $\mathcal{M}$ $(|\mathcal{M}|>k)$, for each  vertex $v \in G$, \knnindex records the  top-$k$ nearest neighbors of $v$ in $\mathcal{M}$, namely $\knnalg(u)$,  in the increasing order of their shortest distances from $v$.  
\end{definition}

\begin{example}
	Given the graph $G$ in \reffig{roadnetwork}, assume  the candidate object set is all vertices in $G$ and $k = 5$, the \knnindex of $G$ is shown in \reffig{VCGKNN}. Take $v_8$ as an example, $V_5(v_8) = \{ v_8, v_{20}, v_2, v_9, v_1 \}$, with shortest distance 0, 3, 3, 4 and 4 respectively.
\end{example}

\begin{figure}[htb]
	{
	\footnotesize{
		\begin{center}
		 \setlength{\tabcolsep}{0.1em}
  	     \begin{tabular}{c | l | c| l}
		  \hline 
			\multicolumn{1}{c|}{\cg  {$v$}} & \multicolumn{1}{c|}{\cg  { \knnindex }} & \multicolumn{1}{c|}{\cg {$v$}} & \multicolumn{1}{c}{\cg {\knnindex}} \\
		    \hline
			\cg $v_1$ & $(v_1, 0) (v_2, 1) (v_6, 2) (v_7, 4) (v_8, 4)$ & \cg $v_{11}$ & $(v_{11}, 0) (v_{12}, 1) (v_{19}, 1) (v_5, 2) (v_{18}, 2)$  \\
			 \hline
			 \cg $v_2$ & $(v_2, 0) (v_1, 1) (v_6, 3) (v_8, 3) (v_7, 5)$ & \cg $v_{12}$ & $(v_{12}, 0) (v_{11}, 1) (v_5, 1) (v_4, 2) (v_{19}, 2)$ \\
			 \hline
			 \cg $v_3$ & $(v_3, 0) (v_{12}, 3) (v_5, 4) (v_{11}, 4) (v_4, 5)$ & \cg $v_{13}$ & $(v_{13}, 0) (v_{18}, 3) (v_{19}, 3) (v_{11}, 4) (v_{20}, 4)$ \\
			 \hline
			 \cg $v_4$ & $(v_4, 0) (v_{12}, 2) (v_{5}, 3) (v_{11}, 3) (v_{19}, 4)$ & \cg $v_{14}$ & $(v_{14}, 0) (v_{10}, 1) (v_{16}, 1) (v_{15}, 3) (v_{20}, 3)$ \\
			 \hline
			 \cg $v_5$ & $(v_5, 0) (v_{12}, 1) (v_{11}, 2) (v_{17}, 2) (v_{4}, 3)$ &\cg  $v_{15}$ & $(v_{15}, 0) (v_{16}, 2) (v_{14}, 3) (v_{17}, 3) (v_{10}, 4)$ \\
			 \hline
			 \cg $v_6$ & $(v_6, 0) (v_1, 2) (v_7, 2) (v_2, 3) (v_{20}, 4)$ & \cg $v_{16}$ & $(v_{16}, 0) (v_{14}, 1) (v_{10}, 2) (v_{15}, 2) (v_{20}, 4)$ \\
			 \hline
			 \cg $v_7$ & $(v_7, 0) (v_6, 2) (v_{20}, 2) (v_9, 3) (v_1, 4)$ & \cg $v_{17}$ & $(v_{17}, 0) (v_{5}, 2) (v_{12}, 3) (v_{15}, 3) (v_{11}, 4)$ \\
			 \hline
			 \cg $v_8$ & $(v_8, 0) (v_{20}, 2) (v_2, 3) (v_9, 3) (v_1, 4)$ & \cg $v_{18}$ & $(v_{18}, 0) (v_{11}, 2) (v_{12}, 3) (v_{13}, 3) (v_{19}, 3)$ \\
			 \hline
			 \cg $v_9$ & $(v_9, 0) (v_{20}, 1) (v_{19}, 2) (v_{7}, 3) (v_8, 3)$ & \cg $v_{19}$ & $(v_{19}, 0) (v_{11}, 1) (v_9, 2) (v_{12}, 2) (v_{5}, 3)$ \\
			 \hline
			 \cg $v_{10}$ & $(v_{10}, 0) (v_{14}, 1) (v_{16}, 2) (v_{15}, 4) (v_{20}, 4)$ & \cg $v_{20}$ & $(v_{20}, 0) (v_9, 1) (v_7, 2) (v_8, 2) (v_{14}, 3)$ \\
			 \hline
		  \end{tabular}
		\end{center}
		\vspace{-0.2cm}
	  \caption{\knnindex of $G$ ($k =5$)}
	  \vspace{-0.1cm}
	  \label{fig:VCGKNN}
	  }
	  }
\end{figure}

\stitle{Query Processing.} Based on our \knnindex, for a $k$NN query regarding a vertex $v$, we can answer the query directly by retrieving the corresponding items of $v$ in the \knnindex.

\subsection{Theoretical Analysis}
\label{sec:theoretical}

Following the index structure and query processing algorithm, we have the following theoretical results. 

\stitle{{Optimal Query Processing.}} Since our query processing algorithm can answer the query directly by scanning the corresponding items of the query vertex in the \knnindex, the following theorem exists obviously:

\begin{theorem}
\label{thm:optimal}
	Given a $k$NN query, our algorithm takes $O(k)$ time to process the query.
\end{theorem}

To answer a $k$NN query, any algorithm needs to output the $k$ results at least, which takes $O(k)$ time. On the other hand, \refthm{optimal} shows the time complexity of our query processing algorithm is $O(k)$. Therefore, the optimality holds.

\stitle{{Incremental Polynomial Query Processing.}} Consider an algorithm that returns several results. Let $k$ be the number of results in the output. An algorithm is said to have incremental polynomial if for all $i \leq k$, the output time of the first $i$ results is bounded by a polynomial function of the input size and $i$ \cite{DBLP:journals/isci/ChangWL94}. Since the items for each vertex $v$ in the \knnindex are recorded in the increasing order of their distance from $v$, we have:

\begin{theorem}
\label{thm:progressive}
Given a $k$NN query regarding $v$, for every $1 \leq i \leq k$, our algorithm outputs the top $i$-th nearest neighbor in $O(i)$ time.
\end{theorem}

\refthm{progressive} shows that our query processing algorithm is incremental polynomial, indicating that it  progressively provides results for a query within a bounded delay. The capability of incremental polynomial query processing is considered as a significant technical contribution of \tenindex \cite{ouyang2020progressive} and \refthm{progressive} confirms that our algorithm also possesses this desirable theoretical guarantee.

\stitle{{Bounded Index Space.}} Since \knnindex only stores the top-$k$ nearest neighbors of each vertex in the road network, we have:

\begin{theorem}
    Given a road network $G$ and an integer $k$, the size of \knnindex is bounded by $O(n \cdot k)$.
\end{theorem}

\stitle{Remark.} According to the above analysis, \knnindex processes a query in optimal time while \tenindex takes $O(h \cdot k)$  time ($h$ is the height of the tree decomposition), which means \knnindex surpasses \tenindex at the query processing. For the index size, the size of \knnindex is $O(n \cdot k)$  while that of \tenindex is $O(n \cdot h)$. As introduced in \refsec{intro}, the $k$ value of $k$NN search in real road network applications is not large, therefore, the index size of \knnindex is advantageous compared with \tenindex in practice. \emph{Note that although \cite{ouyang2020progressive} claims TEN-Index is parameter-free index, both TEN-Index and KNN-Index constructs their index based on a specific $k$,} which means we needs to construct different indices for different values of $k$. A compromise solution to address this problem for \tenindex and our index is that we can construct the index based on a moderately large $k$ value, and the $k$NN search with smaller $k$ values can be answered based on the constructed index directly.

\section{Index Construction}
\label{sec:indexcon}

Based on the structure of \knnindexcnt, it can be constructed straightforwardly by computing the top $k$ nearest neighbors of each vertex through \kw{Dijkstra's} algorithm or \tenindexno. However, these approaches are time-consuming and inefficient to handle large road network. In this section, we present our new approach to construct the \knnindexcnt.

\subsection{Key Properties of $\knnalg(u)$ }
\label{sec:mi}

The above discussed direct approaches using \kw{Dijkstra's} algorithm or \tenindex  compute the $k$ nearest neighbors for each vertex independently, which misses the potential opportunities to re-use the intermediate results during the index construction. In this section, we introduce two important properties  regarding the distance computation, which lays the foundation for our computation-sharing index construction algorithms. We first define:

\begin{definition}
    \textbf{(Bridge Neighbor Set)}
	\label{def:ev}
	Given a vertex $ u \in V(G)$,  the bridge neighbor set of $u$, denoted by $\ev(u)$, is the set of $u's$ neighbors $v$ such that the weight of the edge $(u,v)$ is equal to the  distance between $u$ and $v$ in $G$, i.e.,  $\ev(u) = \{ v | v \in \nbr(u, G) \wedge \phi((u, v), G) = \sd((u,v), G) \}$. 	
\end{definition}

\begin{example}
	 Consider $v_8$ in \reffig{roadnetwork}. $\nbr(v_8, G) = \{ v_2, v_6, v_7, v_{20}, \\v_9 \}$. The shortest path between $v_8$ and $v_9$ is $(v_8, v_{20}, v_9)$, and the distance is $\sd((v_8,v_9), G) = 3$. As $\sd((v_8,v_9),G) \neq \phi((v_8,v_9), G)$, $v_9$ is not in $\ev(v_8)$. Similarly, $v_6$ and $v_7$ also does not belong to $\ev(v_8)$. For the graph $G$ in \reffig{roadnetwork}, the bridge neighbor set of $v_8$ is $\ev(v_8) = \{ v_2, v_{20} \}$. 
\end{example}

Based on \refdef{ev}, we have following property regarding the bridge neighbor set of $u$ and its  $k$ nearest neighbors:

\begin{property}
	\label{pro:vcknn}
	Given a vertex $u \in V(G)$, $\knnalg(u) \subseteq \cup_{v \in \ev(u)} \knnalg(v)$.
\end{property}
	
{	
\proofsketch
	We prove this property by contradiction.  Assume that $w \in \knnalg(u)$ but $w \notin \cup_{v \in \ev(u)} \knnalg(v)$. According to \refdef{ev}, the shortest path between $w$ and $u$ must pass through one vertex $v \in \ev(u)$ such that for all $v_i \in \knnalg(v)$, $i \in [1,k]$, we have $\sd(v,v_i) < \sd(w,v)$. Therefore, $\sd(u, v) + \sd(v,v_i) < \sd(u,v) + \sd(v,w)$. This implies that there are at least $k$ vertices whose distance to $u$ are smaller than the distance between $u$ and $w$, which contradicts $w \in \knnalg(u)$. The proof completes.
\eop }

Following \refpro{vcknn}, we have: 

\begin{property}
\label{pro:sd}
     Given a vertex $ u \in V(G)$, $\sd((u, w), G) = \kw{min}_{ v \in \ev(u)} \\ \{ \sd((u, v), G)  + \sd((v,w), G) \}$ where $w \in \knnalg(u)$.
\end{property}

\proofsketch
	According to \refdef{ev}, for $\forall w \in \knnalg(u)$, each shortest path between $w$ and $u$ must pass through at least one vertex $v \in \ev(u)$, so we have $\sd(u,w) = \sd(u,v) + \sd(v,w)$. 
\eop

Based on \refpro{vcknn}, when a vertex $w \in \knnalg(u)$, $w$ must be in $\cup_{v \in \ev(u)} \knnalg(v)$. Moreover, when the bridge neighbor set $\ev(u)$ of $u$, the distance $\sd(u, v)$ and $\knnalg(v)$ ($\sd((v, w), G)$ accordingly where $w \in \knnalg(v)$) for all $v \in \ev(u)$ have been computed, we can compute $\sd((u, w), G)$ for each $w \in \cup_{v \in \ev(u)}\knnalg(v)$ efficiently following \refpro{sd}. Obviously, $\knnalg(u)$ just selects $k$ vertices from $\cup_{v \in \ev(u)} \knnalg(v)$ with the smallest distance values. Therefore, if we process the vertices in $G$ in a certain order, and when processing each vertex $u$, the vertices $v \in \ev(u)$ and $\knnalg(v)$ have been computed, then $\knnalg(u)$ and thereby \knnindex can be computed efficiently by sharing the computed results. The remaining problem is how to make this idea practically applicable. In  next section, we present a bottom-up computation-sharing algorithm, which paves the way to our final index construction algorithm.

\subsection{A Bottom-Up Computation-Sharing   Algorithm}
\label{sec:ug}

To compute the bridge neighbor set and share the computation effectively, we construct the index based on the bridge neighbor preserved graph $G'$  of the  road network $G$,  which is defined as:

\begin{definition}
\label{def:sdg}
\textbf{(BN-Graph)} Given a road network $G$, a graph $G'$ is  a \underline{b}ridge \underline{n}eighbor preserved \underline{graph} (\sdgraph) of $G$ if (1) $V(G') = V(G)$; (2) for each edge $(u,v) \in E(G')$, $\phi((u,v), G') = \sd((u,v), G)$; (3) for any two vertices $u, v \in V(G')$, $\sd((u,v), G') = \sd((u,v), G)$. 
\end{definition}

{\setlength{\textfloatsep}{2pt}
\setlength{\intextsep}{9pt}
\SetInd{0.6em}{0.6em}
\small
\begin{algorithm}[h]
        \caption{$\sdgraphgen(G, \pi)$ }
        \label{alg:evgen}
        {
        \DontPrintSemicolon
        $G' \gets G$; \\
        \For{\textbf{each} $w \in V(G)$ in increasing order of $\pi(w)$}{
        	$\mathcal{N} \gets \{ v | v \in \nbr(w,G') \wedge \pi(v) > \pi(w) \}$; \;
        	\For{\textbf{each} pair of vertices $u,v \in \mathcal{N}$}{
               	\If{$(u,v) \notin E(G')$}{
               		{insert} $(u,v)$ into $G'$; \\ $\phi((u,v), G') \gets \phi((u,w), G') + \phi((w,v), G')$; \;
            		}\ElseIf{$\phi((u,w), G') + \phi((w,v), G') < \phi((u,v), G')$}{
                		$\phi((u,v), G') \gets \phi((u,w), G') + \phi((w,v), G')$; \;
            		}
        	}
        }
        \For{\textbf{each} $w \in V(G)$ in decreasing order of $\pi(w)$ }{
        	$\mathcal{N} \gets \{ v | v \in \nbr(w,G') \wedge \pi(v) > \pi(w) \}$; \;
        	\For{\textbf{each} pair of $u,v \in \mathcal{N}$}{
        		\If{$\phi((w,v), G') + \phi((v,u), G') < \phi((w,u), G')$}{ 
        			$\phi((w,u), G') \gets \phi((w,v), G') + \phi((v,u), G')$; \;
        		  mark $(w, u)$ as removed;\;
        		}
        	}
        }
         remove all the marked edges  in $G'$;\;
        \For{\textbf{each} $v \in V(G')$}{
        	$\ev(v) \gets \nbr(v, G')$; \;
        }}
\end{algorithm} 
}

Based on \refdef{sdg}, we propose \refalg{evgen} to compute the \sdgraph of an input road network and obtain the bridge neighbor set for each vertex accordingly. Intuitively, a \sdgraph of $G$ with larger bridge neighbor set for each vertex has more potential possibility to share the computation following the analysis of \refsec{mi}. Meanwhile, the construction of \sdgraph should not be costly.  Following this idea, for a given road network $G$ and a total vertex order $\pi$ (the order used in our paper is discussed at the end of this section), our algorithm (\refalg{evgen}) contains two steps to construct \sdgraph: (1) Edge insertion, it aims to add edges to connect vertices to enlarge the bridge neighbor set. (2) Edge deletion, it deletes edges to guarantee that the bridge neighbor set is enlarged correctly. Specifically,

\noindent $\bullet$ \underline{\emph{Step 1. Edge Insertion:}} Given a graph $G$ and a rank  over all vertices in $G$, it initializes $G'$ as $G$, and  iterates every vertex in the increasing order of $\pi(w)$ (line 1-2). For every pair of vertices $u, v$  among the neighbors of $w$ in $G'$ with higher ranks than $w$, if $(u, v) \notin E(G')$, a new edge $(u, v)$ with weight $\phi((u,v), G') = \phi((u,w), G')+\phi((v,w), G')$ is inserted into $G'$ (line 5-7). Otherwise, if $\phi((u,w), G')+\phi((w,v), G') < \phi((u,v), G')$, it updates $\phi((u, v), G')$  as $\phi((u,w), G')$ $+\phi((w,v), G')$ (line 8-9). 

\noindent$\bullet$ \underline{\emph{Step 2. Edge Deletion:}} After the edge insertion step, it further iterates the vertex in the decreasing order of $\pi(w)$ (line 10). For every pair of vertices $u, v$  among the neighbors of $w$ in $G'$ with higher ranks than $w$ (line 11-12), if $\phi((w,v), G') + \phi((v,u), G') < \phi((w,u), G')$,  it updates $\phi((w,u), G')$ as $\phi((w,v), G') + \phi((v,u), G')$ and marks the updated edge as removed (line 13-15). At last, the marked edges in $G'$ are removed (line 16), and $\ev(w)$ for each vertex is set as $\nbr(w, G')$ (line 17-18).

\begin{figure}
    \centering
    \includegraphics[width=0.90\linewidth]{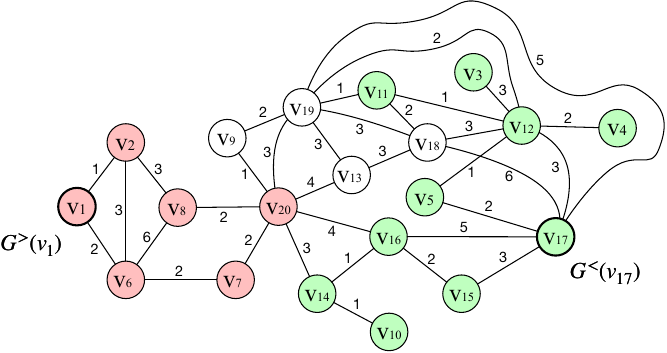}
    \caption{\sdgraph $G'$ of $G$}
    \label{fig:sdg}
\end{figure}
 
\begin{example} 
	Consider the road network $G$ in \reffig{roadnetwork} and assume the vertex order $\pi = (v_1, v_2, ..., v_{20})$, the \sdgraph $G'$ of $G$ is shown in \reffig{sdg}. To construct $G'$, we first conduct the edge insertion step. For $v_1$, its $\mathcal{N}$ is $\{ v_2, v_6 \}$. There exists no edge $(v_2, v_6)$ in $G'$ currently, then $(v_2, v_6)$  with  $\phi((v_2, v_6),G') = 3$ is added into $G'$. The procedure continues until all vertices are processed. In the edge deletion step, vertices are processed in the reverse order of $\pi$. Take $v_7$ as an example. When processing $v_7$, its $\mathcal{N}$ is $\{ v_8, v_{20} \}$. Since $\phi((v_7,v_{20}),G') + \phi((v_{20},v_8),G') = 2+2 < \phi((v_7,v_8),G')= 5$, $(v_7,v_8)$ is marked. When all the vertices are processed, the marked edges are removed, and \reffig{sdg} shows the final $G'$. 
\end{example}

Based on the procedure of \refalg{evgen}, we have:

\begin{lemma}
	\label{lem:sdgraph}
 The graph $G'$ generated at the end of \refalg{evgen} is a \sdgraph of $G$.
\end{lemma}

\proofsketch
Based on \refalg{evgen}, it is direct that $V(G') = V(G)$, meeting the condition (1) of \sdgraph. For any two vertices $u,v \in G$, $\sd((u,v),G) = \sum_{i=1}^{n}\phi((v_{i-1},v_i),G)$, where $\phi((v_{i-1},v_i),G) = \sd((v_{i-1},v_i),G)$. Clearly, $G'$ retains all edges with $\phi((u,v),G) = \sd((u,v),G)$ in $G$ and includes all inserted edges with $\phi((u,v),G') = \sd((u,v),G)$ in $G'$. Therefore, for any two vertices $u,v \in G'$, $\sd((u,v),G') = \sd((u,v),G)$, satisfying the condition (3) of \sdgraph. 
Next, we prove that $G'$  satisfies condition (2) via induction. Obviously, for $v_{n-1}$, $\phi((v_{n-1},v_n),G') = \sd((v_{n-1},v_n),G)$. Assume that for $v_{n-k}$, $\phi((v_{n-k},v),G') = \sd((v_{n-k},v),G)$ for $\forall v \in \mathcal{N}(v_{n-k})$, $\mathcal{N}(v_{n-k}) = \{ v | v \in \nbr(v_{n-k},G') \wedge \pi(v) > \pi(v_{n-k-1}) \}$. Now, we prove it for $v_{n-k-1}$. 
Suppose $v_{n-k}$ connects to $v_{n-k-1}$. From the insertion step, $v_{n-k-1}$ and $\mathcal{N}(v_{n-k-1})$ form a clique. Thus, the shortest path from $v_{n-k-1}$ to $v$ either only contains $v_{n-k-1}$ and $v$, or passes a vertex in $\mathcal{N}(v_{n-k-1})\setminus\{v_{n-k-1},v\}$, i.e, $\sd((v_{n-k-1},v),G') = \phi((v_{n-k-1},v),G')$ or $\sd((v_{n-k-1},v),G') = \min_{u \in \mathcal{N}}( \phi((v_{n-k-1}, u), G') + \sd((u, v),G'))$. Since the $\mathcal{N}(v_{n-k-1})$ includes $v_k$ and $\mathcal{N}(v_k)$, we have $\phi((u, v),G') = \sd((u,v),G')$ for any two vertices $u,v \in \mathcal{N}(v_{n-k-1})$. Thus, $\sd((v_{n-k-1},v),G') = \min_{u \in \mathcal{N}(v_{n-k-1})}( \phi((v_{n-k-1}, u),G') + \phi((u, v),G'))$. As line 10-15 of \refalg{evgen} can guarantee that $\phi((u,v),G') = \sd((u,v),G')$, and $G'$ satisfies condition (3) of \sdgraph, i.e. $\sd((u,v),G') = \sd((u,v),G)$, we have $\phi((u,v),G') = \sd((u,v),G)$.  Therefore, $G'$ is a \sdg of $G$.
\eop

Following \reflem{sdgraph}, it is clear that for each vertex $v \in V(G)$, its $k$NN in $G$ is the same as that in $G'$ based on the condition (3) of \refdef{sdg}. Moreover,  $\nbr(w, G')$ is the bridge neighbor set of $w$ in $G'$ based on the condition (2) of \refdef{sdg}. The following problem is how to compute $\knnalg(u)$ for each vertex $u$ via $G'$ and $\ev(u)$. According to the discussion in \refsec{mi}, to fully utilize  the intermediate computed results during the \knnindex construction, we define a special type of path based on the given total vertex order as follows:

\begin{definition}
\label{def:monotonic}
\textbf{(Monotonic Rank Path)} Given the \sdgraph $G'$ of a road network $G$, for a  vertex $u \in V(G')$, a path $p(u, v) = (u=v_1, v_2, \dots, v_j = v)$ in $G'$ is a  decreasing rank path of $u$ if $\pi(v_j) < \pi(v_{j-1}) < \dots <  \pi(v_1) $, and it is an  increasing  rank path of $u$ if $\pi(v_j) > \pi(v_{j-1}) > \dots> \pi(v_1)$.
\end{definition}

\begin{definition}
	\label{def:lg}
	\textbf{(Monotonic Rank Path Subgraph)} Given  the \sdgraph $G'$ of a road network $G$, for a vertex $u$, the  decreasing rank path subgraph of $u$, denoted by $\sdgl(u)$, is the subgraph induced by all   decreasing rank paths of $u$ in $G'$. The increasing rank path subgraph, denoted by $\sdgu(u)$, is the subgraph induced by all   increasing rank paths of $u$ in $G'$.
\end{definition}

\begin{example}
	Given the \sdg $G'$ in \reffig{sdg}, for vertex $v_1$, increasing rank paths of $v_1$ contain $p(v_1,v_2) = (v_1,v_2), p(v_1,v_6) = (v_1,v_6), p(v_1,v_8) = (v_1,v_2,v_8)$ or $(v_1,v_6,v_8), p(v_1,v_7) = (v_1,v_6,v_7)$ or $(v_1,v_2,v_6,v_7), p(v_1,v_{20}) = (v_1,v_2,v_8,v_{20})$, $(v_1,v_6,v_8,v_{20})$, $(v_1,v_2,\\v_7,v_{20})$, $(v_1,v_2,v_6,v_7,v_{20})$, $(v_1,v_2,v_6,v_8,v_{20})$. The increasing rank path subgraph of $v_1$, i.e., $\sdgu(v_1)$, is the subgraph induced by these paths, which is shown  in pink in \reffig{sdg}. The decreasing rank path subgraph of $v_{17}$, i.e., $\sdgl(v_{17})$, can be obtained similarly, which is shown in green in \reffig{sdg}.
\end{example}

\begin{definition}
	\label{def:pknn}
	\textbf{(Decreasing Rank Partial $k$NN)} Given a vertex $u \in V(G)$ and a set of candidate objects $\mathcal{M}$, the decreasing rank  partial $k$NN of $u$,  denoted by $\evl(u)$, is the $k$NN of $u$ in $\sdgl(u)$. 
	\end{definition}

\begin{lemma}
	\label{lem:naiveknn}
	Given a vertex $u \in V(G)$ in a road network $G$,  $\knnalg(u) \subseteq \cup_{w \in V(\sdgu(u))} \pknnalg(w)$.
\end{lemma}

\proofsketch
	This lemma can be proved directly based on \refpro{vcknn}.
\eop

Therefore, if we can obtain $\pknnalg(w)$ for each vertex, we can obtain $\knnalg(u)$ following \reflem{naiveknn}. Moreover, we have: 

\begin{lemma}
	\label{lem:first}
	Given a road network of $G$ and a set of candidate objects $\mathcal{M}$, let $u_1$ be the vertex with the lowest rank, we have $\pknnalg(u_1) = \{\mathcal{M} \cap \{ u_1 \}\}$.  
\end{lemma} 

\proofsketch
	From \refdef{lg}, $V(\sdgl(u_1)) = \{u_1\}$. Based on \refdef{pknn}, $\pknnalg(u_1) = \{\mathcal{M} \cap V(\sdgl(u_1)) \} = \{\mathcal{M} \cap \{ u_1 \}\}$.
\eop
 
Based on \reflem{first},  the decreasing rank partial $k$NN for the vertex with the lowest rank can be computed directly. Regarding the remaining vertices, we further divide $\ev(u)$ into two parts: $\keyL(u)$ which contains the neighbors of $u$ in $G'$ with lower rank than $u$, i.e., $\keyL(u) = 
\{ v | v \in \ev(u) \wedge \pi(v) < \pi(u) \}$ and $\keyU(u)$ which contains the neighbors of $u$ in $G'$ with higher rank than $u$, i.e., $\keyU(u) = 
\{ v | v \in \ev(u) \wedge \pi(v) > \pi(u) \}$. We have: 

\begin{lemma}
	\label{lem:pknn}
	Given a vertex $u \in V(G)$ in a road network $G$, $\pknnalg(u) \subseteq \{\mathcal{M} \cap \{ u \}\} \cup_{v \in \keyL(u)} \pknnalg(v)$.  
\end{lemma} 

\proofsketch This lemma can be proved directly based on \refpro{vcknn} and \refdef{pknn}.
%	We prove the lemma by contradiction. We assume that there exists a vertex $v \in \pknnalg(u)$, but $v \notin \cup_{w \in \keyL(u)} \pknnalg(w)$. Following \refdef{pknn}, there exists one $w \in \keyL(u)$ that $\pspath(u,v)$ pass through, i.e., $\pspath(u,v) = \spath(u,w) + \spath(w,v)$. For $\forall v_i \in \pknnalg(w)$, $i \in [1,k]$, we have $\psd(w,v_i) < \psd(w,v)$. Then $\phi'(u, w) + \psd(w,v_i) < \phi'(u,w) + \psd(w,v)$. It means there are at least $k$ vertices whose distance to $u$ is smaller than the shortest peak distance between $u$ and $v$, which contradicts $v \in \pknnalg(u)$. The proof completes.
\eop

\reflem{pknn} indicates the scope of  $\pknnalg(u)$ for each vertex. To obtain $\pknnalg(u)$, we only need to compute the distance between $u$ and  $w \in \{\mathcal{M} \cap \{ u \}\} \cup_{v \in \keyL(u)} \pknnalg(v)$, and retrieve the top $k$ objects. To avoid the expensive \kw{Dijkstra's} algorithm, we define: 

\begin{definition}
\label{def:dshortest}
\textbf{(Decreasing Rank Shortest Path)} Given  the \sdgraph $G'$ of a road network $G$, for two vertices $u, v \in V(G')$, the decreasing rank shortest path between $u$ and $v$ is the rank decreasing path from $u$ to $v$ with the smallest length in $G'$. 
\end{definition}

In \sdg $G'$ of $G$, for any two vertices $u,v \in G'$, one shortest path between $u$ and $v$ is a decreasing rank shortest path. We call the length of decreasing rank shortest path between $u$ and $v$ as decreasing rank distance and denote it as $\psd(u, v)$. We have:

\begin{lemma}
	\label{lem:pknnDist}
	Given the \sdg $G'$ of a road network $G$,  for a vertex $u \in V(G')$, $\psd(u,v) = \kw{min}_{ w \in \keyL(u)} \{ \phi((u,w), G') + \psd(w,v) \}$, where $v \in \pknnalg(u)$.  
\end{lemma} 

\proofsketch
	Based on \refdef{lg} and \refdef{pknn}, we have $\pknnalg(u) \subseteq \{ \mathcal{M} \cap V(\sdgl(u)) \}$. According to \refdef{dshortest}, for $\forall v \in \sdgl(u)$, there is one decreasing rank shortest path between $u$ and $v$, which passes through one vertex $w \in \keyL(u)$. Therefore, $\psd(u,v) = min_{ w \in \keyL(u)} \{ \phi((u,w), G') + \psd(w,v) \}$.
\eop

\begin{lemma}
\label{lem:pknntopk}
	Given the \sdg $G'$ of a road network $G$, for a vertex $u$, let $v \in \pknnalg(u) \cap \knnalg(u)$, if  $\psd(u,v) = \sd((u,v),G')$, there is a shortest path between $u$ and $v$ in $G'$, which is also a decreasing rank shortest path.
\end{lemma}

\proofsketch 
According to \refdef{lg} and \refdef{pknn}, if $v \in \pknnalg(u)$, we know $v \in V(\sdgl(u))$. Based on \refdef{dshortest}, there is one decreasing rank shortest path between $u$ and $v$. When $\psd(u,v) \\= \sd((u,v),G')$, there is a shortest path between $u$ and $v$ in $G'$, which is also a decreasing shortest path.
\eop

Based on \reflem{first}, \reflem{pknn}, and \reflem{pknnDist}, to obtain $\pknnalg(u)$ for each vertex,  we can adopt a bottom-up strategy based on the increasing order of $\pi(u)$, and the computed distance  for a lower rank vertex can be re-used to compute the distance for a higher rank vertex. However,  $\pknnalg(u)$ only contains the vertices $v \in \knnalg(u)$ whose shortest paths to $u$ pass through $\keyL(u)$, the vertices $v \in \knnalg(u)$ whose shortest paths to $u$ pass through $\keyU(u)$ does not considered. Unfortunately, these vertices cannot be obtained by only exploring the vertices in $\cup_{v \in \keyU(u)} \pknnalg(v)$ in the similar way as discussed above since this approach only explores the vertices whose ranks are not higher than $\kw{max}_{v \in \keyU(u)}\pi(v)$. On the other hand, we have the following lemmas regarding the distance between $u$ and $v \in \knnalg(u)$  based on \reflem{naiveknn}:

\begin{lemma}
	\label{lem:ugsd}
	Given the \sdg $G'$ of a road network $G$, for a vertex $u \in V(G)$, let 
	 $v, v' \in V(\sdgu(u))$, $\sd((v, v'), \sdgu(u)) = \sd((v, v'), G)$.
\end{lemma}

\proofsketch
This lemma can be proved by \reflem{sdgraph}. 
\eop

\begin{lemma}
	\label{lem:naiveknnDist}
	Given the \sdg $G'$ of a road network $G$, for a vertex $u  \in V(G)$, $ \sd((u,v), G) = \kw{min}_{w \in V(\sdgu(u))} \{ \sd((u,w), \sdgu\\(u)) + \psd(w,v) \}$, where $v \in \knnalg(u)$.
\end{lemma}

\proofsketch
This lemma can be proved directly based on \reflem{naiveknn} and \reflem{ugsd}.
%	Based on \reflem{naiveknn}, for $\forall v \in \knnalg(u)$, $v$ comes from $\cup_{w \in UG(u)} \pknnalg(w)$, so $\psd(w,v)$ can be directly achieved. For $\forall v \in \knnalg(u)$ and $v \notin \pknnalg(u)$, there exists a vertex $w \in UG(u)$ such that $\sd(u,v) = \sd(u,w) + \sd(w,v)$. For each vertex $w \in UG(u)$, $\sd(u,w)$ can be calculated in $UG(u)$ according to \reflem{ugsd}. For $\sd(w,v)$ computation, there is 2 cases: (1) When $\spath(w,v)$ is a peak path, $\sd(w,v) = \psd(w,v)$, which can be from $\pknnalg(w)$ directly. (2) $\spath(w,v)$ is not a peak path, we aim to find the vertex $p$ with highest rank in $\spath(w,v)$. According to \refdef{lg}, $p$ must exists in $UG(u)$. Here, $\sd(u,v)$ can be calculated as $\sd(u,v) = \sd(u,p) + \sd(p,v)$. $\sd(u,p)$ can be calculated in $UG(u)$. For $\sd(p,v)$, $\spath(p,v)$ is definitely a peak path, $\sd(p,v) = \psd(p,v)$, which can be from $\pknnalg(p)$ directly.  
\eop

{\setlength{\textfloatsep}{2pt}
\setlength{\intextsep}{9pt}
\SetInd{0.6em}{0.6em}
\small
\begin{algorithm}[htb]
    \caption{$\kw{KNN}\textrm{-}\kw{Index}\textrm{-}\kw{Cons}(G, \pi, \mathcal{M})$}
    \label{alg:naivecon}
    \DontPrintSemicolon
        $G' \gets  \sdgraphgen(G, \pi)$; \;
        $\mathcal{S} \gets \emptyset$,  $\pknnalg(\cdot) \gets \emptyset$, $\knnalg(\cdot) \gets \emptyset$; \; 
        \For{\textbf{each} $u$ in increasing order of $\pi(u)$}{
            $\mathcal{S} \gets \{\mathcal{M} \cap \{ u \} \} \cup_{w \in \keyL(u)} \pknnalg(w)$; \;
            \For{\textbf{each} $v \in \mathcal{S}$}{
            $ \psd(u,v) \gets \min_{w \in \keyL(u)} \{ \phi((u,w), G') + \psd(w,v)\}$; 
            }
            $\pknnalg(u) \gets$  $k$ vertices in $\mathcal{S}$ with the smallest $\psd(u,v)$; \;
        }
        \For{\textbf{each} $u$ in increasing order of $\pi(u)$}{
        	construct $\sdgu(u)$ by conducting \kw{BFS} search from $u$ on $G'$ following edge $(v, v')$ with $\pi(v) < \pi(v')$; \;
        	\For{\textbf{each} $w \in V(\sdgu(u))$}{
        	compute $\sd((u,w), \sdgu(u))$; \;
        	}
        	$\mathcal{S} \gets \pknnalg(u) \cup_{w \in V(\sdgu(u))} \pknnalg(w)$; \;
        	\For{\textbf{each} $v \in \mathcal{S}$}{
            	$\sd((u,v), G) \gets \min_{w \in \sdgu(u)} \sd((u,w), \sdgu(u)) $+$\psd(w,v)$;
            }
            $\knnalg(u) \gets$ $k$ vertices in $\mathcal{S}$ with the smallest $\sd((u,v), G)$; \;
        }
\end{algorithm}
}

\begin{figure*}[htb]
\centering
\subfigcapskip=0pt
  \hspace{-1.35cm}
  \subfigure[$\mathcal{S}$ for $\kw{V}^<_5(v_{17})$]{\includegraphics[width=0.275\linewidth]{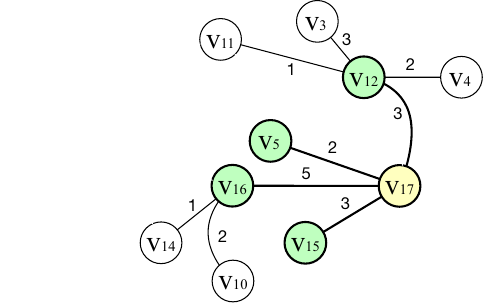}\label{fig:pknn_f}}
  \hspace{0.1cm}
  \subfigure[Computation of $\kw{V}^<_5(v_{17})$]{\includegraphics[width=0.24\linewidth]{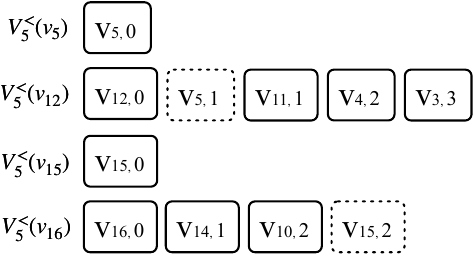}\label{fig:pknn_t}}
  \hspace{0.1cm}
  \subfigure[$\mathcal{S}$ for $\kw{V}_5(v_{17})$]{\includegraphics[width=0.275\linewidth]{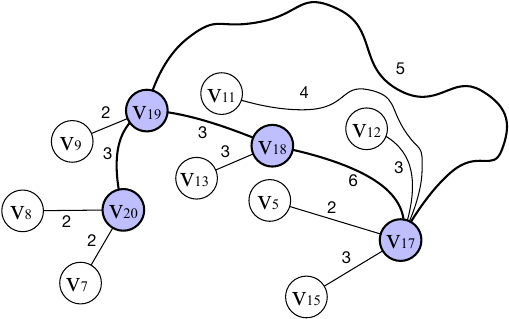}\label{fig:knn_n_f}}
  \hspace{0.1cm}
  \subfigure[Computation of $\kw{V}_5(v_{17})$]{\includegraphics[width=0.24\linewidth]{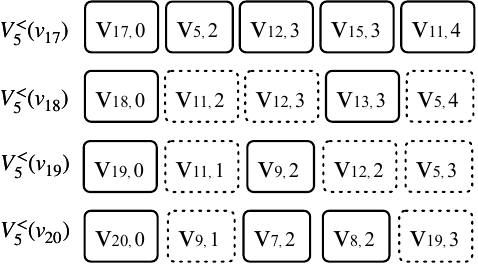}\label{fig:knn_n_t}}
 \caption{Procedure of \refalg{naivecon} to Compute $\kw{V}_5(v_{17})$}
  \label{fig:pknn_naive}
  \vspace{-0.1cm}
\end{figure*}

\stitle{Algorithm.} By combing the above two cases together, our index construction algorithm is shown in \refalg{naivecon}. It first generates the \sdgraph $G'$ using \refalg{evgen} (line 1). Then, it adopts a bottom-up strategy to compute $\pknnalg(u)$ in the increasing order of $\pi(u)$ (line 2-7). Specifically, for each vertex $u$, it retrieves $\{\mathcal{M} \cap \{ u \}\} \cup_{v \in \keyL(u)} \pknnalg(v)$ based on \reflem{pknn} (line 4) and computes $\psd(u,v)$ based on \reflem{pknnDist} (line 5-6). Then, $\pknnalg(u)$ is the $k$ vertices in $\mathcal{S}$ with the smallest $\psd(u,v)$ (line 7). After that,  it constructs $\sdgu(u)$ by conducting \kw{BFS} search from $u$ on $G'$ (line 9). And we compute the single source shortest distance $\sd((u,w), \sdgu(u))$ from $u$ to each vertex $w$ in $\sdgu(u)$ using the Dijkstra's Algorithm (line 10-11). Then, following \reflem{naiveknn}, it retrieves $ \cup_{w \in V(\sdgu(u))} \pknnalg(w)$ (line 12) and computes $\sd((u,v), G)$ based on \reflem{naiveknnDist} (line 13-14). $\psd(w,v)$ can be obtained from $\pknnalg(w)$ directly. $\sd((u,w), \sdgu(u))$ can be computed (line 10-11) after the construction of $\sdgu(u)$ (line $9$) following \reflem{ugsd}. At last, the $k$ vertices in $\mathcal{S}$ with the smallest $\sd((u,v), G)$ is returned as $\knnalg(u)$ in line 15.

\begin{example}
\label{ex:ugcon} 
Following the \sdgraph $G'$ in \reffig{sdg}, \reffig{pknn_naive} takes $v_{17}$ as an example to show the procedure of \refalg{naivecon} to compute $\kw{V}_5(v_{17})$. According to \refalg{naivecon}, we compute $\pknnalgex(v_{17})$ first. Based on $G'$,  $\keyL(v_{17}) = \{ v_5, v_{12}, v_{15}, v_{16} \}$, which is shown in green in \reffig{pknn_naive} (a). Following \refalg{naivecon}, when computing $\pknnalgex(v_{17})$, we already have $\pknnalgex(v_{5}), \pknnalgex(v_{15}), \pknnalgex(v_{12})$ and $\pknnalgex(v_{16})$, which is shown in \reffig{pknn_naive} (b). Consequently, following line 6 of \refalg{naivecon}, we can achieve $\mathcal{S} = \{ \mathcal{M} \cap \{ v_{17} \} \} \cup_{ w \in \keyL(v_{17}) } \pknnalgex(w) = \{ (v_{17},0), (v_{5},2), (v_{12},3), (v_{15},3), (v_{11},4), (v_{4},5), (v_{16},5), (v_{3},6), (v_{14},\\6), (v_{10},7)  \}$. \reffig{pknn_naive} (b) shows this set $\mathcal{S}$ for constructing $\pknnalgex(v_{17})$. After sorting distance, we have $\pknnalgex(v_{17}) = \{ (v_{17},0), (v_{5},2), (v_{12},3), \\(v_{15},3), (v_{11},4) \}$.

 \reffig{pknn_naive} (c) shows the $\sdgu(v_{17})$ in purple with bold lines. Using \kw{Dijkstra's} Algorithm, we compute the distance from $v_{17}$ to each vertex in $\sdgu(v_{17})$. And $\sd((v_{17},v_{18}),  \sdgu(v_{17})) = 6, \sd((v_{17},v_{19}), \\ \sdgu(v_{17})) = 5,$ and $\sd((v_{17},v_{20}), \sdgu(v_{17})) = 8$. Following line 12 of \refalg{naivecon}, when computing $\knnalgex(v_{17})$, we  have $\pknnalgex(v_{18}), \pknnalgex(v_{19})$ and $\pknnalgex(v_{20})$, which is shown in \reffig{pknn_naive} (d). Following line 14 of \refalg{naivecon}, we achieve $\mathcal{S} = \pknnalgex(v_{17}) \cup_{w \in \sdgu(v_{17})} \pknnalgex(w) = \{ (v_{17},0), (v_{5},2), (v_{12},3), (v_{15},3), (v_{11},4), (v_{19},5), (v_{18},6), (v_{9},7), (v_{20}\\, 8), (v_{13},9), (v_{7},10), (v_{8},10) \}$, which is shown in \reffig{pknn_naive} (c). Then, we select $5$ nearest objects from $\mathcal{S} $ as the \knnindex of $v_{17}$, namely, $\knnalg(v_{17}) = \{ (v_{17},0), (v_{5},2), (v_{12},3), (v_{15},3), (v_{11},4) \}$ .

\end{example}

The correctness of \refalg{naivecon} is straightforward following the above discussion. For the efficiency of the algorithm, we have:

\begin{theorem}
	\label{thm:naiveknn} 
	The time complexity of \refalg{naivecon} is bounded by $O(n \cdot (\rho^2 + \eta \cdot \tau \cdot log(\eta) + (\tau+\eta) \cdot k))$, where $\rho$ represents the maximum degree of vertices in the graph generated by \refalg{evgen} when Step 1 finishes, $\eta = \kw{max}_{v \in V(G)}|\sdgu(v)|$ and $\tau = \kw{max}_{v \in V(G)}|\keyU(v)|$.
\end{theorem}

\proofsketch  
\refalg{evgen} requires $O(n \cdot \rho^2)$ time (line 1 of \refalg{naivecon}). Specifically, in the for loop (line 2-9 of \refalg{evgen}), for each vertex $w$, line 4-9 of \refalg{evgen} takes $O(\rho^2)$ time and the for loop terminates at $n$ iterations. Therefore, the edge insertion step (line 2-9 of \refalg{evgen}) requires $O(n \cdot \rho^2)$ time. Similarly, the edge deletion step requires $O(n \cdot \rho^2)$ (line 10-16 of \refalg{evgen}). Scanning all vertices to achieve $\ev(\cdot)$ is bounded by $O(n \cdot \tau)$ (line 17-18 of \refalg{evgen}). Obviously, for $\forall u \in V(G)$, $\tau \leq \rho$. Therefore, the time complexity of \refalg{evgen} is $O(n \cdot (\rho^2 + \tau)) = O(n \cdot \rho^2)$. In the for loop from line 3 to line 7 of  \refalg{naivecon}, line 4 of  \refalg{naivecon} takes $O(\tau \cdot k)$, since each vertex $u$ is only explored by the vertex $w \in \keyU(u)$. At the same time with obtaining $\mathcal{S}$ (line 4 of \refalg{naivecon}), line 5-7 of \refalg{naivecon} could be done. Therefore, $\pknnalg(\cdot)$ construction (line 3-7 of  \refalg{naivecon}) requires $O(n \cdot \tau \cdot k)$ time. In the for loop (line 8-15 of \refalg{naivecon}), constructing $\sdgu(u)$ by conducting \kw{BFS} search requires $O(\eta \cdot \tau)$ time (line 9 of \refalg{naivecon}). Computing $\sd(u, v)$ for $\forall v \in V(\sdgu(u))$ via \kw{Dijkstra's} algorithm (line 10-11 of \refalg{naivecon}) consumes $O(\eta \cdot \tau \cdot log(\eta))$ time. Obtaining $\mathcal{S}$ and distance computation require $O(\eta \cdot k)$ (line 12-15 of \refalg{naivecon}). Therefore, $\knnalg(\cdot)$ construction (line 8-15 of \refalg{naivecon}) requires $O(n \cdot (\eta \cdot \tau \cdot log(\eta) + (\tau + \eta) \cdot k))$. In summary, the time complexity of \refalg{naivecon} is $O(n \cdot (\rho^2 + \eta \cdot \tau \cdot log(\eta) + (\tau+\eta) \cdot k))$.
\eop

%$\knnalg(\cdot)$ construction (lines 8-15) costs $O(n \cdot \eta \cdot \tau \cdot log(\eta) + n \cdot \eta \cdot k)$. Obtaining $\sdgu(u)$ by BFS (line 9) and computing $\sd(u, v)$ for $\forall v \in V(\sdgu(u))$ by \kw{Dijkstra's} algorithm (line 11-12) require $O(\eta \cdot \tau \cdot log(\eta))$ time. Obtaining $\mathcal{S}$ and distance computation is bounded by $O(\eta \cdot k)$ (line 12-14).

\stitle{Remark.} Based on \refthm{naiveknn}, we prefer the generated \sdgraph with smaller $\rho$ and $\tau$. Thus, we use the following heuristic total order $\pi$ in this paper: (1) The vertex with the minimum degree in $G$ has the lowest rank (the vertex with a smallest \kw{id} has the lowest rank if more than one vertices have the minimum degree); (2) for two unprocessed vertices $u$ and $v$ in line 2 of \refalg{evgen}, $\pi(u) > \pi(v)$ if the number of unprocessed neighbors of $u$ is bigger than that of $v$ in $G'$. Note  this order can be obtained incidentally in \refalg{evgen}, and does not affect the time complexity of \refalg{evgen}.

\subsection{A Bidirectional Construction Algorithm}
\label{sec:utl}

\refalg{naivecon} adopts a bottom-up strategy to construct the \knnindex with which the computation regarding $\pknnalg(u)$ is well shared. However, it still needs to invoke \kw{Dijkstra's} algorithm to compute the distance between $u$ and $w \in V(\sdgu(u))$ in line 11-12, which is costly. To address this problem,  we propose a new  algorithm to further improve the index construction efficiency. Instead of following the sole bottom-up direction which adopted in \refalg{naivecon},  the new algorithm constructs the index in a bidirectional manner, which totally avoids   the invocation of \kw{Dijkstra's} algorithm. Before introducing our algorithm, we have:

\begin{lemma}
	\label{lem:finalv}
	Given a road network $G$, let $u_n$ be the vertex with the highest rank, $\knnalg(u_n) = \pknnalg(u_n)$. 
\end{lemma}

\proofsketch
Following \refdef{lg}, $V(\sdgu(u_n)) = \{ u_n \}$. Based on \reflem{naiveknn}, $\knnalg(u_n) \subseteq \cup_{w \in V(\sdgu(u_n))} \pknnalg(w) = \pknnalg(u_n)$.
\eop

\begin{lemma}
	\label{lem:knn}
	Given the \sdgraph $G'$ of a road network $G$, for a vertex $u \in V(G)$,  $\knnalg(u) \subseteq \pknnalg(u) \cup_{w \in \keyU(u)} \knnalg(w)$.
\end{lemma}

\proofsketch
	This lemma can be proved directly based on \refpro{vcknn}.
\eop

\reflem{finalv} and \reflem{knn} imply that if we process the vertices in the decreasing order of their ranks when computing $\knnalg(u)$, it can re-use the computed information of vertices with higher ranks in the computation of the $k$NN for vertices with lower ranks. Moreover, we have:

\begin{lemma}
	\label{lem:knnDist}
	Given the \sdgraph $G'$ of a road network $G$, for a vertex $u \in V(G)$,  $\sd((u,v), G) = \kw{min} \{ \kw{min}_{w \in \keyU(u)} \{ \phi((u,w), G') \\+ \sd((w,v), G) \}, \psd(u,v) \}$, where $v \in \knnalg(u)$. 
\end{lemma}

\proofsketch
For $v \in \knnalg(u)$, there are two parts. The one part contains all $v$ whose shortest paths to $u$ pass through $\keyU(u)$, this distance computation can be proved based on \refpro{sd}. The other part contains all $v$ whose shortest paths to $u$ pass through $\keyL(u)$, $\psd(u,v)$ can be directly obtained from $\pknnalg(u)$ based on \reflem{pknntopk}.
\eop

{\setlength{\textfloatsep}{2pt}
\setlength{\intextsep}{9pt}
\SetInd{0.6em}{0.6em}
\small
\begin{algorithm}[tb]
    \caption{$\kw{KNN}\textrm{-}\kw{Index}\textrm{-}\kw{Cons}^+(G, \pi, \mathcal{M})$}
    \label{alg:utl}
    \DontPrintSemicolon
        $G' \gets \sdgraphgen(G, \pi)$ ; \;
        $\pknnalg(\cdot) \gets$ line 3-7 of \refalg{naivecon}; \; 
        $\mathcal{S} \gets \emptyset$, $\knnalg(\cdot) \gets \emptyset$; \; 
        \For{\textbf{each} $u$ in decreasing order of $\pi(u)$}{
        	$\mathcal{S} \gets \pknnalg(u) \cup_{w \in \keyU(u)} \knnalg(w)$;  \;
        	\For{\textbf{each} $v \in \mathcal{S}$}{
        		$d \gets  \min_{w \in \keyU(u)} \{ \phi((u,w), G') + \sd((w,v), G) \};$\;
        		$\sd(u,v) \gets \min\{d, \psd(u,v)\}$; \;
            }
        	$\knnalg(u) \gets$ $k$ vertices in $\mathcal{S}$ with the smallest $\sd((u,v), G)$; \;
        }
\end{algorithm}
}

\stitle{Algorithm.} Following \reflem{knnDist}, our new bidirectional construction algorithm is shown in \refalg{utl}. It first generates the \sdgraph $G'$ of $G$ using \refalg{evgen} (line 1) and computes $\pknnalg(u)$ in the same way as \refalg{naivecon} (line 2). After that, it processes the vertices in the decreasing order of their ranks (line 4-9). For each vertex $u$, it retrieves  $\pknnalg(u) \cup_{w \in \keyU(u)} \knnalg(w)$ based on \reflem{knn} and stores them in $\mathcal{S}$ (line 5). Then, the distance between $u$ and $v \in \mathcal{S}$ is computed following \reflem{knnDist} (line 6-8). Since the index construction procedure follows the decreasing order of $\pi(u)$, $\knnalg(w)$ for $\forall w \in \keyU(u)$ has been computed before computing $\knnalg(u)$. $\sd((w,v), G)$ can be obtained from $\knnalg(w)$ directly. And $\phi((u,w), G')$ can be achieved from \sdg directly. At last, the $k$ vertices in $\mathcal{S}$ with the smallest $\sd((u,v), G)$ is returned as $\knnalg(u)$ in line 9.

\begin{figure}
    \centering
    \subfigcapskip=0pt
    \subfigcaptopadj=0pt  
    \subfigure[$\mathcal{S}$ for $\kw{V}_5(v_{17})$]{\includegraphics[width=0.50\linewidth]{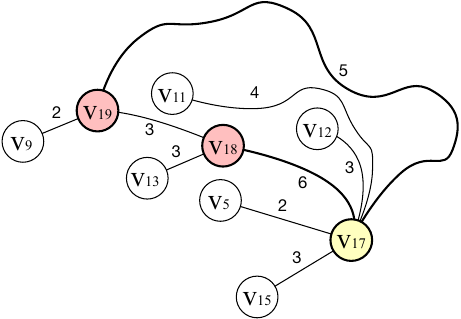}\label{fig:knn_f}}
    \hspace{0.15cm}
    \subfigure[Computation of $\kw{V}_5(v_{17})$]{\includegraphics[width=0.45\linewidth]{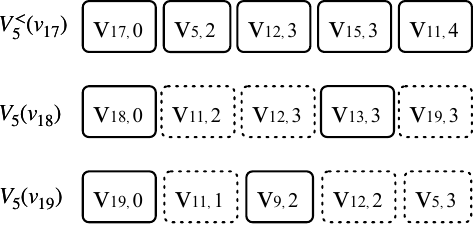}\label{fig:knn_t}}
    \caption{Procedure of \refalg{utl} to Compute $\knnalgex(v_{17})$}
    \label{fig:knn}
    \vspace{-0.3cm}
\end{figure}

\begin{example}

\reffig{knn} shows the $\knnalgex(v_{17})$ construction procedure following \refalg{utl}. Based on the \sdgraph $G'$ in \reffig{sdg}, for  $v_{17}$,  $\keyU(v_{17}) = \{ v_{18}, v_{19} \}$, which is shown in pink in \reffig{knn} (a).  $\pknnalgex(v_{17})$ can be constructed in the same way as shown in \refex{ugcon}.  Following line 5 of \refalg{utl}, when computing $\pknnalgex(v_{17})$, we already have $\pknnalgex(v_{17}), \knnalgex(v_{18})$, and $\knnalgex(v_{19})$, which is shown in \reffig{knn} (b). According to line 7-8 of \refalg{utl}, we have $\mathcal{S} = \pknnalgex(v_{17}) \cup_{w \in \keyU(v_{17})} \knnalgex(w) = \{ (v_{17},0), (v_{5},2), (v_{12},3), (v_{15},\\3), (v_{11},4), (v_{19},5), (v_{18},6), (v_{9},8), (v_{13},9) \}$. After sorting distance, the $5$ nearest neighbors for $v_{17}$ is selected from the set $\mathcal{S}$, namely, $\knnalgex(v_{17}) = \{ (v_{17}, 0), (v_{5}, 2), (v_{12}, 3), (v_{15}, 3), (v_{11}, 4) \}$.
\end{example}

\begin{theorem}
	\label{thm:tc_utl}
	Given a road network $G$, the time complexity of \refalg{utl} is bounded by $O(n \cdot \rho^2 + n \cdot \tau \cdot k)$  where $\rho$ represents the maximum degree of vertices in the graph generated by \refalg{evgen} when Step 1 finishes, and $\tau = \kw{max}_{v \in V(G)}|\keyU(v)|$.
\end{theorem}

\proofsketch
As proved in \refthm{naiveknn}, \refalg{evgen} requires $O(n \cdot \rho^2)$ time (line 1 of \refalg{utl}) and $\pknnalg(\cdot)$ construction requires $O(n \cdot \tau \cdot k)$ (line 2 of \refalg{utl}).  In the for loop from line 4 to line 9 of \refalg{utl}, obtaining $\mathcal{S}$ and distance computation require $O(\tau \cdot k)$ (line 5-9 of \refalg{utl}) and the loop terminates in $n$ iterations. Therefore, the for loop takes $O(n \cdot \tau \cdot k)$ (line 4-9 of \refalg{utl}). In summary, the bidirectional $\knnindex$ construction (\refalg{utl}) requires $O(n \cdot \rho^2 + n \cdot \tau \cdot k)$.
\eop

Compared with \refthm{naiveknn}, \refthm{tc_utl} shows that the time complexity  of our new bidirectional  algorithm to construct the index is significantly improved, which is also verified by the experimental results illustrated in \refsec{exp}.

\section{Candidate Object Update}
\label{sec:update}

In some cases, the candidate objects $\mathcal{M}$ may be updated by inserting new objects or deleting existing objects. Straightforwardly, we can reconstruct the index from scratch by \refalg{utl} to handle the update. However, this approach is inefficient as the update of a candidate object may not affect the $k$NN results of all the vertices. In this section, we discuss how to maintain the \knnindex incrementally when the candidate objects are updated. 

Obviously, when a candidate object $u$ is inserted or deleted, the update of $u$ will not  affect the $k$NN results of a vertex $v$ if $u$ and $v$ are far away from each other. Specifically, let $v_k$ be the vertex in $\knnalg(v)$ with the largest distance to $v$. If $\sd(u,v) > \sd(v,v_k)$, then $u$ cannot be in $\knnalg(v)$, which means deleting or inserting $u$ will not affect $\knnalg(v)$. Moreover, we have the following lemma based on \refpro{vcknn} and \refpro{sd}: 

\begin{lemma}
	\label{lem:stopCon} Given the \sdg $G'$ of the road network $G$, for a vertex $v \in V(G))$,  when an object $u$ is inserted/deleted, $\knnalg(v)$ could be affected if and only if there exists at least one vertex $w \in \ev(v)$ whose $\knnalg(w)$ changes due to the update of $u$. 
\end{lemma}

\proofsketch
	This lemma can be directly proved based on  \refpro{vcknn} and \refpro{sd}.
\eop

Therefore, we can maintain the \knnindex starting from the vertex of the updated object $u$. Based on the definition of $k$NN, it is clear that $\knnalg(u)$ will be changed. Following \reflem{stopCon}, the change of $\knnalg(u)$ will possibly lead to the change of $\knnalg(v)$ where  $v \in \ev(u)$. Then, we check whether $\knnalg(v)$ needs to be updated based on $\sd(u,v)$ and $\sd(v,v_k)$ as discussed above. We continue to repeat the above procedure recursively, and it is obvious that the \knnindex is correctly maintained when no more vertices whose $k$NN results change. Based on the above idea, our algorithms to handle the candidate object insertion and deletion are shown in \refalg{insert} and \refalg{delete}, respectively.

{\setlength{\textfloatsep}{2pt}
\setlength{\intextsep}{9pt}
\SetInd{0.6em}{0.6em}
\small
\begin{algorithm}[!htb]
        \caption{$\kw{KNN}\textrm{-}\kw{Index}\textrm{-}\kw{Ins}(G', \knnalg(\cdot), u)$}
        \label{alg:insert}
        \DontPrintSemicolon 
           	$\sd[\cdot] \gets +\infty$; $\mathcal{S} \gets \{ \emptyset \}$; $Q \gets \emptyset$; \;
            $\sd[u] \gets 0$; $\mathcal{S} \gets \{ u \}$; $ Q.push(u)$; \;
           	\While{$Q \neq \emptyset$}{
           		$w \gets Q.\kwnospace{pop}()$; \;
           		\For{\textbf{each} $v \in \ev(w)$}{
           			$\sd[v] \gets \min\{\sd[v], \sd[w] + \phi((w,v),G')\}$; \;
           			\If{$v \notin \mathcal{S} \wedge \kw{checkIns}(v,  \knnalg(v), \sd[v])$}{
           				$Q.\kw{push}(v)$; $\mathcal{S} \gets \mathcal{S} \cup \{ v \}$;\;
           			}
           		}
           	}
           	\For{\textbf{each} $v \in \mathcal{S}$}{
           	remove $v_k$ from $\knnalg(v)$; 	insert $u$ into $\knnalg(v)$;\;
           	}
    \BlankLine
	\SetKwProg{myproc}{Procedure}{}{}
	\myproc{\kwnospace{checkIns}$(v, \knnalg(v), d)$}{
		$v_k \gets$ the vertex with the largest distance to $v$ in $\knnalg(v)$;\;
		\textbf{if} $ \sd(v,v_k) \leq d$ \textbf{then} \KwRet{$\kw{False}$};\;
		\textbf{else} \KwRet{$\kw{True}$};
	}
\end{algorithm} 
}

\stitle{Object Insertion.} \refalg{insert} shows the algorithm for candidate object insertion. An array $\sd[\cdot]$ stores the distance between $u$ and other vertices, a set $\mathcal{S}$ stores vertices whose  $k$NN results should be updated, and a queue $Q$ stores the vertices whose bridge neighbor sets should be checked (line 1). Then, it initializes $\sd[u]$ as $0$ and adds $u$ in $\mathcal{S}$ and $Q$ (line 1). After that, it pops a vertex $w$ from $Q$ (line 4), and for each $v \in \ev(w)$, it computes the distance between $u$ and $v$, which can be obtained based on the fact $\sd(u, v) = \kw{min}_{ w' \in \ev(v)}  \{ \sd(u, w')  + \sd(w',v) \}$ (line 6). Note that  instead of visting all  $w' \in \ev(v)$, only the vertices whose $\knnalg(w)$ changes due to the update of $u$ need to be explored following \reflem{stopCon}, which is captured  by $Q$.  If $v$ is not in $\mathcal{S}$ and $\sd[v]$ is smaller than the distance between $v$ and $v_k$, where $v_k$ is the vertex with the largest distance to $v$ in $\knnalg(v)$ which can be obtained directly based on \knnindex, it adds $v$ into $Q$ and $\mathcal{S}$ (line 7-8).  The procedure terminates when $Q$ becomes empty (line 3). At last, for each vertex $v \in \mathcal{S}$,  it removes $v_k$ from $\knnalg(v)$ and inserts $u$ into $\knnalg(v)$ (line 9-10). When inserting $u$ into $\knnalg(u)$, $\sd[v]$ is the shortest distance between $u$ and $v$, which can be guaranteed by \reflem{stopCon}.

\begin{theorem}
Given the \sdg $G'$ and its corresponding \knnindex of a road network $G$, \refalg{insert} maintains the \knnindex correctly when an object $u$ is inserted.	
\end{theorem}

\proofsketch
\refalg{insert} (line 5-6) can guarantee $\sd[v]$ for $\forall v \in \mathcal{S}$ is the distance between $u$ and $v$ before inserting $u$ to $\knnalg(v)$ for $v \in \mathcal{S}$  (line 9-10). Even though when using $\kw{checkIns}(v,\knnalg(v),d)$ (line 7), $d$ may not be the distance between $u$ and $v$ and $\kw{checkIns}(v,\knnalg(v),d)$ returns \kw{True}, the final result can not be affected. Since $\sd(u,v) \leq d$, $\kw{checkIns}$ returning \kw{True} denotes that $d < \sd(v,v_k)$. Therefore, we have $\sd(u,v) < \sd(v,v_k)$. Overall, \refalg{insert} maintains the \knnindex correctly when an object $u$ is inserted.	 
\eop

\begin{theorem}
Given the \sdg $G'$ and its corresponding \knnindex of a road network $G$, when an object $u$ is inserted, \refalg{insert} maintains the \knnindex in  $O(\Delta \cdot \tau')$, where $\Delta = |\mathcal{S}|$ and $\tau' = \kw{max}_{v \in V(G)}|\ev(v)|$. 
\end{theorem}

\proofsketch
The time complexity of $\kw{checkIns}(v,\knnalg(v),d)$ (line 11-14) is $O(1)$. In the for loop (line 3-8), for each vertex $w$, line 5-8 requires $O(\tau')$ time and the loop terminates in at most $\Delta$ iterations. Therefore, the for loop (line 3-8) requires $O(\Delta \cdot \tau')$ time. In the for loop (line 9-10), removing $v_k$ from $\knnalg(v)$ requires $O(1)$, inserting $u$ into $\knnalg(v)$ in the correct position needs $O(k)$ and the loop stops in $\Delta$ iterations. Therefore, the for loop (line 9-10) requires $O(\Delta \cdot k)$ time. In summary, the overall time complexity of \refalg{insert} is $O(\Delta \cdot (\tau' + k) =  O(\Delta \cdot \tau')$, since in real applications the parameter $k$ is not large and $k<\tau'$. 
\eop 

% Since $\knnalg(v)$ is stored in the doubly linked list, $v_k$ and $\sd(v,v_k)$ can be directly obtained from $\knnalg(v)$.
%Since we store $\knnalg(\cdot)$ in the bidirectional linked lists, the last vertex $v_k$ and $\sd(v,v_k)$ can be directly obtained from $\knnalg(v)$. $\kw{checkIns}(v,\knnalg(v),d)$ only costs $O(1)$ (line 11-14). The loop will conduct $O(\Delta)$ time, the time complexity is $O(\Delta \cdot \tau')$ (line 3-8). Inserting $u$ into $\knnalg(v)$ needs $O(k)$ and deleting the last one in $\knnalg(v)$ only takes $O(1)$, the time can be bounded by $O(\Delta \cdot k)$ (line 9-10). The overall time complexity is bounded by $O(\Delta \cdot (\tau' + k)) \leq O(\Delta \cdot \tau' \cdot 2) = O(\Delta \cdot \tau')$, since in real applications the parameter $k$ is not large.

{\setlength{\textfloatsep}{2pt}
\setlength{\intextsep}{9pt}
\SetInd{0.6em}{0.6em}
\small
\begin{algorithm}[htb]
        \caption{$\kw{KNN}\textrm{-}\kw{Index}\textrm{-}\kw{Del}(G', \knnalg(\cdot), u)$}
        \label{alg:delete}
        \small
        \DontPrintSemicolon   
            $\sd[\cdot] \gets +\infty$; $\mathcal{S} \gets \{ \emptyset \}$; $Q \gets \emptyset$; \; 
            $\sd[u] \gets 0$; $\mathcal{S} \gets \{ u \}$; $ Q.push(u)$; \;       
           	\While{$Q \neq \emptyset$}{
           		$w \gets Q.\kw{pop}()$; \;
           		\For{\textbf{each} $v \in \ev(w)$}{
           			$\sd[v] \gets \min\{\sd[v], \sd[w] + \phi((w,v),G')\}$; \;
           			\If{$v \notin \mathcal{S} \wedge \kw{checkDel}(u, v, \knnalg(v), \sd[v])$}{
           				$Q.\kw{push}(v)$; $\mathcal{S} \gets \mathcal{S} \cup \{ v \}$;\;
           			}

           		}
           	}
           	\For{\textbf{each} $v \in \mathcal{S}$ in decreasing order of $\pi(v)$}{
           		$\kw{processDel}(v,\ev(v),\knnalg(\cdot))$; delete $u$ from $\knnalg(v)$; \;
           	}
    \BlankLine
	\SetKwProg{myproc}{Procedure}{}{}
	\myproc{\kwnospace{checkDel}$(u, v, \knnalg(v), d)$}{
		$v_k \gets$ the vertex with the largest distance to $v$ in $\knnalg(v)$;\;
		\textbf{if} {$ \sd(v,v_k) < d \vee u \notin \knnalg(v)$}			\textbf{then} \KwRet{\kw{False}};\;
		
		\textbf{else} \KwRet{$\kw{True}$};\;
	}
	\BlankLine
	\myproc{\kwnospace{processDel}$(v,\ev(v), \knnalg(\cdot))$}{
		$\mathcal{S'} \gets \{\cup_{w \in \ev(v)}\knnalg(w)\} \setminus \knnalg(v)$; \;
		
			$v' \gets \kw{argmin}_{v' \in \mathcal{S}'} \sd(v',v)$;\;
			insert $v'$ into $\knnalg(v)$; \;
		}
\end{algorithm} 
}

\stitle{Object Deletion.} \refalg{delete} shows the algorithm for candidate object deletion, which follows a similar framework as \refalg{insert}. The  main difference is in line 10. When the vertices whose $\knnalg(v)$ need to be updated are determined, \refalg{delete} finds a new vertex $v'$ to replace $u$ in $\knnalg(v)$ by procedure \kw{processDel} and  deletes $u$ from $\knnalg(v)$. For procedure \kw{processDel}, it is easy to know that $v'$ must be the vertex in  $\{\cup_{w \in \ev(v)}\knnalg(w)\} \setminus \knnalg(v)$ with the smallest distance to $v$ according to \refpro{vcknn}, thus it first retrieves such set of vertices, namely $\mathcal{S'}$ (line 16), and finds the vertex in $\mathcal{S'}$ with the smallest distance to $v$ (line 17, since the vertices are processed in decreasing order of $\pi(v)$,  $\sd(v', v)$ can be obtained in the similar way as line 7-8 of \refalg{utl} following the same idea). At last, it inserted $v'$ into $\knnalg(v)$ in line 18.

\begin{theorem}
Given the \sdg $G'$ and its corresponding \knnindex of a road network $G$, \refalg{delete} maintains the \knnindex correctly when an object $u$ is deleted.	
\end{theorem}

\proofsketch
\refalg{delete} (line 5-6) can guarantee $\sd[v]$ for $\forall v \in \mathcal{S}$ is the distance between $u$ and $v$, before processing each $v \in \mathcal{S}$ (line 9-10). Even though when using $\kw{checkDel}(v,\knnalg(v),d)$ (line 7), $d$ may not be the distance between $u$ and $v$ and $\kw{checkDel}(v,\knnalg(v),d)$ returns \kw{True}, the final result can not be affected. Since $\sd(u,v) \leq d$, $\kw{checkDel}$ returning \kw{True} denotes that $d < \sd(v,v_k)$. Therefore, we have $\sd(u,v) < \sd(v,v_k)$. Overall, \refalg{delete} maintains the \knnindex correctly when an object $u$ is deleted.	
\eop 

\begin{theorem}
Given the \sdg $G'$ and its corresponding \knnindex of a road network $G$, when an object $u$ is deleted, \refalg{delete} maintains the \knnindex in  $O(\Delta \cdot \tau' \cdot k)$, where $\Delta = |\mathcal{S}|$ and $\tau' = \kw{max}_{v \in V(G)}|\ev(v)|$. 
\end{theorem}

\proofsketch 
The time complexity of $\kw{checkDel}(u,v,\knnalg(v),d)$ (line 11-14) is $O(1)$. In the for loop (line 3-8), for each vertex $w$, line 5-8 requires $O(\tau')$ time and the loop terminates in at most $\Delta$ iterations. Therefore, the for loop (line 3-8) requires $O(\Delta \cdot \tau')$ time. For the procedure $\kw{processDel}(v,\ev(v),\knnalg(\cdot)$ (line 15-18), retrieving $\mathcal{S}'$ from $\{\cup_{w \in \ev(v)}\knnalg(w)\} \setminus \knnalg(v)$ needs $O(|\ev(v)| \cdot |\knnalg(w)|) = O(\tau' \cdot k)$ time (line 16). At the same time with retrieving $\mathcal{S}'$, the vertex $v'$ with the smallest distance from $\mathcal{S}'$ can be achieved (line 17). Line 18 requires $O(1)$ time, since $\sd(v,v') \leq \sd(v,v_k)$ and $v'$ should be inserted into the end of $\knnalg(v)$. Therefore, the time complexity of $\kw{processDel}(v,\ev(v),\knnalg(\cdot)$ (line 15-18) is $O(\tau' \cdot k)$. In the for loop (line 9-10), deleting $u$ from $\knnalg(v)$ requires $O(k)$, and the loop stops in $\Delta$ iterations. Therefore, the for loop (line 9-10) requires $O(\Delta \cdot \tau' \cdot k)$ time. In summary, the overall time complexity of \refalg{insert} is $O(\Delta \cdot \tau' \cdot (1 + k)) =  O(\Delta \cdot \tau' \cdot k)$. 
\eop 

%Since $\knnalg(\cdot)$ is stored in the bidirectional linked lists, the last vertex $v_k$ and $\sd(v,v_k)$ can be directly obtained from $\knnalg(v)$. $\kw{checkDel}(v,\knnalg(v),d)$ only costs $O(1)$ (line 11-14). $\kw{processDel}(v,\ev(v),\knnalg(\cdot))$ costs $O(\tau' \cdot k)$ (line 15-18) since $|\mathcal{S}'| \leq (\tau' \cdot k)$ line 16. The big loop costs $O(\Delta)$, line 5-8 take $O(\tau')$, the time complexity is $O(\Delta \cdot \tau')$ (line 3-8). For processing each affected vertex in $\mathcal{S}$, the time complexity is $O(\Delta \cdot \tau' \cdot k)$ (line 9-10). Overall, the time complexity of \refalg{delete} is $O(\Delta \cdot \tau' \cdot k)$. 

\section{Experiments}
\label{sec:exp}

In this section, we compare our algorithms with the state-of-the-art method. All experiments are conducted on a machine with an Intel Xeon CPU and 384 GB main memory running Linux.

\begin{table}[htb]
    {
    \small{
    \setlength{\tabcolsep}{0.35em}
        \begin{center}
          \begin{tabular}{c| c | r | r | r | r | r}
            \hline
             \multicolumn{1}{c|}{\cg {Dataset}} & \cg {Name} & \multicolumn{1}{c|}{\cg $n$} & \multicolumn{1}{c|}{\cg $m$} & \multicolumn{1}{c|}{\cg $\eta$} & \multicolumn{1}{c|}{\cg $\tau$} & \multicolumn{1}{c}{\cg $\rho$} \\
           \hline
           {New York City} & \kw{NY} & 264,346 & 733,846 & 725 & 56 & 116\\
            \hline
            {San Francisco Bay Area}& \kw{BAY} & 321,270 & 800,172  & 388 & 45 & 100\\
             \hline
            {Colorado} & \kw{COL} & 435,666 & 1,057,066 & 524 & 65 & 122\\
             \hline
            {Florida} & \kw{FLA} & 1,070,376 & 2,712,798 & 556 & 49 & 85\\
             \hline
            {Northwest USA} & \kw{NW} & 1,207,945 & 2,840,208 & 619 & 49 & 119\\ 
             \hline
            {Northeast USA} & \kw{NE} & 1,524,453 & 3,897,636 & 1096 & 81 & 149\\
            \hline
            {California and Nevada} & \kw{CAL} & 1,890,815 & 4,657,742 & 795 & 93 & 204\\
             \hline
            {Great Lakes} & \kw{LKS} & 2,758,119 & 6,885,658 & 1674 & 124 & 327\\
             \hline
            {Eastern USA} & \kw{EUS} & 3,598,623 & 8,778,114 & 1089 & 102 & 233\\
             \hline
            {Western USA} & \kw{WUS} & 6,262,104 & 15,248,146 & 1356 & 128 & 276\\
             \hline
            {Central USA} & \kw{CTR} & 14,081,816 & 34,292,496 & 2811 & 234 & 531\\
             \hline
            {Full USA} & \kw{USA} & 23,947,347 & 58,333,344 & 3315 & 257 & 587\\
             \hline
          \end{tabular}
        \end{center}
        }
      }
      \vspace{0.2cm}
      \caption{{Datasets in Experiments}}
      \label{tab:dataset}
      \vspace{-0.5cm}
\end{table}

\begin{figure*}[!htb]
  \centering
  \subfigcapskip=0pt
  \subfigure[NY]{\includegraphics[width=0.16\linewidth]{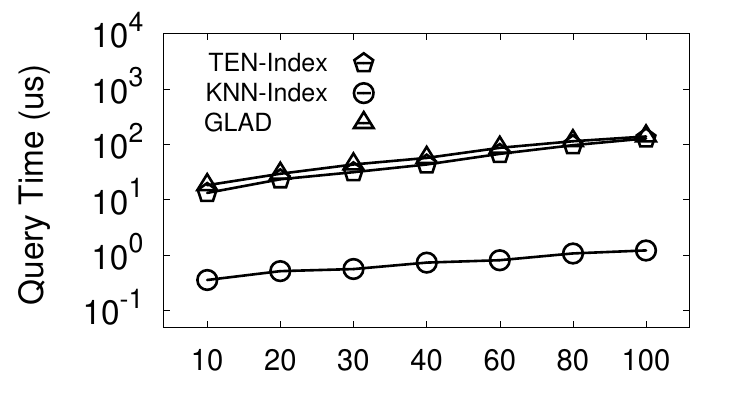}}\label{fig:nyqvk}
  \subfigure[BAY]{\includegraphics[width=0.16\linewidth]{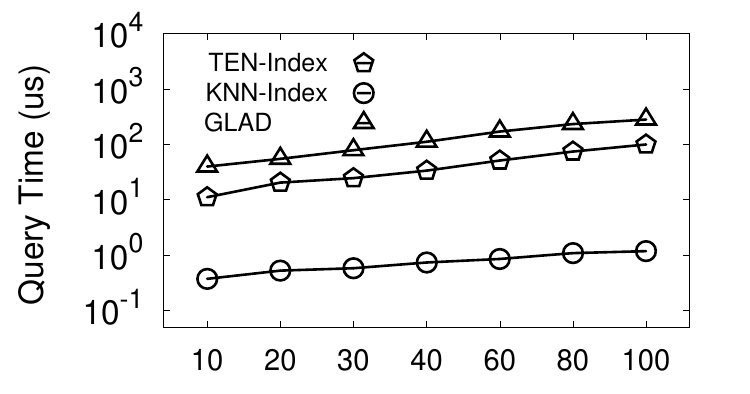}}\label{fig:bayqvk}
  \subfigure[COL]{\includegraphics[width=0.16\linewidth]{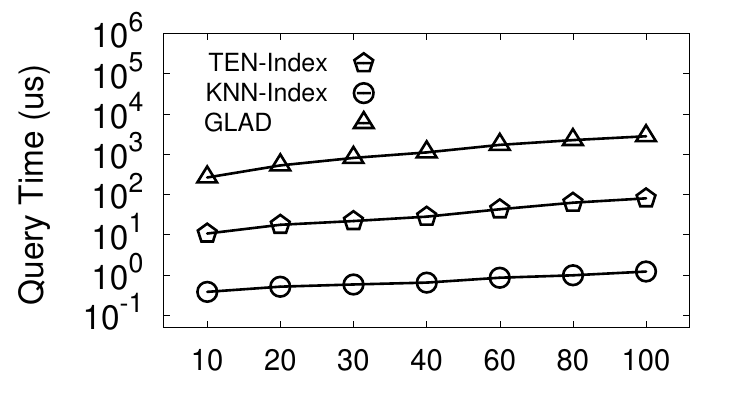}}\label{fig:colqvk}
  \subfigure[FLA]{\includegraphics[width=0.16\linewidth]{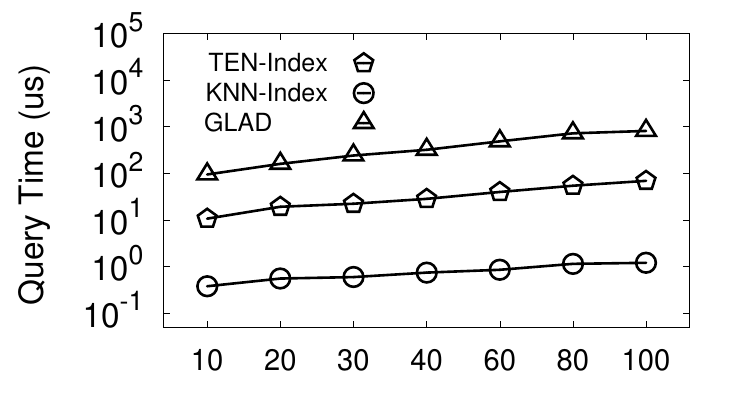}}\label{fig:flaqvk}
  \subfigure[NW]{\includegraphics[width=0.16\linewidth]{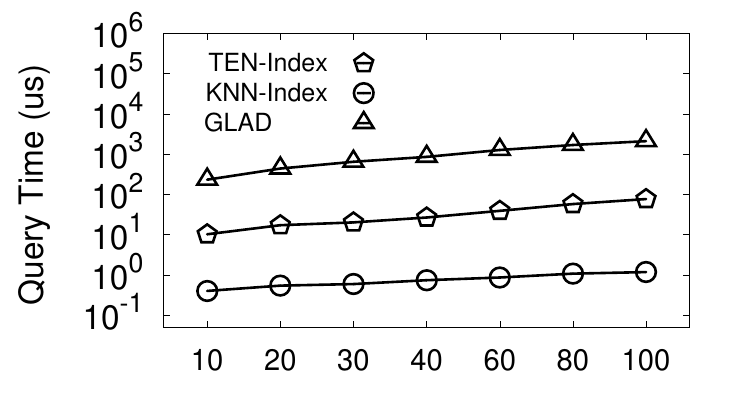}}\label{fig:nwqvk}
  \subfigure[NE]{\includegraphics[width=0.16\linewidth]{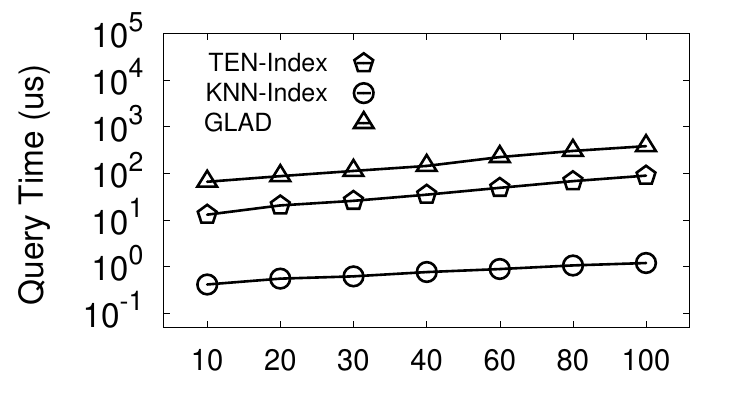}}\label{fig:neqvk}\\
  \subfigure[CAL]{\includegraphics[width=0.16\linewidth]{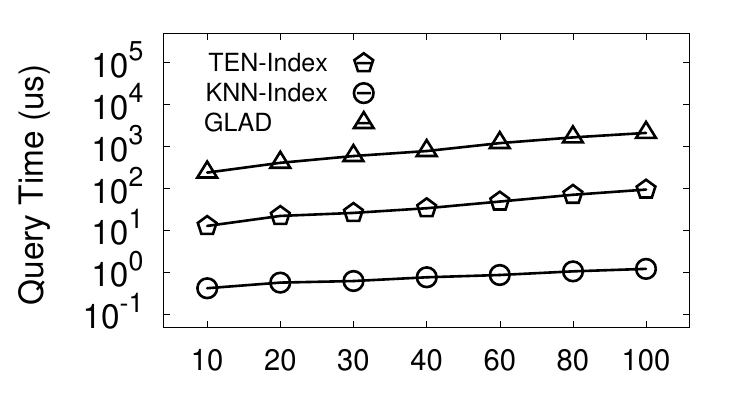}}\label{fig:calqvk}
  \subfigure[LKS]{\includegraphics[width=0.16\linewidth]{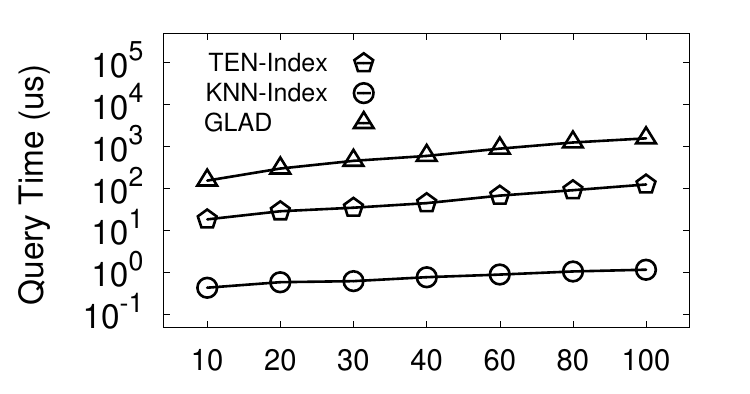}}\label{fig:lksqvk}
  \subfigure[EUS]{\includegraphics[width=0.16\linewidth]{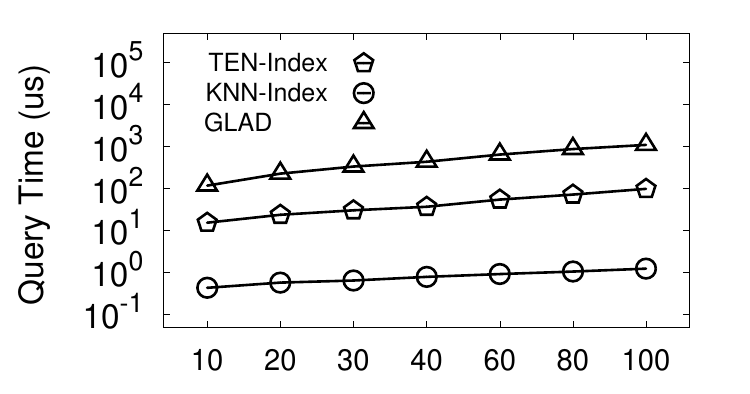}}\label{fig:eqvk}
  \subfigure[WUS]{\includegraphics[width=0.16\linewidth]{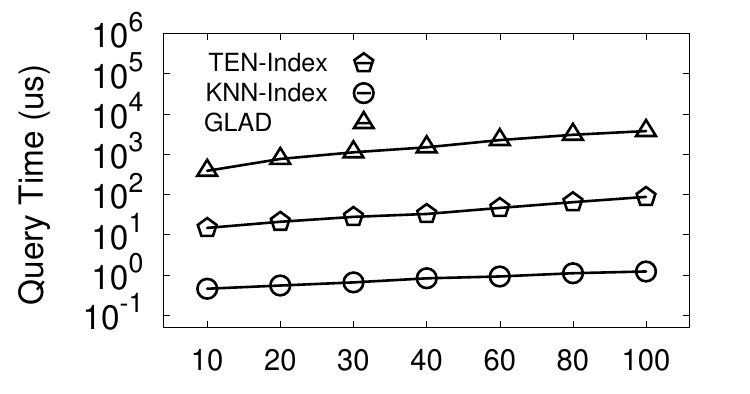}}\label{fig:wqvk}
  \subfigure[CTR]{\includegraphics[width=0.16\linewidth]{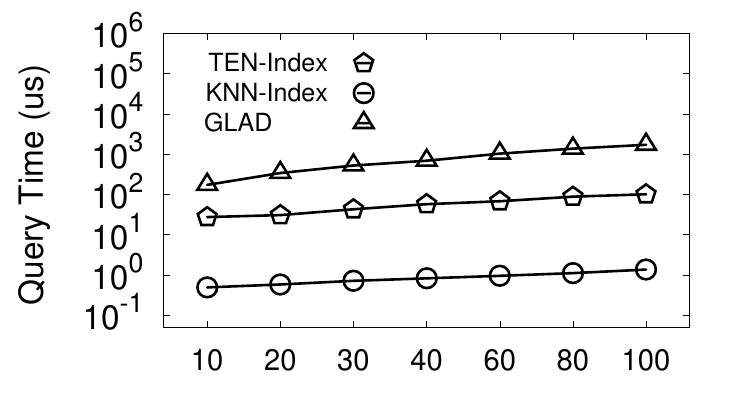}}\label{fig:ctrqvk}
  \subfigure[USA]{\includegraphics[width=0.16\linewidth]{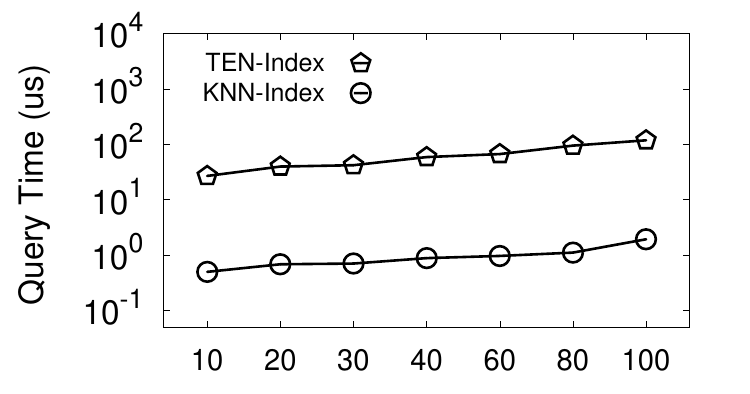}}\label{fig:usaqvk}\\
  \vspace{-0.2cm}
  \caption{Query Processing Time by Varying $k$}
   \vspace{-0.2cm}
  \label{fig:queryvk}
\end{figure*}

\stitle{Datasets.} We use twelve publicly available real road networks from DIMACS \footnote{http://users.diag.uniroma1.it/challenge9/download.shtml}. In each road network, vertices represent intersections between roads, edges correspond to roads or road segments, the weight of an edge is the physical distance between two vertices.  \reftable{dataset} provides the details about these datasets. \reftable{dataset} also shows the value of $\eta$, $\tau$ and $\rho$ for each road network. Clearly, $\eta$, $\tau$ and $\rho$ are small in practice. 

\stitle{Algorithms.} We compare the following algorithms: 
\begin{itemize}[leftmargin=*]
	\item \kwnospace{\tenindexcnt}: The state-of-the-art algorithm for $k$NN queries queries, which is introduced in \refsec{existing}.
	\item \knnindexcnt: Our proposed algorithms for $k$NN queries. For the index construction algorithms, we further distinguish between \naivecon (\refalg{naivecon}) and \optconcnt (Algorithm 3) for comparison.
	\item \kw{GLAD}: Another algorithm for $k$NN queries proposed in \cite{luo2018toain}, which is introduced in \refsec{related}.
	 \item \kw{\dijk}: Using Dijkstra's Algorithm to compute top-$k$ nearest neighbors for all vertices in a given graph $G$ to construct the \knnindex as discussed in \refsec{indexcon}.
	\item \kwnospace{\tenknncnt}: Using \tenindexcnt to compute top-$k$ nearest neighbors for all vertices in a given graph $G$ to construct the \knnindex as discussed in \refsec{indexcon}.
	%\item \hl{\kw{\glad}: The other state-of-the-art algorithm for top-$k$ nearest neighbors queries, which is introduced in \refsec{related}}. 
\end{itemize}

 All the algorithms are implemented in C++ and compiled in GCC with -O3. 
 The time cost is measured as the amount of wall-clock time elapsed during the program's execution. If an algorithm cannot finish in 6 hours, we denote the processing time as \kw{NA}. 
 
\stitle{Parameter Settings.} Following previous $k$NN works \cite{ouyang2020progressive, he2019efficient, luo2018toain}, we randomly select candidate objects in each dataset with a density $\mu = |\mathcal{M}|/|V|$. The candidate density $\mu$ and the query parameter $k$ settings are shown in \reftable{para}, default values display in bold and italic font.

\begin{table}[htb]
	\vspace{0.2cm}
    {    
    \setlength{\tabcolsep}{0.55em}
        \begin{center}
          \begin{tabular}{c| c }
            \hline
             \multicolumn{1}{c|}{\cg {Parameters}} & \multicolumn{1}{c}{\cg Values}\\
           \hline
           {$\mu$} & {0.5, 0.1, 0.05, 0.01, \emph{\textbf{0.005}}, 0.001, 0.0005, 0.0001}\\
            \hline
            {$k$}& {100, 80, 60, 40, 30, \emph{\textbf{20}}}, 10\\
            \hline
          \end{tabular}
        \end{center}
        }
      \caption{Parameter Settings}
      \label{tab:para}
      \vspace{-0.8cm}
\end{table}

\stitle{Exp-1: Query Processing Time when Varying $k$.} In this experiment, we evaluate the query processing time of our algorithms \knnindex, the SOTA solutions \tenindex  and \glad by varying the parameter $k$. We randomly generate $10,000$ queries and report average running time of each algorithm in \reffig{queryvk}.

As shown in \reffig{queryvk}, our algorithm is the most efficient one compared with \tenindex and \glad and the growth for query processing time of \tenindex and \glad is sharper than that of \knnindexcnt with increase of  $k$. For example, on \kw{NY} dataset, when $k$ increases from $10$ to $100$, the query processing time of \tenindex grows from $13.217$us to $127.025$us and \glad raises from $18.172$us to $138.212$us. However, our \knnindex processing time grows from $0.395$us to $1.210$us. This is consistent with our analysis in \refsec{theoretical}. \underline{For \kw{USA}}, \underline{\kw{GLAD} is out of memory, the query processing time for \kw{USA} could} \\\underline{not be tested at this and following experiments.}

\begin{figure*}[htb]
  \centering
  \subfigcapskip=0pt
  \subfigure[NY]{\includegraphics[width=0.16\linewidth]{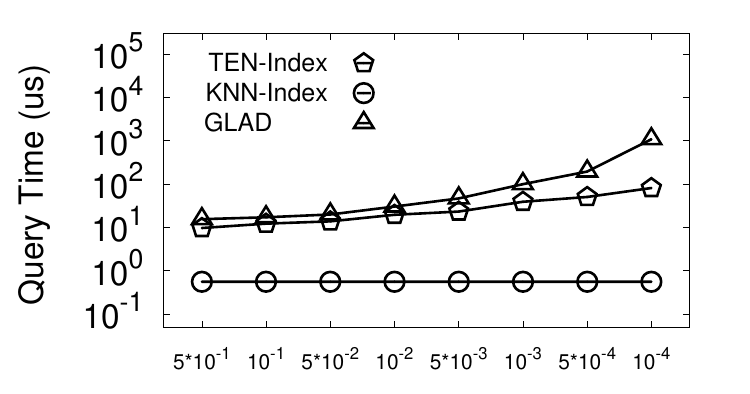}}\label{fig:nyqvden}
  \subfigure[BAY]{\includegraphics[width=0.16\linewidth]{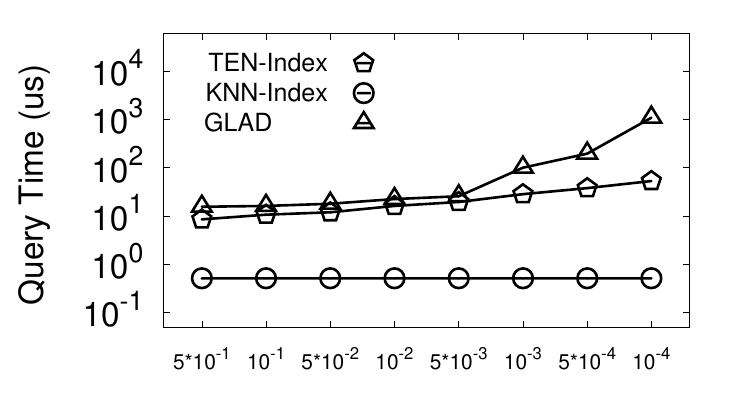}}\label{fig:bayqvden}
  \subfigure[COL]{\includegraphics[width=0.16\linewidth]{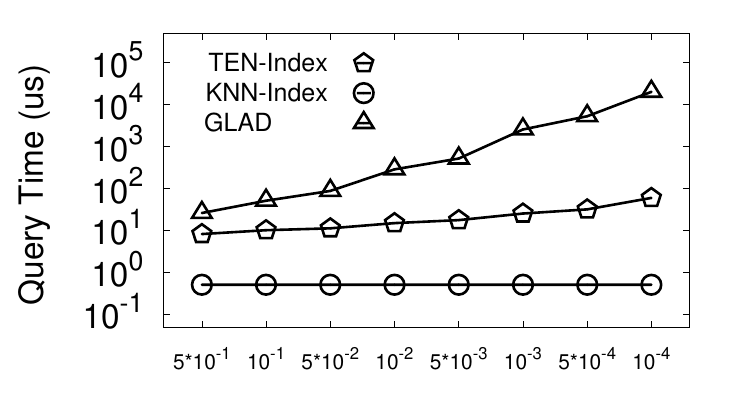}}\label{fig:colqvden}
  \subfigure[FLA]{\includegraphics[width=0.16\linewidth]{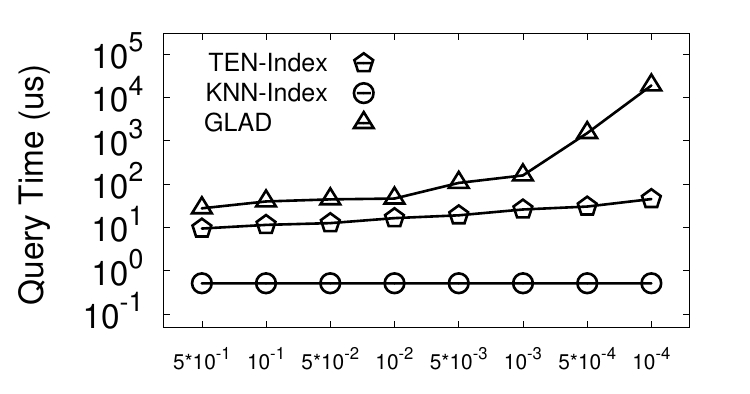}}\label{fig:flaqvden}
  \subfigure[NW]{\includegraphics[width=0.16\linewidth]{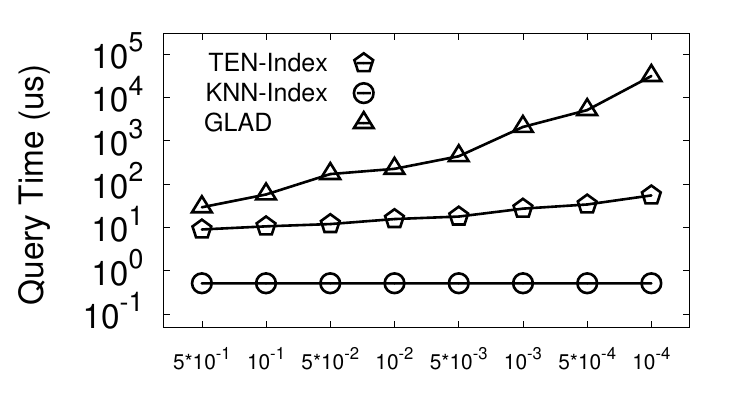}}\label{fig:nwqvden}
  \subfigure[NE]{\includegraphics[width=0.16\linewidth]{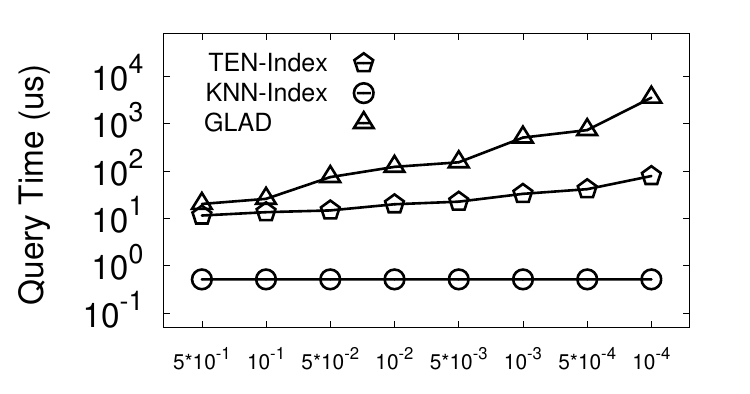}}\label{fig:neqvden}\\
  \subfigure[CAL]{\includegraphics[width=0.16\linewidth]{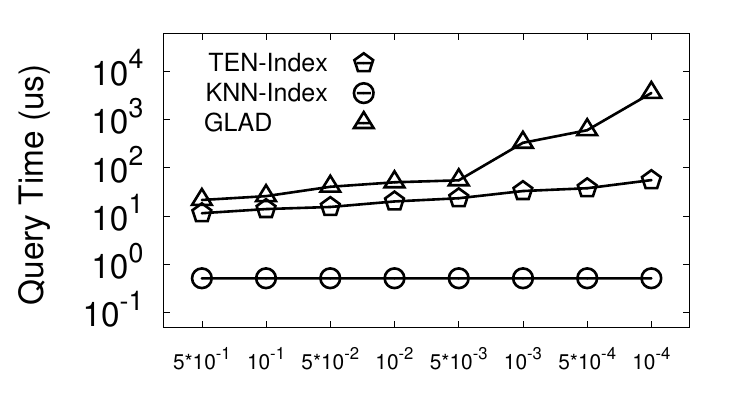}}\label{fig:calqvden}
  \subfigure[LKS]{\includegraphics[width=0.16\linewidth]{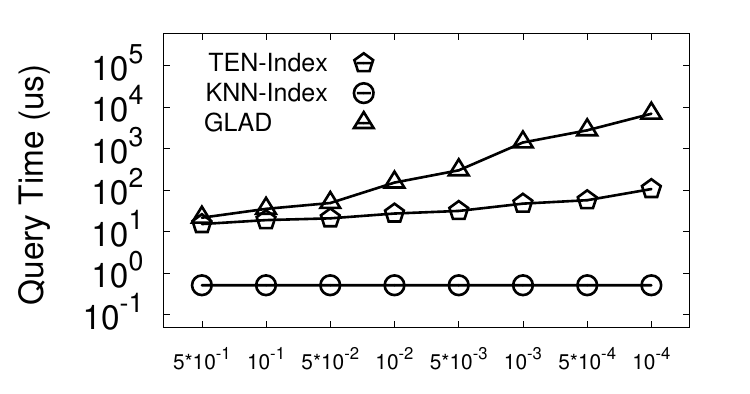}}\label{fig:lksqvden}
  \subfigure[EUS]{\includegraphics[width=0.16\linewidth]{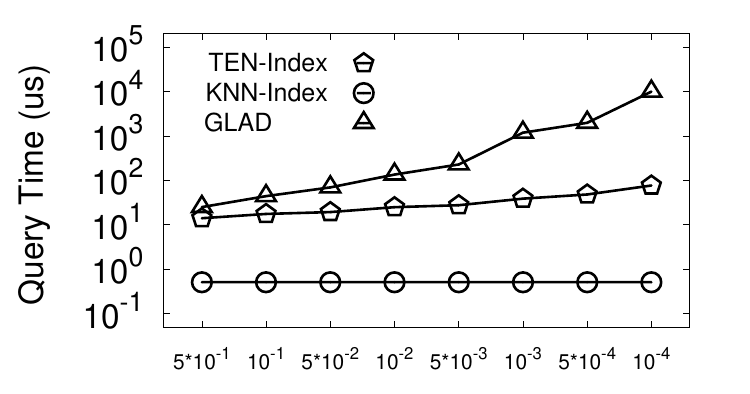}}\label{fig:eqvden}
  \subfigure[WUS]{\includegraphics[width=0.16\linewidth]{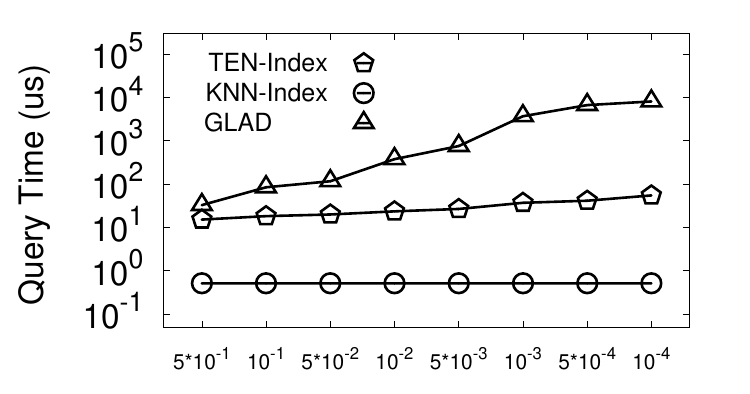}}\label{fig:wqvden}
  \subfigure[CTR]{\includegraphics[width=0.16\linewidth]{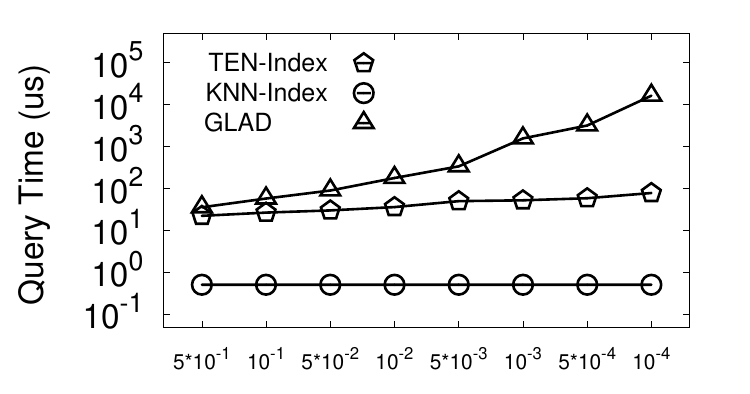}}\label{fig:ctrsqden}
  \subfigure[USA]{\includegraphics[width=0.16\linewidth]{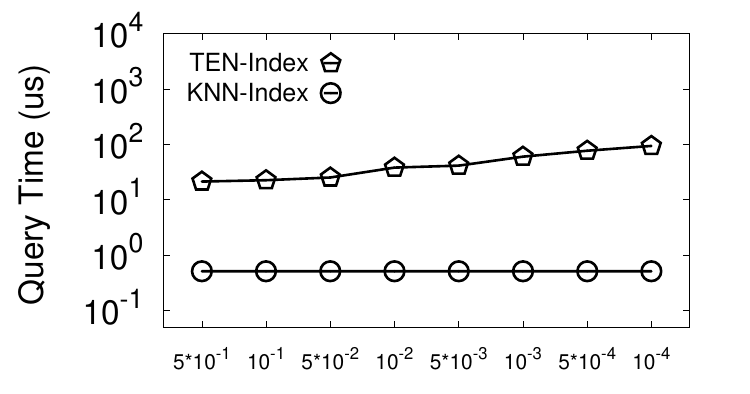}}\label{fig:usasqden}\\
  \vspace{-0.2cm}
  \caption{Query Processing Time by Varying $\mu = |\mathcal{M}|/|V|$}
  \label{fig:queryvd}
  \vspace{-0.2cm}
\end{figure*}

\stitle{Exp-2: Query Processing Time when Varying $\mathcal{M}$.} We also compare our \knnindex with the SOTA solutions \tenindex and \kw{GLAD} by varying object $\mathcal{M}$ (the density $\mu = |\mathcal{M}|/|V|$, therefore, we vary $\mathcal{M}$ by changing $\mu$). We randomly generate $10,000$ queries for every dataset. We report the average processing time of each algorithm in \reffig{queryvd}.

 As shown in \reffig{queryvd}, the query processing time of our algorithm is  stable with the decrease of candidate object $\mathcal{M}$. However, the query processing time of \tenindex and \kw{GLAD} increases significantly with the decrease of candidate density $\mu$. For example, when $\mu = 0.0001$ \knnindex achieves 2 orders of magnitude speedup compared with \tenindex in all datasets, and \knnindex achieves 4 orders of magnitude speedup compared with \kw{GLAD}. Moreover, the more sparsely the object set distributes, the larger speedup is. This is because our proposed algorithm is optimal regarding query processing as analyzed in \refsec{theoretical}.

\begin{figure*}[htb]
  \centering
  \subfigcapskip=0pt
  \subfigure[NY]{\includegraphics[width=0.16\linewidth]{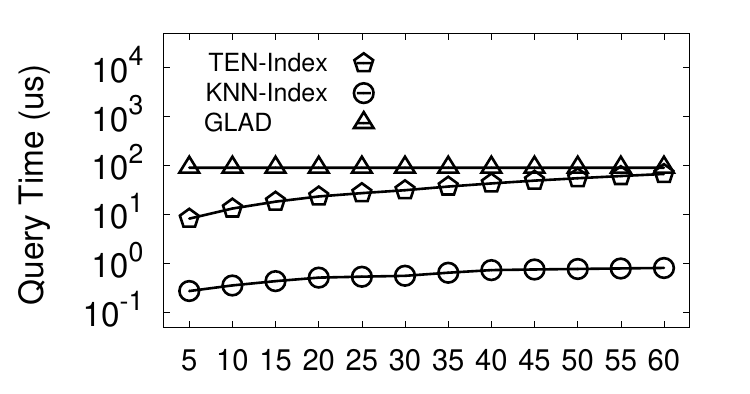}}\label{fig:nyqvden}
  \subfigure[BAY]{\includegraphics[width=0.16\linewidth]{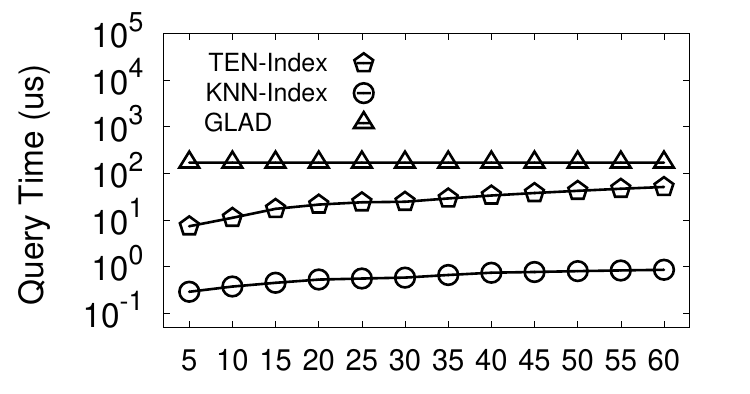}}\label{fig:bayqvden}
  \subfigure[COL]{\includegraphics[width=0.16\linewidth]{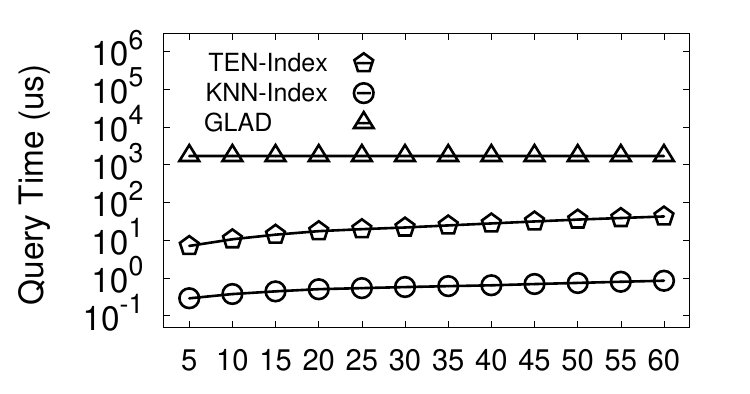}}\label{fig:colqvden}
  \subfigure[FLA]{\includegraphics[width=0.16\linewidth]{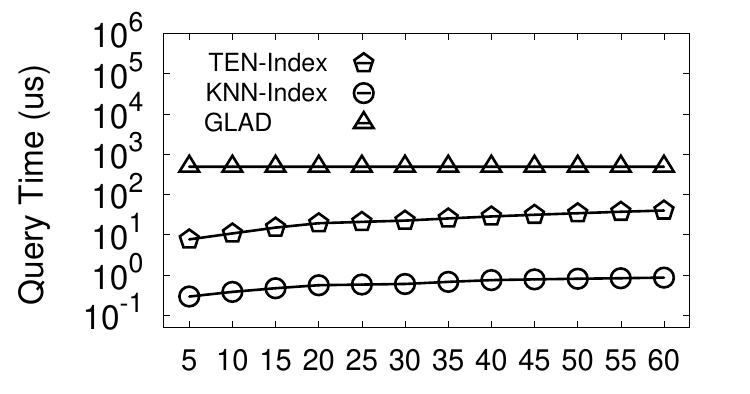}}\label{fig:flaqvden}
  \subfigure[NW]{\includegraphics[width=0.16\linewidth]{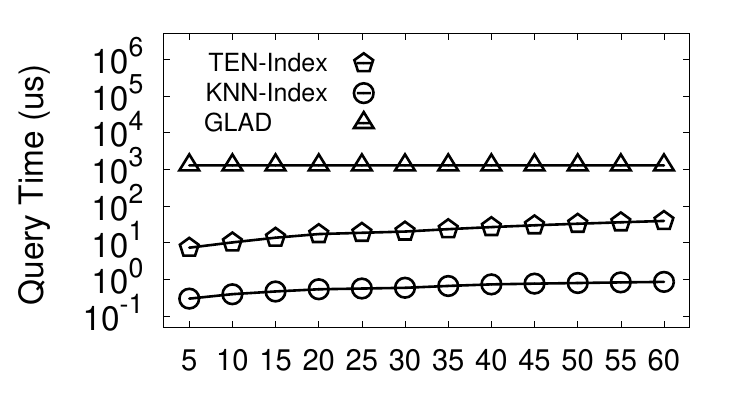}}\label{fig:nwqvden}
  \subfigure[NE]{\includegraphics[width=0.16\linewidth]{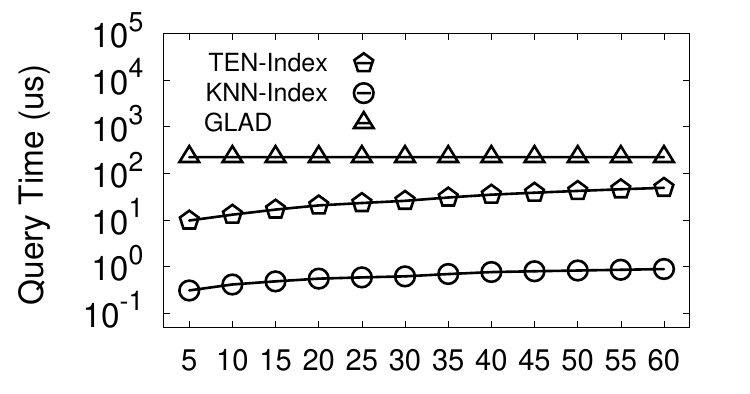}}\label{fig:neqvden}\\
  \subfigure[CAL]{\includegraphics[width=0.16\linewidth]{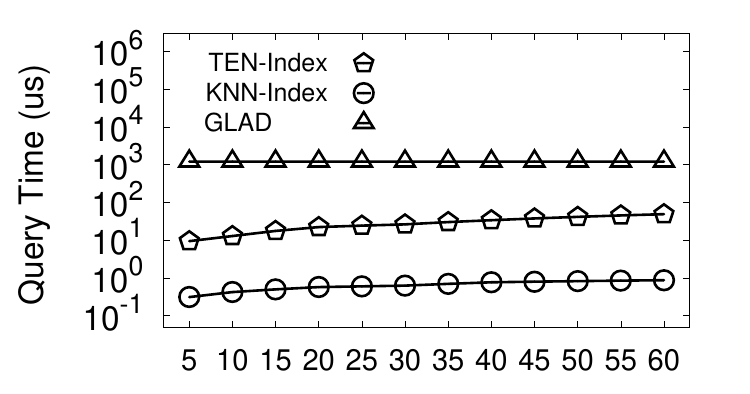}}\label{fig:calqvden}
  \subfigure[LKS]{\includegraphics[width=0.16\linewidth]{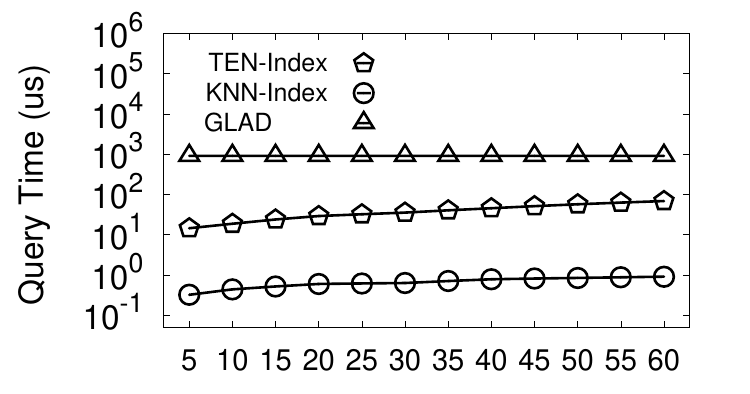}}\label{fig:lksqvden}
  \subfigure[EUS]{\includegraphics[width=0.16\linewidth]{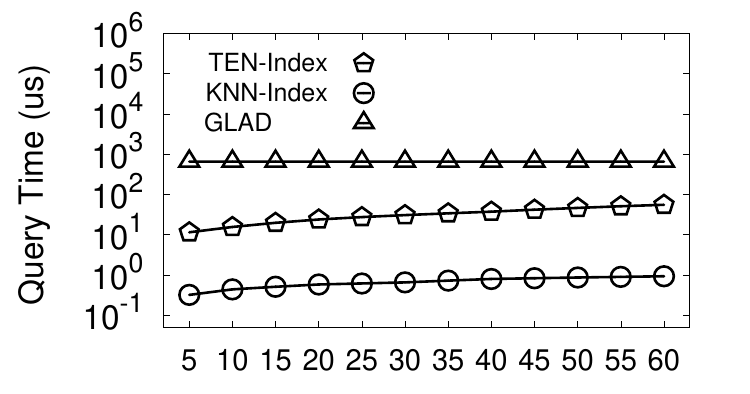}}\label{fig:eqvden}
  \subfigure[WUS]{\includegraphics[width=0.16\linewidth]{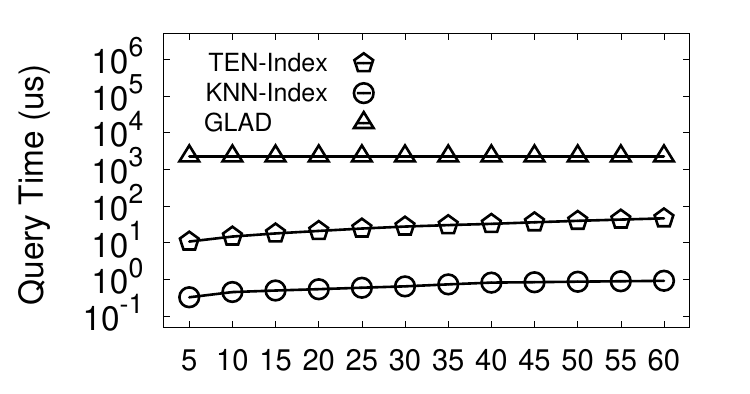}}\label{fig:wqvden}
  \subfigure[CTR]{\includegraphics[width=0.16\linewidth]{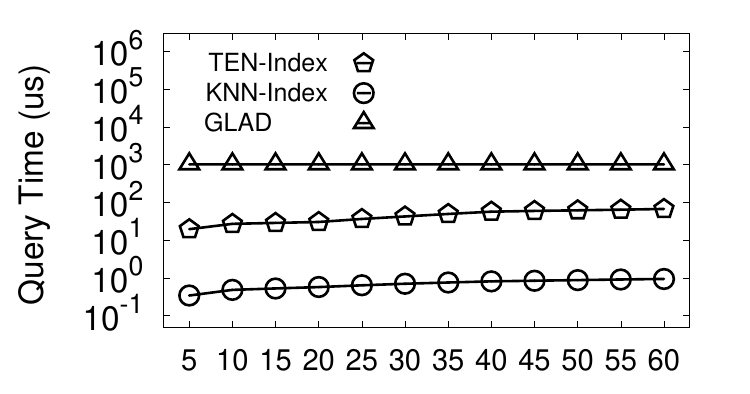}}\label{fig:ctrsqden}
  \subfigure[USA]{\includegraphics[width=0.16\linewidth]{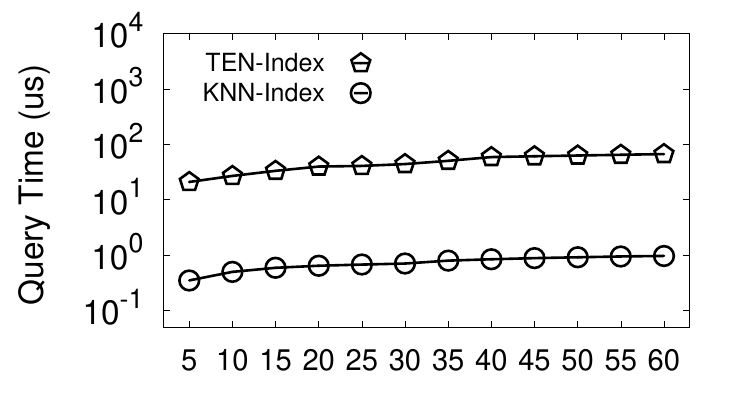}}\label{fig:usasqden}\\
  \vspace{-0.2cm}
  \caption{Query Processing Time for Different Outputs}
  \label{fig:queryvo}
  \vspace{-0.2cm}
\end{figure*}

\stitle{Exp-3: Progressive Query Processing.} In this experiment, we evaluate the progressive query processing strategy of \knnindex, \tenindex and \kw{GLAD} by outputing every $5$ outputs in all datasets with $k=60$. As shown in \reffig{queryvo}, the lines of \tenindex and \knnindex are linear in all datasets, and the total time of \knnindex is always smaller than that of \tenindexno. The reasons are similar as explained in Exp-1 and Exp-2. \kw{GLAD} does not support incremental polynomial query processing, so it has the same the total time for different outputs.

\begin{figure*}[!htb]
\vspace{0.1cm}
  \centering
  \includegraphics[width=0.95\linewidth]{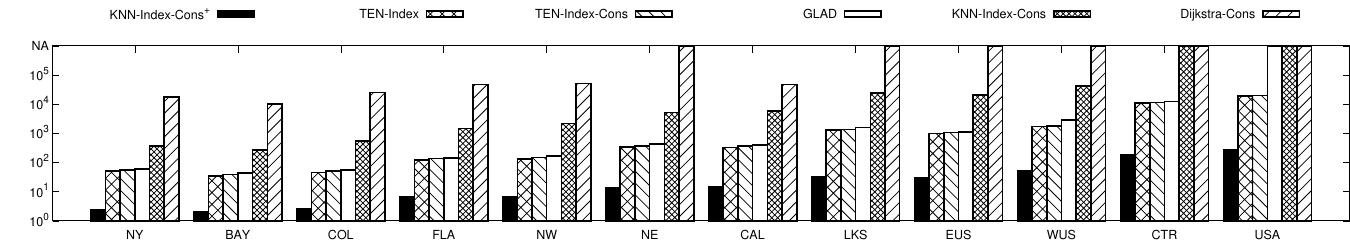}
  \vspace{-0.3cm}
  \caption{Indexing Time (s)}
  \label{fig:indextime}
 \end{figure*}
 
 \begin{figure}[tb]
  \centering
  \includegraphics[width=0.95\linewidth]{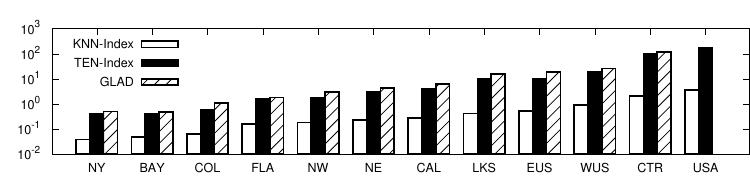} 
  \vspace{-0.3cm}
  \caption{Index Size (GB)}
  \label{fig:indexsize}
  %\vspace{-0.3cm}
\end{figure}

\stitle{Exp-4: Indexing Time.} In this experiment, we evaluate the indexing time for \optconcnt, \tenindexcnt, \tenknncnt, \glad, \naivecon and \dijk. \reffig{indextime} shows that \optcon is the fastest in all datasets, and achieves up to 2 orders of magnitude speedup compared with \tenindex and \glad. For example, \optcon only takes $283.78$s for \kw{USA} while \tenindex costs $19655.68$s. \tenknn and \tenindex takes the similar indexing time as \tenknn depends on the \tenindexcnt. They both rely on \hhindex. Also, the indexing time of \glad and \tenindex are similar, since \glad constructs the additional grid index on the basis of \hhindexcnt.  As shown in \reffig{indextime}, \naivecon cannot complete the index construction within $6$ hours for \kw{CTR} and \kw{USA}. And for \kw{USA} \glad is out of memory. \dijk cannot finish index construction within $6$ hours for \kw{NE}, \kw{LKS}, \kw{EUS}, \kw{WUS}, \kw{CTR} and \kw{USA}. Although index construction frameworks in  \naivecon and \optcon are similar, \naivecon consumes much more time compared with \optconcnt. For example, for \kw{WUS}, \optcon only costs $50.15$s, but \naivecon costs $42061.20$s. This is because \naivecon first uses BFS to construct $\sdgu(u)$ for each vertex $u \in V(G)$, and then uses Dijkstra's Algorithm to compute $\sd(u,v)$ for $\forall v \in V(\sdgu(u))$ when constructing the index. However, \optcon adopts a bidirectional construction strategy to avoid the time-consuming BFS search and the computation of Dijkstra's Algorithm during the index construction. The experimental results demonstrate the efficiency of our proposed algorithm regarding index construction. 
%Even though \naivecon and \optcon are similar methods to construction \knnindex, the time consumption has a huge difference. When \naivecon constructs total \knnindex, it uses BFS to construct $\sdgu(u)$ for each vertex $u \in V(G)$, and then uses Dijkstra's Algorithm to compute $\sd(u,v)$ for $\forall v \in V(\sdgu(u))$, which costs huge time. However, \optcon adopts a bottom-down direction to avoid the invocation of BFS search and Dijkstra's Algorithm.

\stitle{Exp-5: Index Size.} In this experiment, we evaluate the index size for \knnindex, \tenindex and \kw{GLAD}. The experimental results for the $12$ road networks are shown in \reffig{indexsize}. \reffig{indexsize} shows the index size of \knnindex is much smaller than that of \tenindex and \kw{GLAD}. For example, for the dataset \kw{USA}, the \knnindex size is only $3.57$ GB while \tenindex size is $169.28$ GB, which is  $47.42$ times smaller than  that of \tenindexcnt.

\begin{table*}[htb]
\vspace{-0.2cm}
    {
    \small{
    \setlength{\tabcolsep}{0.35em}
        \begin{center}
          \begin{tabular}{c| r | r | r | r | r | r | r | r }
            \hline
            \multicolumn{1}{c|}{\cg {}} & \multicolumn{4}{c|}{\cg USA} &  \multicolumn{4}{c}{\cg  CTR} \\
            \hline
            \multirow{2}*{\cg $k$} 
             	& \multicolumn{1}{c|}{\cg \tenindex} &  \multicolumn{1}{c|}{\cg \knnindex} & \multicolumn{1}{c|}{\cg  \tenindex} & \multicolumn{1}{c|}{\cg  \knnindex} & \multicolumn{1}{c|}{\cg \tenindex} & \multicolumn{1}{c|}{\cg \knnindex} & \multicolumn{1}{c|}{\cg  \tenindex} & \multicolumn{1}{c}{\cg  \knnindex} \\ 
             	\multirow{-2}*{\cg $k$}& \cg {Indexing Time (s)} & \multicolumn{1}{c|}{\cg Indexing Time (s)} & \multicolumn{1}{c|}{\cg  Index Size (GB)} & \multicolumn{1}{c|}{\cg  Index Size (GB)} & \cg {Indexing Time (s)} & \multicolumn{1}{c|}{\cg Indexing Time (s)} & \multicolumn{1}{c|}{\cg  Index Size (GB)} & \multicolumn{1}{c}{\cg  Index Size (GB)} \\
           \hline
            {$10$} & 19669.787 & 266.005 & 169.265 & 1.784 & 10877.527 & 169.578 & 98.021 & 1.049 \\
            \hline
            {$20$} & 19670.198 & 283.850 & 169.277 & 3.568 & 10878.179 & 179.480 & 98.028 & 2.908 \\
             \hline
            {$30$} & 19670.336 & 297.224 & 169.286 & 5.353 & 10881.515 & 186.859 & 98.034 & 3.148 \\
             \hline
            {$40$} & 19671.470 & 321.845 & 169.293 & 7.1369 & 10882.434 & 201.225 & 98.039 & 4.197 \\
             \hline
            {$60$} & 19672.287 & 345.169 & 169.305 & 10.705 & 10884.793 & 215.737 & 98.046 & 6.295 \\ 
             \hline
            {$80$} & 19672.839 & 382.074 & 169.315 & 14.274 & 10885.981 & 238.944 & 98.053 & 8.393 \\
            \hline
            {$100$} & 19710.726 & 418.251 & 169.324 & 17.842 & 10887.369 & 253.219 & 98.058 & 10.492 \\
            \hline
          \end{tabular}
        \end{center}
        }
      }
      \vspace{0.1cm}
      \caption{Indexing Time and Index Size when Varying $k$} 
      \label{tab:varyk}
      %\vspace{-0.6cm}
\end{table*}

\stitle{Exp-6: Indexing Time and Index Space when Varying $k$.} We evaluate the performance of \knnindex and \tenindex when varying $k$ on \kw{USA} and \kw{CTR} with $k = 10, 20, 30, 40, 60, 80, 100$. The results on the other datasets are omitted due to similar  trends.  As shown in \reftable{varyk}, the index size and the indexing time of \tenindex and \knnindex both increases with the growth of $k$ slightly. For example, when $k$ increases from $10$ to $100$, the indexing time for \kw{USA} increases by $40.940$s and the index size for \kw{USA} increases by $0.059$ GB. This is consistent with our analysis  in \refsec{theoretical}.

\begin{figure}[htb]
\vspace{-0.2cm}
  \centering
   \subfigcapskip=0pt
   \subfigcaptopadj=0pt
   \includegraphics[width=0.95\linewidth]{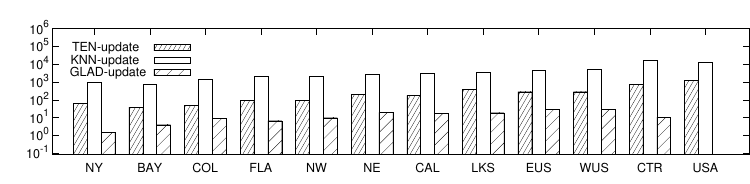}
   \vspace{-0.2cm}
 \caption{Update Time (us)}
  \label{fig:up}
  %\vspace{-0.3cm}
\end{figure}

\begin{figure}[htb]
  \centering
   \subfigcapskip=0pt
   \subfigcaptopadj=0pt
 \subfigure[Index Size]{ \includegraphics[width=0.45\linewidth]{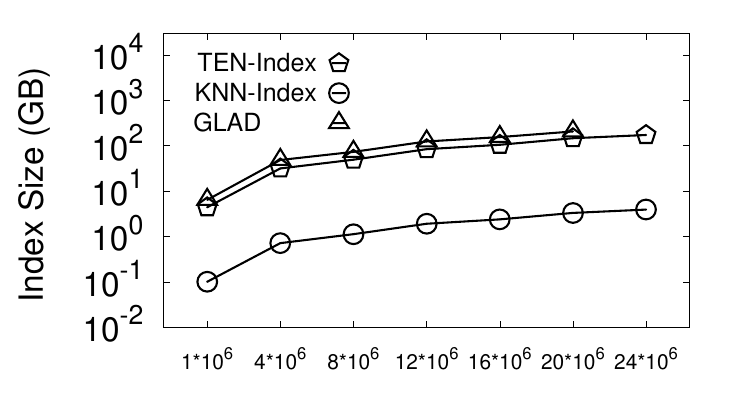}
     \label{fig:sc_size}}
 \subfigure[Indexing Time]{ \includegraphics[width=0.45\linewidth]{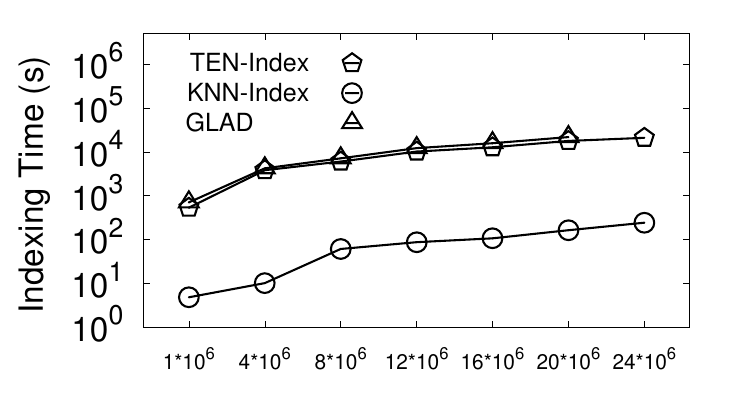}
     \label{fig:sc_time}}
     \vspace{-0.3cm}
 \caption{Scalability}
  \label{fig:scale}
  %\vspace{-0.5cm}
\end{figure}

\stitle{Exp-7: Scalability when Varying Graph Size.} In this experiment, we evaluate the scalability of \knnindex, \tenindex and \kw{GLAD}. To test the scalability for indexing time and index size,  we divide the map of the whole US into $10 \times 10$ grids. We select a $1 \times 1$ grid in the middle and generate an induced network by all vertices falling the grid. Using this method, we generate $10$ datasets. We report the indexing time and index size for these ten networks in \reffig{scale}. The labels on $x$-axis represent the number of vertices. For the largest datasets, \glad is out of memory. As shown in \reffig{scale}, when the dataset increases from $10^6$ to $24 * 10^6$, the indexing time increases stably for all algorithms, which verifies our \knnindex has a good scalability. Moreover, our \knnindex always outperforms \tenindex and \glad. The reasons are similar as explained in Exp-4 and Exp-5.

\stitle{Exp-8: Object Update.} In this experiment, we evaluate the performance of our update algorithms. To generate updated objects, we randomly select an object $u$ with either insertion or deletion. We skip the update if $u \notin \mathcal{M}$ for deletion and $u \in \mathcal{M}$ for insertion. For each dataset, we repeat this step until $10,000$ updates are generated. The average time for each update is reported in \reffig{up}. The update time of \knnindex is slower than that of \tenindex and that of \kw{GLAD}, since our update algorithm needs more time to compute the distance between each  vertex and the updated objects. As analyzed in \refsec{existing}, \tenindex contains \hhindex, \hhindex can compute the distance between any two vertices efficiently. Therefore, based on \hhindex, \tenindex can finish insertion or deletion in the shorter time. Since \kw{GLAD} only needs to update objects' grid index, the update operation is easier and costs shorter time.

\begin{figure}[htb]
 \vspace{-0.3cm}
  \centering
   \subfigcapskip=0pt
   \subfigcaptopadj=0pt
 \subfigure[\kw{BUA}+\kw{QF} (vary $\mu)$]{ \includegraphics[width=0.45\linewidth]{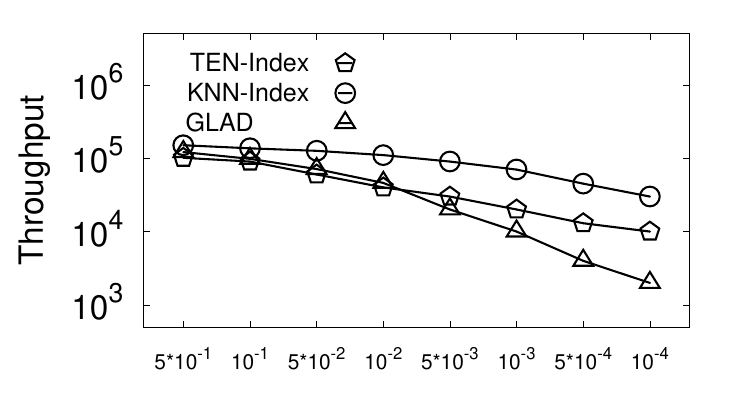}
     \label{fig:qd}}
 \subfigure[\kw{BUA}+\kw{QF} (vary $k$)]{ \includegraphics[width=0.45\linewidth]{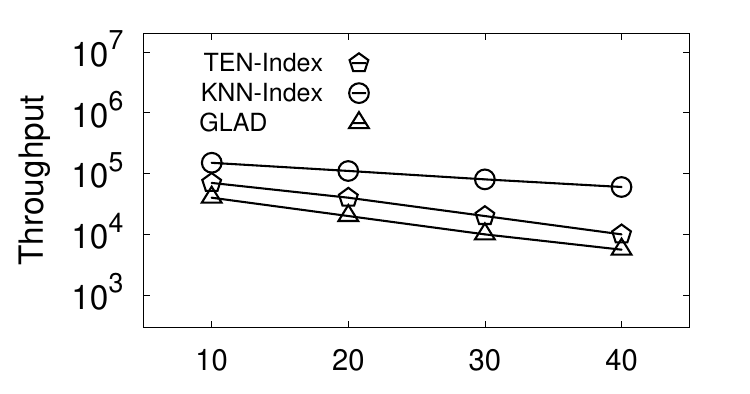}
     \label{fig:qk}}
 \subfigure[\kw{RUA}+\kw{FCFS} (vary $\mu$)]{ \includegraphics[width=0.45\linewidth]{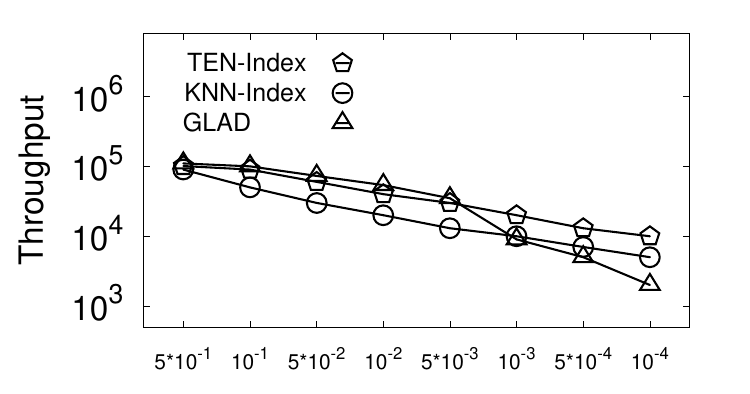}
     \label{fig:ud}}
 \subfigure[\kw{RUA}+\kw{FCFS} (vary $k$)]{ \includegraphics[width=0.45\linewidth]{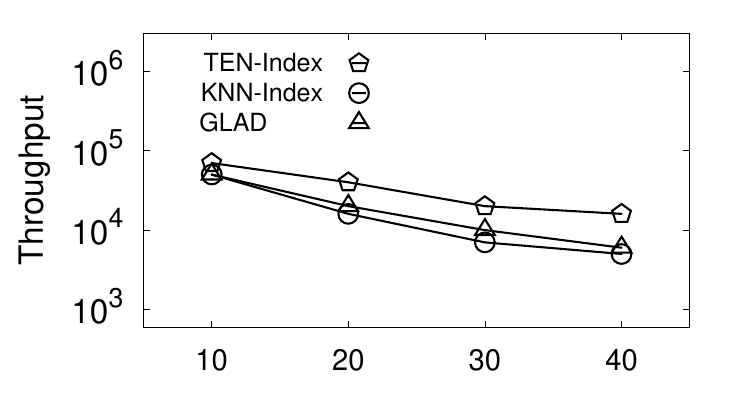}
     \label{fig:uk}}
     \vspace{-0.2cm}
 \caption{Throughput for \kw{NY}}
  \label{fig:tp}
  \vspace{-0.3cm}
\end{figure}

\stitle{Exp-9: System Throughput.} We evaluate the throughput of answering queries mixed by $k$NN queries and object updates. Following \cite{ouyang2020efficient}, the throughput is calculated based on two models, which are (1) Batch Update Arrival + Query First (\kw{BUA}+\kw{QF}) and (2) Random Update Arrival + First Come First Served (\kw{RUA}+\kw{FCFS}) \cite{luo2018toain}.  We report the throughput of each algorithm under different $k$ and $\mu$ for a representative dataset \kw{NY}, which is also used in \cite{ouyang2020efficient}. The results are shown in \reffig{tp}. We can see that the throughput of \knnindex is larger for the \kw{BUA}+\kw{QF} model as our query processing algorithm is very fast. However, in the \kw{RUA}+\kw{FCFS} model, the throughput of \knnindex is smaller than that of \tenindex and \kw{GLAD}. This is because our update algorithm costs more time than the update algorithms for \tenindex and \kw{GLAD}. Also, the throughput of \knnindex, \tenindex and \kw{GLAD} all will go down with the decrease of $\mu$. When the candidate density is relatively high, our update algorithms are faster. When $\mu$ decreases in \reffig{ud}, our update algorithms cost more time. Even if our query processing time is not affected by $\mu$, our throughput still decrease with the fall of $\mu$. For \tenindex and \kw{GLAD}, the query processing time and updating time both will be longer with decrease of $\mu$. Therefore, the throughput of \tenindex and \kw{GLAD} have a similar trend with that of \knnindex. In summary, \knnindex is good at handling the \kw{BUA}+\kw{QF} workload while \tenindex and \kw{GLAD} are good at handling \kw{RUA}+\kw{FCFS} workload, and \knnindexcnt, \tenindex and \kw{GLAD} have their respective advantages regarding throughput.

\begin{figure}[htb]
\vspace{0.1cm}
  \centering
  \includegraphics[width=1\linewidth]{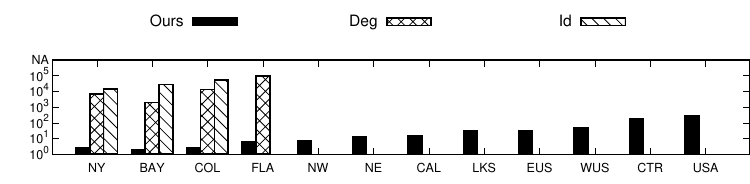}
  \vspace{-0.5cm}
  \caption{Indexing Time of Different Vertex Orders (s)}
  \label{fig:order}
 \end{figure}

\stitle{Exp-10: Indexing Time of Different Vertex Total Orders.} We evaluate index construction performance using different total orders. We adopt three total orders: (1) degree-based total order in which the vertex with the smallest degree is processed first; (2) id-based total order in which the vertex with the smallest id is processed first;  (3) and our proposed total order. As shown in \reffig{order}, degree-based order can finish \knnindex construction for four small datasets, and id-based order only can  construct \knnindex for 3 small datasets within $6$ hours. Our order can finish constructing \knnindex for all datasets, and the construction time in our order is 4 orders of magnitude faster than the other two orders. %For example, for \kw{COL}, our order only needs $2.63$s to finish constructing \knnindex, but degree-based order costs $13313.79$s and id-based order consumes $54789.89 $s. 
The experimental results are also consistent with our analysis in \refsec{ug}.

\section{Related Work} 
\label{sec:related}

The direct approach to answer a $k$NN query is the \kw{Dijkstra}'s algorithm \cite{dijkstra2022note}. Nevertheless, this approach is inefficient obviously. Therefore, a plethora of index based  \kwnospace{Dijkstra}-search enhanced solutions \cite{papadias2003query, demiryurek2009efficient, lee2010road, zhong2015g, shen2017v, luo2018toain} are proposed in the literature, which generally adopts the following search framework for a given query vertex $u$: (1) Initialize the distance for vertices $v$ it connected as their edge weights and  other vertices as $+\infty$. (2) Maintain two vertex sets $S$ and $T$. $S$ contains vertices whose distance to $u$ is computed. $T$ contains vertices whose  distance to $u$ is not computed yet, but have neighbors in $S$. Initially,  $u$ is inserted in $S$ and the neighbors of $u$ are inserted in $T$. (3) Select one node $v$ with the smallest distance to $u$ from $T$, and add it to $S$. Then, the neighbors of $v$ are inserted into $T$. Here, different indexing methods add different restrictions, pruning unnecessary vertices to be inserted in $T$, to improve the query processing performance. (4) Repeat (3) until $|S| = k$.

Specifically, \kw{IER} \cite{papadias2003query} uses Euclidean distance as a pruning bound to acquire the $k$NN results.  \kw{INE}\cite{papadias2003query} improves \kw{IER}'s Euclidean distance bound by expending searching space from the query location. \cite{demiryurek2009efficient} adapts a Euclidean restriction-based method to deal with continuous $k$ nearest neighbor problem. \cite{demiryurek2009efficient} divides the map into $N \times N$ grids and records which vertices and edges belong to some grid. Given a query vertex,  the fixed distance between grids is used to filter a proximate range. \kw{ROAD} \cite{lee2010road}  separates the input graph $G$ into many subgraphs hierarchically and skips the subgraphs without candidate objects  to speedup $k$NN query processing. \kw{G}-\kw{tree} \cite{zhong2015g} adapts a binary tree division method to divide a graph into two disjoint subgraphs recursively until the number of vertices in a tree node is smaller than a predefined parameter. In each subgraph, \kw{G}-\kw{tree} maintains a distance matrix which stores distance between borders and vertices, which is used to prune unnecessary vertex exploration during the \kw{Dijkstra} search. \kw{V}-\kw{tree} \cite{shen2017v} constructs a similar structure as \kw{G}-\kw{tree} but adds additional $k$ nearest objects for borders, which leads to a faster query processing than \kw{G}-\kw{tree}. Based on the contraction hierarchy (\kw{CH}) \cite{geisberger2008contraction}, \kw{TOAIN} \cite{luo2018toain} constructs a $k$DNN index recording the top-$k$ nearest neighbors for each vertex $u$ from objects whose ranks are lower than $u$, where the rank is defined by the contraction hierarchy. To answer a $k$NN query with vertex $u$, \kw{TOAIN} performs \kw{Dijkstra} search from $u$ following the \kw{CH} and maintains a candidate result set $R$. When visiting a vertex $v$,  if there is a vertex $w$ in the $k$DNN of $v$ such that the distance of $w$ and $u$ is smaller than the $k$-th distance to $u$ in $R$, \kw{TOAIN} updates $R$. The processing finishes when the \kw{Dijkstra} search is far enough or all vertices are explored. Although the methods  design different pruning algorithms to reduce  the \kw{Dijkstra} search space in step (3), the number of explored vertices cannot be well-bounded. In worst case, these methods degenerate into \kw{Dijkstra}'s algorithm, which leads to long query processing delay unavoidably. For \kw{TOAIN}, asit constructs $k$DNN based on \kw{CH}, which causes a relatively huge index size. Additionally, the vertex ranking method in \kw{TOAIN} employs \kw{Dijkstra}'s Algorithm, which incurs an expensive time cost regarding index construction. The experimental results of \cite{ouyang2020progressive} also verify above discussions.

Apart from the \kwnospace{Dijkstra}-search enhanced solutions, \cite{DBLP:conf/icde/Li0QHDY18} exploits the massive parallelism of GPU to accelerate the $k$NN query processing. \kw{GLAD} \cite{he2019efficient} partitions the road network into $2^x \times 2^x$ girds based on the geographical coordinate of each vertex. When answering a $k$NN query, it starts the search from  the grid containing the query vertex and updates the candidate result via probing vertices in neighbor grids iteratively. It avoids the exploration to the vertices in a grid  if the minimum Euclidean distance between any vertex inside the grid and the query vertex is not less than the largest distance in the candidate result. As \kw{GLAD} needs to use \hhindex to compute the exact shortest distance to select the final exact $k$NN results, the query processing is long. Moreover, since \kw{GLAD} depends on \hhindex, the index size of \kw{GLAD} is huge and the indexing time of \kw{GLAD} is long, which are similar to \tenindex \cite{ouyang2020progressive}. \tenindex \cite{ouyang2020progressive} is the state-of-the-art approach to $k$NN query in road network, which has been discussed in \refsec{existing}. \cite{li2023efficient} extend \tenindex \cite{ouyang2020progressive} and \kw{GLAD} \cite{he2019efficient} onto time-dependent road networks. \cite{jiang2023graph} extends tree decomposition method \cite{ouyang2018hierarchy} to deal with $k$NN search on flow graph.

Besides, continuous $k$NN query problem on road network is also studied in the literature \cite{DBLP:journals/geoinformatica/ShahabiKS03,DBLP:journals/geoinformatica/KolahdouzanS05,cho2005efficient,DBLP:conf/vldb/MouratidisYPM06,DBLP:conf/ssd/DemiryurekKS09,DBLP:journals/mis/ChoKC13,DBLP:conf/icde/ZhengZXSYZL16,DBLP:conf/vldb/KolahdouzanS04, jiang2023graph, jiang2021tree}.  Different from our setting,  these studies generally assume that the query vertex is moving on the road network, and  thus are orthogonal to ours. As a result,  the proposed techniques in these studies  cannot be used to address our problem.

\section{Conclusion}
\label{sec:conclusion}

Motivated by existing complex-index-based approaches for classical top $k$ nearest neighbors search in road networks suffers from the  long query processing delay, oversized index space, and prohibitive indexing time, we embrace minimalism and design a simple index for $k$NN query. The index has a \emph{well-bounded} space and supports progressive and optimal query processing. Moreover, we further design efficient algorithms to support the index construction. Experimental results demonstrate the significant superiority of our index over the state-of-the-art approach.

\balance
\bibliographystyle{ACM-Reference-Format}

\bibliography{KNN}

\end{document}